\documentclass[twocolumn,times]{aastex631}

\usepackage{threeparttablex}
\usepackage{amsmath}

\shorttitle{The SEDs and bolometric luminosities of local AGN}
\shortauthors{Spinoglio, Fern\'andez-Ontiveros \& Malkan}
\usepackage{subfigure}
\usepackage{url}

\usepackage{afterpage}

\graphicspath{{./}{figures/}}

\begin{document}

\title{The spectral energy distributions and the bolometric luminosities of local AGN:\\ study of the complete 12$\mu$m AGN sample}


\author[0000-0001-8840-1551]{Luigi Spinoglio}
\affiliation{Istituto di Astrofisica e Planetologia Spaziali (INAF--IAPS), Via Fosso del Cavaliere 100, I--00133 Roma, Italy}

\author[0000-0001-9490-899X]{Juan Antonio Fern\'andez-Ontiveros}
\affiliation{Centro de Estudios de F\'isica del Cosmos de Arag\'on (CEFCA), Plaza San Juan 1, E--44001, Teruel, Spain}
\affiliation{Istituto di Astrofisica e Planetologia Spaziali (INAF--IAPS), Via Fosso del Cavaliere 100, I--00133 Roma, Italy}

\author[0000-0001-6919-1237]{Matthew A. Malkan}
\affiliation{Department of Physics and Astronomy, UCLA, Los Angeles, CA 90095-1547, USA}

\begin{abstract}
We measure the bolometric luminosity of a complete and unbiased 12$\mu$m-selected sample of active galactic nuclei (AGN) in the local Universe. 
For each galaxy we used a 10-band radio-to-X-ray Spectral Energy Distribution (SED) 
to isolate the 
genuine AGN continuum in each band, including sub-arcsecond measurements where available, and correcting those contaminated by the host galaxy. We derive the median SED of Seyfert type 1 AGN, Seyferts with hidden broad-lines (HBL), Seyferts of type 2, and LINER nuclei in our sample. The median Seyfert 1 SED shows the characteristic blue bump feature in the UV, but nevertheless the largest contribution to the bolometric luminosity comes from the IR and X-ray continua. The median SEDs of both HBL and type 2 AGN are affected by starlight contamination in the optical/UV. The median SED of HBL AGN is consistent with that of Seyfert 1's, when an extinction of $A_V\sim1.2\,\rm{mag}$ is applied. The comprehensive SEDs allowed us to measure accurate bolometric luminosities and derive robust bolometric corrections for the different tracers. The 12$\mu$m and the \textit{K}-band nuclear luminosities have good linear correlations with the bolometric luminosity, similar to those in the X-rays. We derive bolometric corrections for either continuum bands (K-band, 12$\mu$m, 2-10\,keV and 14-195\,keV) and narrow emission lines (mid-IR high ionization lines of [OIV] and [NeV] and optical [OIII]5007\AA) as well as for combinations of IR continuum and line emission. A combination of continuum plus line emission accurately predicts the bolometric luminosity up to quasar luminosities ($\sim\,10^{46}\,\rm{erg\, s^{-1}}$).
\end{abstract}


\section{Introduction} \label{sec:intro}
The determination of the bolometric luminosity of active galactic nuclei (AGN) is important, to study the AGN phenomenon in itself, which is sustained by an accreting supermassive black hole \citep[SMBH; see, e.g.,][]{rees1984} and, on the other hand, to understand the role of AGN in galaxy evolution.

The need to place the Seyfert nucleus phenomenon in the framework of galaxy evolution was already clear more than 40 years ago.
It was realized that the most universal major contributions to the bolometric luminosity of AGN come from the hard X-ray and the mid-IR bands. For the former, we know that the X-ray band emission is characteristic of AGN: X-ray emission is ubiquitous in AGN and is thought to be produced by comptonization of UV/optical disc photons by a corona of hot electrons located above the black hole.
For the latter, in fact, the mid-IR, and in particular the IRAS 12$\mu$m band has been used to select a complete sample of AGN \citep{spinoglio1989, rush1993, spinoglio1995} because it was realized that this band was carrying a constant fraction of the bolometric luminosity of different types of low redshift AGN, including Seyfert galaxies of type 1 and type 2, and quasars.
This result arises from the fact that AGN energy distributions are roughly described by a one-parameter family: a relatively blue quasar continuum altered by varying amounts of nuclear dust. 
The dust preferentially absorbs continuum at the shortest wavelengths and re-emits it in the far-infrared. These two processes make the resulting energy distribution redder in the optical/ultraviolet, and in the IR. However, there is a pivot wavelength, at which the absorption of the original continuum is balanced by the thermal re-emission. This occurs in the mid-IR and in fact, 7-12 $\mu$m is the only wavelength range where both of these effects (absorption and re-emission) are relatively small \citep{spinoglio1989}. 

This study was based on the large-aperture IRAS data \citep{Neugebauer1984}, possibly contaminated by emission from the host galaxy. It therefore needed to be re-examined using mid-IR observations of the intrinsic nuclear emission of the AGN. 

The first determination of the bolometric luminosity of low redshift quasars has been done by \citet{Sanders1989} who presented multi-wavelength observations (10$^{9.7}$--10$^{18}\, \rm{Hz}$) of 109 bright quasars from the Palomar-Green (PG) survey \citep{Schmidt1983} and showed that the PG quasars emit the bulk of their luminosity (typically more than 90\%) between 3000\AA \ and 300$\mu$m. A few years later, 
\citet{elvis1994} integrated the energy distributions of 47 quasars selected mainly from the PG, 3C, and Parkes catalogs, using a simple linear interpolation through the data points in log $\nu$L$_{\nu}$ space, i.e., connecting the individual points with a power law and derived bolometric corrections for UV, visible, and infrared monochromatic luminosities. They lacked complete data from the hard X-ray bands, and their sample was primarily an ultraviolet-excess selected sample, and biased toward relatively X-ray loud quasars.

Following that pioneering work 
there have been a few determinations of the bolometric luminosity of AGN, either from selective measurements of the continuum in various spectral ranges and assuming templates to reproduce the spectral energy distributions (SED) of AGN \citep[e.g.][]{marconi2004, Hopkins2007, lusso2012}, or from observations of spectral lines \citep[e.g.][]{melendez2008, spinoglio2022}. These estimates assume that the measured radiation is isotropic and that simple bolometric corrections can recover the total luminosities of most AGN.


\citet{marconi2004} used the following approach to derive the bolometric corrections, i.e. the corrections to be applied to an {\it observed} luminosity to derive the {\it total intrinsic}, i.e. bolometric, luminosity of the AGN. 
Their argument is that although the observed SED of an AGN provides the total observed luminosity (L$_{obs}$), 
this does not give an accurate estimate of the BH mass accretion rate because it often includes reprocessed radiation (i.e. radiation absorbed along other lines of sight and re-emitted isotropically). In their view, the accretion rate is related to the total luminosity directly produced by the accretion process, which they call the total intrinsic luminosity L$_{int}$. In AGN, L$_{int}$ is given by sum of the optical–ultraviolet and X-ray luminosities radiated by the accretion disc and hot corona, respectively. 
They consider that, in order to estimate L$_{int}$, one has to remove the IR bump 
(2.5--60 microns) from the observed SED of unobscured AGN to avoid double counting the source luminosity, i.e. extinction-corrected X-ray emission and reprocessed IR emission. 
\citet{marconi2004} used a template spectrum for the AGN, 
based on the observed optical-UV and X-ray spectra, 
excluding the observed IR bump, following the above argument. 
From the assumed template they compute the bolometric corrections with respect to the optical B-band, and the soft and hard X-ray bands. 

With the aim of determining the bolometric quasar luminosity function, \citet{Hopkins2007} have taken a similar approach to the one of \citet{marconi2004} in  calculating the template spectrum, but they included in the template the observed IR bump, giving a more detailed treatment of the optical/IR. In particular, they used the \citet{Richards2006} template spectrum in the B, \textit{g} and \textit{i} optical bands and a different determination of the spectral index between optical and X-rays ($\alpha_{ox}$).

They also found a dependence of the bolometric correction on the luminosity. 
They used a compilation of many AGN samples, selected from the optical, the soft X-rays, the hard X-rays and also at 8-15 $\mu$m, including the $12\, \rm{\micron}$ AGN sample \citep{rush1993}. 

\citet{Vasudevan2007} showed that the variations in the disc emission in the ultraviolet (UV) are important by construction of optical-to-X-ray SED for 54 AGN, using Far Ultraviolet Spectroscopic Explorer (FUSE) UV and X-ray data from the literature to constrain the disc emission as well as possible. They also found evidence for very significant spread in the bolometric corrections, with no simple dependence on luminosity being evident. Populations of AGN such as narrow-line Seyfert 1 nuclei \citep{Osterbrock1985}, 
radio-loud and X-ray-weak AGN may have bolometric corrections which differ systematically from the rest of the AGN population. 

\citet{lusso2012} obtained accurate estimates of bolometric luminosities, bolometric corrections and Eddington ratios of a large X-ray selected sample of broad-line (Type-1) and narrow-line (Type-2) AGN from the COSMOS \citep{scoville2007} XMM–Newton survey for which extensive data from the far-IR to the optical and ultraviolet regimes were available. They assumed that the intrinsic nuclear luminosity is the sum of the IR and X-ray luminosities ($L_{\rm bol}$ = L$_{IR}$ + L$_{X}$).
They used a SED fitting code to disentangle the various contributions (using starburst, AGN and host-galaxy templates) in the observed SED by using a standard $\chi^2$ minimization procedure and they integrated the nuclear component from 1 to 1000\,$\mu$m to obtain the total IR luminosity L$_{IR}$.
To derive the nuclear accretion disc luminosity from this value, they applied a geometric correction factor based on the models by \citet{pier1992} to account for the torus geometry and their associated anisotropy \citep[
see ][]{pozzi2010}. 
The total X-ray luminosity L$_{X}$ is estimated by integrating the X-ray SED in the 0.5–100 keV range. They found that the bolometric correction is significantly lower at high luminosities with respect to the previous estimates of \citet{marconi2004} and \citet{Hopkins2007}. The main limitation of the study of \citet{lusso2012} is that it is entirely based on an X-ray selected sample of AGN, and therefore can be biased towards X-ray bright AGN with most of the sources showing relatively low hydrogen column densities (N$_H\lesssim 10^{23.5}\, \rm{cm^{-2}}$), thus missing heavily absorbed -- Compton thick -- nuclei. 

Other attempts have been made to derive the AGN bolometric luminosity from  optical and IR lines originated in the narrow line regions (NLR) of AGN. \citet{melendez2008} have studied the relations of the [OIV]26\,$\mu$m and [OIII]5007\,\AA \ luminosities with the 2-10\,keV and 14-195\,keV luminosities. They concluded that [OIV] is a good indicator of the AGN power. \citet{rigby2009} calibrate the [OIV]26\,$\mu$m line as a measure of AGN intrinsic luminosity. Finally, recently  \citet{spinoglio2022} have computed the AGN bolometric luminosities using the \citet{lusso2012} bolometric correction and the corrected (2–10)\,keV luminosities as a function of the luminosities of the three mid-IR high-ionization lines of [OIV]26\,$\mu$m, [NeV] 14.3\,$\mu$m, and [NeV]24.3\,$\mu$m.

Our approach to measure the bolometric luminosity of AGN is different from the other derivations summarized above: 

(i) we do not assume an {\it a priori} template AGN SED as was done in the other derivations \citep[e.g.,][]{marconi2004,Hopkins2007}, but we have defined ten photometric bands from the radio to the X-rays to cover the SED of AGN of the chosen local AGN sample and we computed the correlations first among the various bands and then, after having integrated a total {\it bolometric} nuclear luminosity for as many AGN in the sample as possible, we computed the correlations of the various bands, and combinations of them, with the estimated bolometric luminosity;

(ii) with respect to the work of \citet{marconi2004}, we now include sub-arcsecond measurements of the IR bump, which might be {\it also} due to reprocessed radiation re-radiated by hot dust, because --- even if reprocessed --- this radiation ultimately originates in the accretion process in the AGN;

(iii) compared to the other works, we separated the nuclear emission, due to the AGN, from the extended emission, due to stellar emission in the galaxy; 

(iv) 
we also use the mid-IR spectroscopy of the lines of [OIV]26$\mu$m and [NeV]14.3$\mu$m and 24.3$\mu$m, which are mainly emitted in the narrow line regions excited by the accreting SMBH. 

(v) our sample includes only local AGN, avoiding possible cosmic evolution in the bolometric corrections derived when different samples obtained at different redshift intervals are combined.

\section{Sample of local Active Galactic Nuclei and multifrequency dataset}

The sample of local 
AGN used to compute the bolometric luminosity is the $12\, \rm{\micron}$ AGN sample (hereafter 12MAGN), 
originally selected from the IRAS catalog from \citet{spinoglio1989} and updated in \citet{rush1993}.
The chosen sample is essentially a bolometric flux-limited survey outside the galactic plane, and therefore largely unbiased, given the empirical fact that galaxies emit an approximately constant fraction of their total bolometric luminosity at $12\, \rm{\micron}$. This fraction is $\sim 15\%$ for AGN, including Seyfert type 1, type 2 and quasars  \citep{spinoglio1989}, and $7\%$ for normal and starburst galaxies, independent of star formation activity \citep{spinoglio1995}. The sample is given in Table \ref{tab:sample0} and it is the same as the one used in \citet{spinoglio2022}. 
This sample has been selected from the 12$\mu$m flux-limited survey of 893 galaxies \citep{rush1993} extracted from the IRAS Faint Source Catalog, Version 2 \citep[FSC-2;][]{Moshir1990} with the classification of galaxy nuclear activity with catalogs of active galaxies available at the time of the selection \citep{Veron1991, Hewitt1991, Helou1991}. 
It contains in total 117 AGN, divided into 48 Seyfert type 1, 27 Hidden Broad Line galaxies (hereafter HBL), 30 Seyfert type 2 and 10 Low Ionization Emission Line galaxies (hereafter LINER) and two other galaxies, one classified as ``non-Seyfert'' (NGC\,1056) and one as a starburst galaxy (NGC\,6810). We have adopted here the class of HBL galaxies because they are indistinguishable in their IR properties from the Seyfert type 1 galaxies, and therefore belong to the broader class of type 1 AGN \citep[see, e.g.][]{tommasin2010}. 
The 10 radio-loud objects in the sample are flagged in Table \ref{tab:sample0}.
The average redshift(s) and dispersion(s) for the whole sample is 
$<z>=0.025\pm0.029 $, while for the four classes of galaxies considered are: for Sy1: $<z>=0.032\pm0.039 $, for HBL: $<z>=0.023\pm0.019 $, for Sy2: $<z>=0.019\pm0.016 $ and for LIN: $<z>=0.016\pm0.021 $.


To compute the bolometric luminosity of local AGN, we have considered the following photometric bands, from low to high frequencies: radio 8.4\,GHz, mid-IR 12\,$\mu$m, 7\,$\mu$m, 5.8\,$\mu$m and 4.5\,$\mu$m bands, near-IR 2.2\,$\mu$m, the near ultraviolet (NUV) Galaxy Evolution Explorer \citep[GALEX,][]{morrissey2007} band at an effective wavelength of 2316\,\AA \ and the far ultraviolet (FUV) GALEX band at 1539\,\AA, and the 2-10\,keV and 14-195\,keV hard X-ray bands.

Our main concern in this  analysis has been to derive, from the available observations, the {\it nuclear} fluxes of the AGN of our sample, because many local AGN are {\it contaminated} by the galactic component, which at some frequencies can be dominant. 
In order to isolate the nuclear emission from the total observed emission in the various mid-IR bands that we have considered, we have used the sub-arcsecond observations at 12$\mu$m to correct the other observations and we have corrected the near-IR K-band observations as well as the ultraviolet fluxes. However, the 8.4\,GHz flux densities are already detecting the nuclei at sub-arcsecond resolution and the hard X-ray emission is intrinsically produced mainly by the AGN.

From the low frequency data (radio) to the high energy data (X-rays), we have adopted the following methodology:\\

(i): the radio data at 8.4\,GHz have been taken from the VLA \citep{Thompson1980} with a resolution of 0$^{\prime\prime}$.25 \citep{Thean2000} and therefore are sampling the nuclear emission; at the average distance of the whole AGN sample, the angular distance of 0$^{\prime\prime}$.25 corresponds to a linear  distance of 134 pc, while at the average distance of Seyfert 1 to 142 pc, at the average distance of Seyfert 2 to 102 pc and at the average distance of LINER to 86 pc, thus confirming that the radio VLA observations are mainly sampling the nuclear emission; the presence of a jet core in the data would in any case be due to the AGN;\\

(ii): the 12\,$\mu$m continuum flux density has been taken from VLT-VISIR \citep{Lagage2004} sub-arcsecond resolution observations \citep{Asmus2014}; when these observations were not available, we used the small aperture $\sim$10\,$\mu$m observations from \citet{Gorjian2004} and in only two cases from \citet{Rieke1978} (see Table \ref{tab:sample1});\\

(iii): the 7\,$\mu$m and 12.8\,$\mu$m continuum fluxes have been derived using {\it Spitzer}-IRS \citep{Houck2004} spectroscopy of the [ArII]7.0\,$\mu$m line and of the [NeII]12.8\,$\mu$m line, respectively, using the line intensities together with the measured equivalent widths from \citet{Gallimore2010}, these data include galactic emission and therefore need to be corrected;\\

(iv): the {\it Spitzer}-IRS data at 7\,$\mu$m, and the {\it Spitzer}-IRAC \citep{Fazio2004} data at 5.8\,$\mu$m and 4.5\,$\mu$m have been corrected using the observed ratio of the {\it Spitzer}-IRS 12.8\,$\mu$m continuum to the VLT-VISIR 12\,$\mu$m sub-arcsecond continuum flux density (or the small aperture $\sim$10\,$\mu$m flux density, see the above point ii);\\

(v): the K-band nuclear flux has been derived from Keck and 2MASS observations as reported and fully explained in \citet{spinoglio2022};\\

(vi): the GALEX near-ultraviolet (NUV) and far-ultraviolet (FUV) fluxes have been derived using the GALEX point spread functions of the two UV bands, following the work of \citet{morrissey2007};\\

(vii): the 2-10\,keV and 14-195\,keV hard X-ray data are nuclear and emitted by large by the AGN, so we assume that they do not need to be corrected. The compilation for the 12\,$\mu$m sample has been taken from \citet{spinoglio2022} and the observations have been reported in \citet{ricci2017} and \citet{oh2018}.

\begin{table}
\centering
\begin{ThreePartTable}
\setlength{\tabcolsep}{2.pt}
\setlength{\LTcapwidth}{\columnwidth}
\footnotesize
\caption{The AGN sample: coordinates, redshift, types and radio loudness.}\label{tab:sample0}
\begin{tabular}{rlccccc}
\hline\\[-0.3cm]
\bf n. & \bf Name         & \bf RA (J2000.0)  & \bf dec (J2000.0) & \bf $z$     & \bf Type & Radio \\
       & (1)             & (2)               & (3)               & (4)         & (5)  & (6)    \\
\hline\\[-0.4cm]
  1 &  MRK0335          &  00:06:19.5  & +20:12:10  &  0.0258  & Sy1  & \\  
  2 &  NGC34=Mrk938=N17 &  00:11:06.5  & -12:06:26  &  0.0196  & Sy2 &  \\  
  3 &  IRASF00198-7926  &  00:21:57.0  & -79:10:14  &  0.0728  & Sy2 &  \\  
  4 &  ESO012-G021      &  00:40:47.8  & -79:14:27  &  0.0300  & Sy1 &  \\  
  5 &  NGC0262=MRK348   &  00:48:47.1  & +31:57:25  &  0.0150  & HBL & RL  \\
\hline
\end{tabular}
\begin{tablenotes}
\item {\bf Notes:} (This table is available in its entirety in machine-readable form).\\
The columns give for each AGN in the sample: (1) name; (2) Right Ascension (J2000.0); (3) declination (J2000.0); (4) redshift from NED; (5) AGN type: Sy1 = Seyfert type 1; Sy2 = Seyfert type 2; HBL = Hidden Broad Line Region AGN; LIN = LINER galaxy;  (6) Radio loudness (flagged as RL): an AGN is considered radio loud if the radio flux is larger than ten times the B-band optical flux \citep{Kellermann1989}($F({\rm 5GHz}) \geq 10 \times F_{B-band}$).
\end{tablenotes}
\end{ThreePartTable}
\end{table}

\subsection{Deriving the nuclear fluxes for the mid-IR and UV bands}

To build up the AGN bolometric luminosity, we have compiled, for our sample of galaxies, in Table \ref{tab:sample1} the low-frequency (from radio to near-IR) nuclear fluxes and in Table \ref{tab:sample2} the high-frequency (from UV to X-rays) nuclear fluxes. In Table \ref{tab:sample1} we give, for each galaxy of the 12MAGN sample observed, the 8.4\,GHz radio flux, measured with the VLA \citep{Thean2000} with a 0$^{\prime\prime}$.25 effective aperture, the 12\,$\mu$m nuclear flux density measured with the VLT-VISIR in most cases \citep{Asmus2014} (where these data were not available, we have used the 10\,$\mu$m {\it small aperture} flux taken from \citet{Gorjian2004, Rieke1978} and inserted a note in Table \ref{tab:sample1}).

\begin{table*}
\centering
\begin{ThreePartTable}
\setlength{\tabcolsep}{2.pt}
\setlength{\LTcapwidth}{\textwidth}
\footnotesize
\caption{Low frequency nuclear photometry of the AGN sample. }\label{tab:sample1}
\begin{tabular}{rlcccccccccc}
\hline\\[-0.2cm]    
\bf n. & \bf Name & \bf \bf Redshift  & \bf Type & \bf F(8.4GHz) & \bf F(12$\mu$m) & \bf F(12$\mu$m) & \bf R12          & \bf F(7.0$\mu$m) & \bf F(5.8$\mu$m) & \bf F(4.5$\mu$m) & \bf F(K)  \\[0.05cm]
    &      &      &      &  VLA          &   VLT     &   IRS   &  IRS/VLT & IRS$_{\rm COR}$   & IRS$_{\rm COR}$        & IRS$_{\rm COR}$  & Keck/2MASS    \\[0.05cm]
    &      &      &      &  \multicolumn{3}{c}{(mJy)}          &          & \multicolumn{4}{c}{(mJy)}                                         \\[0.05cm]
    & (1)  & (2)  & (3)  & (4)  & (5)  &  (6)                  & (7)      & (8)           & (9)                & (10)         & (11)   \\[0.1cm]
\hline\\[-0.2cm]
   1 & MRK0335            &  0.0258 & Sy1 &     2.1 &    151.$\dag$  &   190.52 &    1.26   &  100.32 &  77.91 &  67.53 &   16.00    \\  
   2 & NGC34=Mrk938=N17   &  0.0196 & Sy2 &    14.5 &    57.8  &   210.38 &    3.64   &    ---  &   ---  &   ---  &    5.25    \\  
   3 & IRASF00198-7926    &  0.0728 & Sy2 &    ---  &    ---   &    ---   &    ---    &    ---  &   ---  &   ---  &    2.15    \\  
   4 & ESO012-G021        &  0.0300 & Sy1 &    ---  &    ---   &   115.67 &    ---    &    ---  &   ---  &   ---  &    5.10    \\  
   5 & NGC0262=MRK348     &  0.0150 & HBL &   346.0 &    117.  &   304.83 &    2.61   &   47.89 &  37.73 &  28.48 &    2.17    \\  
\hline \\[0.1cm]
\end{tabular}
\begin{tablenotes}
\scriptsize
\item {\bf Notes:} (This table is available in its entirety in machine-readable form).\\
The columns give for each AGN in the sample: (1) name; (2) redshift from NED; (3) AGN type: Sy1 = Seyfert type 1; Sy2 = Seyfert type 2; HBL = Hidden Broad Line Region AGN; LIN = LINER galaxy; (4) 8.4GHz nuclear flux density \citep{Thean2000} ; (5) VLT-VISIR subarcsecond nuclear 12$\mu$m flux density from \citet{Asmus2014}, where not available the small aperture 10$\mu$m flux density has been taken from \citet{Gorjian2004, Rieke1978}; (6) {\it Spitzer}-IRS 12$\mu$m continuum flux density as derived from \citet{Gallimore2010}, using the [NeII]12.8$\mu$m line flux and the corresponding equivalent width; (7) R12 is the ratio between the {\it Spitzer}-IRS continuum and the VLT-VISIR subarcsecond flux density at 12$\mu$m; (8) {\it Spitzer}-IRS 7$\mu$m continuum flux density as derived from \citet{Gallimore2010}, using the [ArII]7.0$\mu$m line flux and the corresponding equivalent width corrected for R12; (9) {\it Spitzer}-IRAC 5.8$\mu$m flux density corrected for R12; (10) {\it Spitzer}-IRAC 4.5$\mu$m flux density corrected for R12. This table is available in its entirety in machine-readable form. $\dag$: the 10.8$\mu$m flux density is from \citet{Gorjian2004}.
\end{tablenotes}
\end{ThreePartTable}
\end{table*}

We have then extracted from the fluxes and equivalent width measurements of the two fine structure lines of [ArII]\,7.0$\mu$m and [NeII]\,12.8$\mu$m from \citet{Gallimore2010}, who made systematic {\it Spitzer} observations of the 12MAGN sample, the corresponding flux densities of the continuum at 7.0\,$\mu$m and 12.8\,$\mu$m.  The spectral information was 
extracted from synthetic, 20$^{\prime\prime}$ diameter circular apertures centered on the brightest, compact IR source, and therefore the derived continuum flux densities are relative to the extended 20$^{\prime\prime}$ emission. In order to remove the extended emission and obtain {\it nuclear} flux densities, we have computed the ratio of the 12.8\,$\mu$m 20$^{\prime\prime}$ diameter flux density from {\it Spitzer}-IRS to the nuclear 12\,$\mu$m flux density measured by VLT-VISIR \citep{Asmus2014}. We ignored the small difference in wavelength between the two observations. Using the same ratio, R12 in Table \ref{tab:sample1}, we have corrected the 4.5\,$\mu$m and 5.8\,$\mu$m flux densities from {\it Spitzer}-IRAC, measured by \citet{Gallimore2010} using a synthetic, 20$^{\prime\prime}$ diameter circular aperture centered on the brightest infrared source associated with the active galaxy. We have also corrected the 7.0\,$\mu$m continuum derived from the {\it Spitzer}-IRS observations of the [ArII]7.0\,$\mu$m line, as described above. We have included in Table \ref{tab:sample1} the corrected 7.0\,$\mu$m, 5.8\,$\mu$m and 4.5\,$\mu$m flux densities only for objects for which the R12 ratio is less than 3 (R12\,$<$\,3), discarding all objects for which this ratio is greater than this threshold. This latter has been chosen because we have estimated reasonable a correction only below this threshold. The number of objects above this threshold (with R12\,$>$\,3) is 14 plus two galaxies with upper limits in the VLT-VISIR data (see Table \ref{tab:sample1}).

The near-IR {\it nuclear} K-band flux density has been derived in \citet{spinoglio2022} and we refer to this work for the details.

The high-frequency (from UV to X-rays) nuclear flux densities reported in Table \ref{tab:sample2} include the NUV and FUV ultraviolet flux densities derived from GALEX observations and the 2-10\,keV and 14-195\,keV band hard X-ray fluxes, together with the photon indexes $\Gamma_1$  and $\Gamma_2$ relative to the 2-10\,keV and 14-195\,keV observations, respectively.
The Galaxy Evolution Explorer (GALEX) \citep{morrissey2007} data have been extracted from the Mikulski Archive for Space Telescopes \citep[MAST, ][]{conti2011} GALEX Catalog Search\footnote{\url{http://galex.stsci.edu/GR6/?page=mastform}}. 

For each AGN in our sample, we have extracted the NUV and FUV fluxes and errors in the 4 smallest apertures and the given values of the galactic interstellar reddening E(B-V). 
The GALEX absolute calibration and extinction coefficients have been taken from \citet{morrissey2007} and \citet{wall2019}, respectively. In order to obtain the best approximation of the {\it nuclear} UV fluxes, we have used the fluxes from APER\_1 to APER\_3, corresponding to radii of 1$^{\prime\prime}$.5 to 3$^{\prime\prime}$.8.
For the 23 AGN for which upper limits were present in the APER\_1 FUV flux, we have taken the APER\_2 flux, for the 9 AGN for which also the APER\_2 flux was an upper limit, we have taken the APER\_3 flux, while when also this flux was an upper limit we have given an upper limit to the FUV flux. The APER\_n (with n=1--3) fluxes have then been corrected using the GALEX photometric {\it curve of growth} published in \citet{morrissey2007}, to derive the nuclear fluxes.

The 2-10\,keV X-ray fluxes, corrected for absorption, and the 14-195\,keV band hard X-ray fluxes have been taken from the references given in Table \ref{tab:sample2}. Because these fluxes are intrinsically produced by the AGN, no further correction should be needed.

\begin{table*}
\centering
\begin{ThreePartTable}
\setlength{\tabcolsep}{2.pt}
\setlength{\LTcapwidth}{\textwidth}
\footnotesize
\caption{High frequency nuclear photometry of the AGN sample and derived bolometric fluxes and luminosities. }\label{tab:sample2}
\begin{tabular}{rlccccccccccccc}
\\ \hline\\[-0.2cm]    
\bf n. &\bf Name   & $z$     & \bf Type &  \bf F$_{NUV}$ & \bf F$_{FUV}$ & \bf F$_{2-10keV}$    & \bf F$_{14-195keV}$ & \bf F$_{bol}$   & \bf $L_{\rm bol}$ & $\Gamma_1$ & $\Gamma_2$  & Ref$_{1}$  & Ref$_{2}$ & Flag \\[0.05cm]
         &                  &            &   &          \multicolumn{2}{c}{(mJy)}   & \multicolumn{3}{c}{($10^{-12} \rm{erg s^{-1} cm^{-2}}$)}   &  log($\rm{erg\,s^{-1}}$) &   &    & \\[0.05cm]
         & (1)             & (2)      & (3)   & (4)  & (5)   &  (6)    & (7)             & (8)                & (9)   & (10)  & (11) & (12) & (13) & (14)   \\[0.1cm]
\hline\\[-0.2cm]
   1 & MRK0335            &  0.0258 & Sy1 &  3.517  & 1.880  &   16.69 &    15.97 &     402.72   &     44.751   &  2.03 & 2.31 &   (1)  & (32) &  \\  
   2 & NGC34=Mrk938=N17   &  0.0196 & Sy2 &  0.072  & 0.246  &    2.35 &      --- &     105.68   &     43.927   &  1.9  &  --- &   (2)  &      &  \\  
   3 & IRASF00198-7926    &  0.0728 & Sy2 &  ---    & 0.133  &    ---  &    21.80 &     148.25   &     45.248   &  ---  &  --- &        & (33) &  \\  
   4 & ESO012-G021        &  0.0300 & Sy1 &  ---    & ---    &    5.51 &     4.00 &      47.86   &     43.959   &  1.21 &  --- &   (3)  & (34) &  \\  
   5 & NGC0262=MRK348     &  0.0150 & HBL &  0.089  & 0.143  &   37.45 &   144.81 &     642.69   &     44.475   &  1.68 & 1.90 &   (1)  & (32) &  \\  
\hline \\[0.1cm]
\end{tabular}
\begin{tablenotes}
\scriptsize
\item \textbf{Notes.} The columns give for each AGN in the sample: (1) name; (2) AGN type: Sy1 = Seyfert type 1; Sy2 = Seyfert type 2; HBL = Hidden Broad Line Region AGN; LIN = LINER galaxy; (3) redshift; (4) (4) Near-ultraviolet flux (NUV) from the Mikulski Archive for Space Telescopes \citep[MAST, ][]{conti2011} GALEX Catalog Search; (5) Far-ultraviolet flux (FUV) from the Mikulski Archive for Space Telescopes \citep[MAST, ][]{conti2011} GALEX Catalog Search; (6) 2-10\,keV absorption corrected X-ray flux from the references of column (10); (7) 14-195\,keV observed X-ray flux from the references of column (11); (8) Computed bolometric flux (see text); (9) Logarithm of the total (bolometric) luminosity; (10) photon index of the 2-10\,keV observations $\Gamma_1$, from either \citet{Brightman2011} or \citet{ricci2017}; (11) photon index of the 14-195\,keV observations $\Gamma_2$, from \citet{oh2018}; (12) reference of the 2-10keV flux, Ref$_{1}$: (1): derived from the absorption corrected luminosity of \citet{Brightman2011}, (2): derived from the absorption corrected luminosity of \citet{Guainazzi2005}, (3): \citet{Ghosh1992}, (4): \citet{Reeves2000}, (5): derived from the absorption corrected luminosity of \citet{Tan2012}, (6): \citet{ricci2017}, (7): derived from the absorption corrected luminosity of \citet{Marinucci2012}, (8): \citet{Bi2020}, (9): derived from the absorption corrected luminosity of \citet{Iyomoto1996}, (10): derived from the absorption corrected luminosity of \citet{Saade2022}, (11): derived from the absorption corrected luminosity of \citet{Levenson2009}, (12): \citet{Rivers2013}, (13): \citet{Walton2021}, (14): \citet{Bassani1999}, (15): \citet{Tanimoto2022}, (16): \citet{DellaCeca2008}, (17): \citet{Iyomoto2001}, (18): \citet{Boissay2016}, (19): \citet{Brightman2008}, (20): derived from the absorption corrected luminosity of \citet{Akylas2009}, (21): \citet{Chen2022}, (22): derived from the absorption corrected luminosity of \citet{Zhou2010}, (23): \citet{Vasylenko2018}, (24): derived from the absorption corrected luminosity of \citet{Yamada2023}, (25): \citet{Lutz2004}, (26): \citet{Osorio-Clavijo2022}, (27): \citet{Cardamone2007}, (28): \citet{Osorio-Clavijo2023}, (29): \citet{Corral2014}, (30): \citet{Braito2009}, (31): \citet{Zhou2010}; (13) reference of the 14-195keV flux, Ref$_{2}$: (32): \citet{oh2018}, (33): \citet{Deluit2004}, (34): \citet{Cusumano2010}; (14) Bolometric flux/luminosity flag: L = lower limit to the bolometric flux and luminosity; N = data not available for a proper integration of bolometric flux and luminosity.
\end{tablenotes}
\end{ThreePartTable}
\end{table*}

\subsection{Building the bolometric luminosity from the available nuclear luminosities}

Once the nuclear fluxes were computed, we have used trapezoidal integration to calculate the total, or {\it bolometric}, luminosity for our AGN sample. 
The integration has been performed in two separate spectral regions: the first interval is from the radio (8.4\,GHz band) to the FUV band, while the second interval includes only the two X-ray bands (2-10\,keV and the 14-195\,keV). For those AGN where only one X-ray band flux was available, we computed the missing flux using the median spectral index between 2-10\,keV and the 14-195\,keV for each of the four AGN classes.  The resulting bolometric luminosity is simply the sum of these two integrations.  No interpolation has been included between the UV and X-rays data, because this spectral region is not observable and a power-law interpolation would likely have over-estimated the bolometric luminosity. 

To assign a bolometric luminosity to an AGN of our sample we require a minimum number of detections over the ten photometric bands considered. We require at least 3 available detections in three bands and that these must include either the 2-10\,keV or the 14-195\,keV X-ray band. When no X-ray data were available, we were nevertheless able to assign a lower limit to the bolometric flux and luminosity, because the low frequency domain, from the radio to the K-band, is carrying a substantial part of the total luminosity, as can be seen from the shape of the SEDs (see Figs. \ref{fig:Sy1_medianSED} - \ref{fig:lin_medianSED}).

The derived bolometric luminosities and bolometric fluxes for our sample are given in Table \ref{tab:sample2}. For 18 objects we have assigned lower limits to the bolometric flux and luminosity, namely ESO\,541-IG012, MRK\,1034NED02, NGC\,1056, NGC\,1241, IRASF\,03362-1642, IRASF\,0345+0055, 
ESO\,253-G003, IRASF\,07599+6508, IRASF\,08752+3915, MCG\,+00-29-023, NGC\,4602, MRK\,0231, NGC\,4922, NGC\,5005, NGC\,5953, ARP\,220, MKR\,0897, NGC\,7496 and CGCG\,381-051. 
For only two objects we were not able to assign even a lower limit to the bolometric luminosity and flux, namely NGC\,1056 and NGC\,3511, because of the lack of observations both in the mid-IR (only IRS large aperture data are available) and at X-rays.  



\begin{figure}[ht!]
\includegraphics[width = \columnwidth]{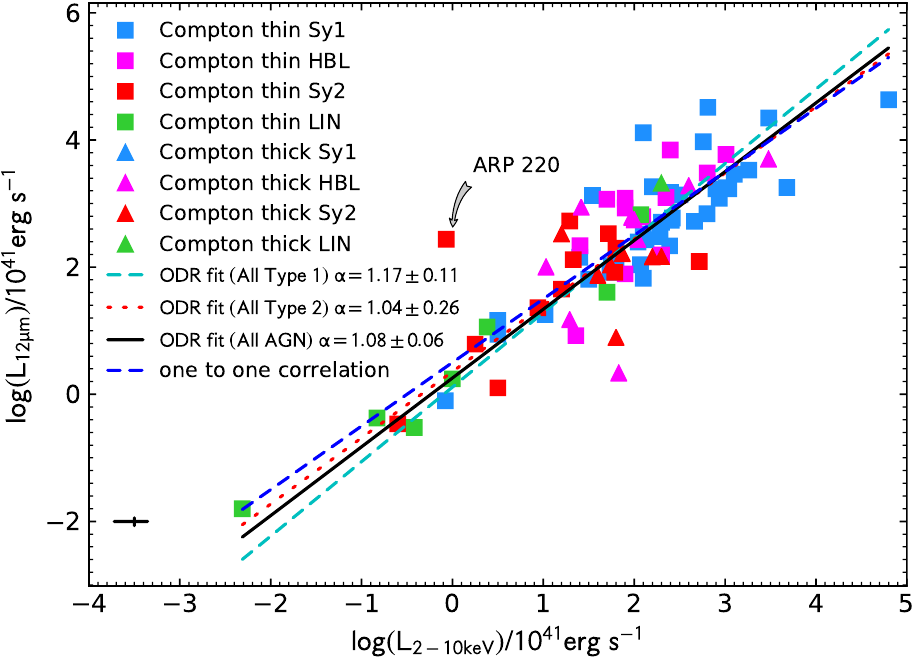}
\caption{
Nuclear 12$\mu$m luminosity as a function of the corrected 2-10\,keV luminosity. The position of the outlier ARP\,220 has been indicated. 
In each figure representing the correlations between physical quantities, the error bar at the lower left corner has been computed using the median value of the relative errors of the plotted data (from Fig. \ref{fig:L12_x2-10} to Fig. \ref{fig:F12_OIV_bolflux} and from Fig. \ref{fig:ourlbol_lusso} to Fig. \ref{fig:compositebolcor}).
\label{fig:L12_x2-10}}
\end{figure}

\begin{figure}[ht!]
\includegraphics[width = \columnwidth]{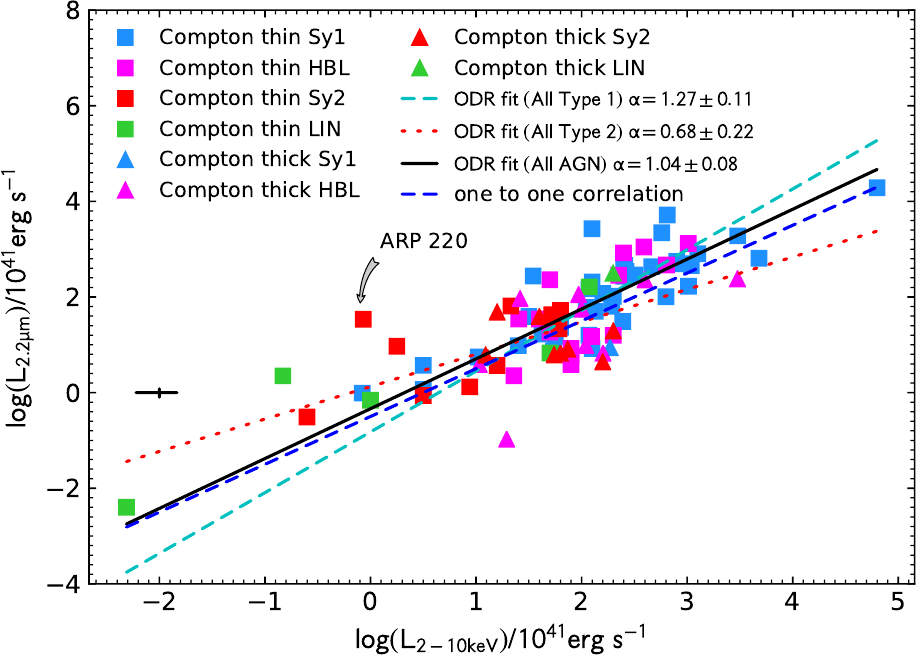}
\caption{Nuclear K-band luminosity as a function of the corrected 2-10\,keV luminosity. The position of the outlier ARP\,220 has been indicated.
\label{fig:LK_x2-10}}
\end{figure}

\begin{figure}[ht!]
\includegraphics[width = \columnwidth]{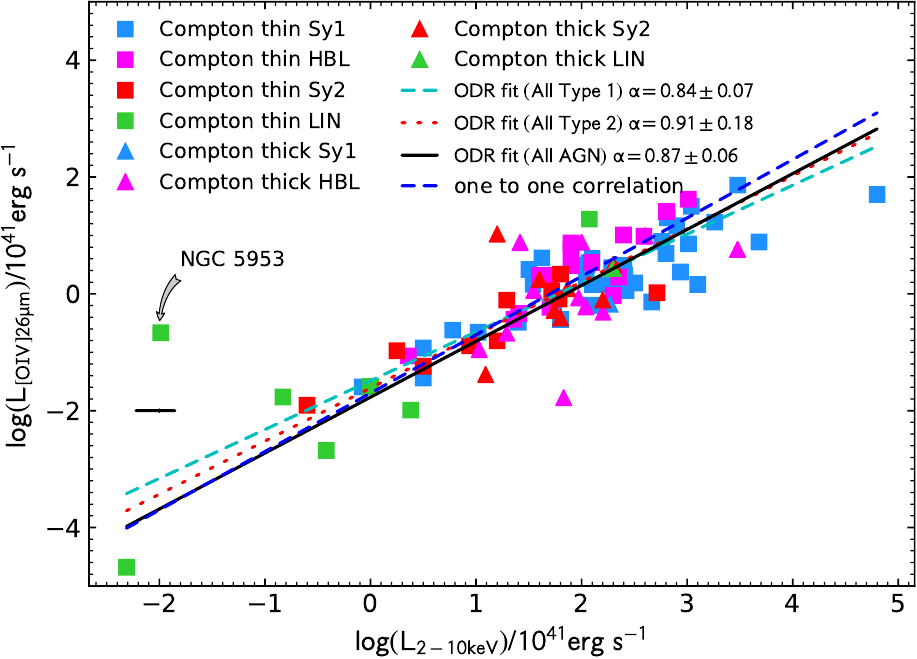}
\caption{[OIV]26$\mu$m line luminosity as a function of the corrected 2-10\,keV luminosity.  The position of the outlier NGC\,5953, which has been recently classified as a non-AGN \citep{Osorio-Clavijo2023}, has been indicated. 
\label{fig:O4_x2-10}}
\end{figure}

\section{Results}\label{results}

\subsection{Monochromatic luminosities}

We have selected three observables as the best AGN bolometric luminosity indicators: the corrected 2-10\,keV hard X-ray luminosity,  the nuclear 12\,$\mu$m luminosity, and the luminosity of the [OIV]\,26\,$\mu$m line. These three physical quantities have been chosen because: (i) X-ray emission is ubiquitous in AGN and is believed to be produced by comptonization of UV/optical disc photons by a corona of hot electrons located near the black hole; (ii) the mid-IR, and in particular the nuclear 12\,$\mu$m emission, is the spectral region which has been found to have the minimum scatter in the ratio of the observed to total luminosity among various types of AGN \citep{spinoglio1989} and therefore the mid-IR flux is considered one of the best general indicators of the bolometric luminosity of all types of AGN; (iii) the [OIV]\,26\,$\mu$m line emission, is mainly produced in the AGN narrow line regions (NLR) and therefore it is indirectly linked to the accretion power of the AGN: it is a good measure of the AGN bolometric luminosity, as recently demonstrated by, e.g., \citet{spinoglio2022}. Shorter wavelengths, from the soft X-rays through the optical, can suffer from large extinctions, which are difficult to determine. 

To quantify the statistical significance of the relations that we present in the following sections, we used an orthogonal distance regression (ODR) fit \citep[see, e.g.,][]{spinoglio2022}. The best empirical correlations that we have found with the corrected 2-10\,keV hard X-ray luminosity are the nuclear 12\,$\mu$m luminosity (Fig. \ref{fig:L12_x2-10}), the nuclear K-band luminosity (Fig. \ref{fig:LK_x2-10}) and the [OIV]\,26\,$\mu$m line luminosity (Fig. \ref{fig:O4_x2-10}). 
The statistics of these correlations are given in Table \ref{tab:cor1}, broken out for each of the subsets of data, i.e. the whole 12\,$\mu$m AGN sample for which are available these observations, the Type 1 objects, which include both Seyfert type 1's and the so-called {\it hidden broad-line} AGN (HBL) and the Type 2 objects, i.e. the {\it pure} Seyfert type 2 galaxies: the number of objects (N), the Pearson  correlation coefficient $\rho$, the null hypothesis probability, the linear regression fit parameters ($\alpha$ and $\beta$) and the residual variance $\sigma$, measuring the goodness of the fit. The linear equation is defined in the notes of Table \ref{tab:cor1}. 




\begin{table*}[ht!!!]
\centering
\setlength{\tabcolsep}{3.pt}
\caption{Correlation of various luminosities with the best bolometric indicators (2-10\,keV, nuclear 12$\mu$m and [OIV]26$\mu$m line luminosities).}\label{tab:cor1}
\footnotesize
\begin{tabular}{llclccc}
    \hline\\[-0.3cm]
\bf Considered variables &\bf Subset  &  \bf N &  \bf $\rho$~~~~~ (P$_{\rm null}$)  & \bf $\beta$ $\pm\ \Delta \beta$ & \bf $\alpha$ $\pm\ \Delta \alpha$  & $\sigma$ \\ [0.02cm]
 (1)            & (2)       & (3)           & (4)     & (5)      &  (6)   & (7)  \\[0.05cm]
\hline\\[-0.25cm]
$L^{\rm nuc}_{\rm 12 \mu m}$\ vs. $L_{\rm 2-10 keV}$    & All   & 87  &  0.85 (3.3 $\times 10^{-25}$) &  0.33 $\pm$ 0.14 & {\bf 1.08 $\pm$ 0.07 } & 0.203\\
 "~~~~~~~~~~~~~~~~~~~~"~~~~~~~~~~~~" & Type~1 & 49 &  0.82 (4.4 $\times 10^{-13}$) &  0.11 $\pm$ 0.25 & {\bf 1.17 $\pm$ 0.11}   & 0.145\\
 "~~~~~~~~~~~~~~~~~~~~"~~~~~~~~~~~~" & Type~2 & 19 &  0.58  (8.4  $\times 10^{-3}$) & 0.36 $\pm$ 0.39 & {\bf 1.04 $\pm$ 0.26}  &  0.316\\
$L^{\rm nuc}_{K}$\ vs. $L_{\rm 2-10 keV}$   & All   & 90  &  0.79 (1.7 $\times 10^{-20}$) &  -0.34 $\pm$ 0.16 & {\bf 1.04 $\pm$ 0.08 }  & 0.242\\
 "~~~~~~~~~~~~~~~~~~~~"~~~~~~~~~~~~" & Type~1 & 55 &  0.81 (6.8 $\times 10^{-14}$) &  -0.82 $\pm$ 0.26 & {\bf 1.27 $\pm$ 0.11}   & 0.145\\
 "~~~~~~~~~~~~~~~~~~~~"~~~~~~~~~~~~" & Type~2 & 18 &  0.45  (5.9  $\times 10^{-2}$) & 0.12 $\pm$ 0.31 & {\bf 0.68 $\pm$ 0.22}  &  0.296\\
$L_{\rm [OIV]26 \mu m}$\ vs. $L_{\rm 2-10 keV}$     & all & 96 &  0.83  (5.0 $\times 10^{-25}$) &  -1.56  $\pm$ 0.12 & {\bf 0.87 $\pm$ 0.06} & 0.202 \\
 "~~~~~~~~~~~~~~~~~~~~"~~~~~~~~~~~~" & Type~1 & 57 &  0.83 (1.0 $\times 10^{-15}$) &  -1.49 $\pm$ 0.16 & {\bf 0.84 $\pm$ 0.07}  &  0.105\\
 "~~~~~~~~~~~~~~~~~~~~"~~~~~~~~~~~~" & Type~2 & 18 &  0.73 (5.9  $\times 10^{-4}$) & -1.61 $\pm$ 0.28 & {\bf 0.91 $\pm$ 0.18}  &  0.167\\
 \hline
\hline
\end{tabular}\\[0.2cm] 
\begin{tablenotes}
\footnotesize
\item \textbf{Notes.} Fit results. The columns give for each correlation: (1) variables; (2) Subset of the sample on which the fit was computed: ``all'' indicates the entire sample and Type~1 and Type~2 the Seyfert type subsets; (3) Number of sources; (4) Pearson  correlation coefficient $\rho$ (1: completely correlated variables, 0: uncorrelated variables) with the relative null hypothesis (zero correlation) probability; (5) and (6): Parameters of the linear regression fit using the equation: 
${\rm log(L_y) = \beta + \alpha \times log(L_x)}$; (7): residual variance of the fit $\sigma$. 
\end{tablenotes}
\end{table*}


\subsection{Monochromatic vs. bolometric luminosities}\label{mono_bol}

\begin{table*}[ht!!!]
\centering
\setlength{\tabcolsep}{3.pt}
\caption{Logarithmic correlations of various continuum and line luminosities and their combinations with the bolometric luminosities.}\label{tab:cortot1}
\footnotesize
\begin{tabular}{llclccc}
    \hline\\[-0.3cm]
\bf Considered variables &\bf Subset  &  \bf N &  \bf $\rho$~~~~~ (P$_{\rm null}$)  & \bf $\beta$ $\pm\ \Delta \beta$ & \bf $\alpha$ $\pm\ \Delta \alpha$  & $\sigma$ \\ [0.02cm]
 (1)            & (2)       & (3)           & (4)     & (5)      &  (6)   & (7)  \\[0.05cm]
\hline\\[-0.25cm]
 $L_{\rm 2-10 keV}$\ vs. $L_{\rm bol}$             & all & 95 &  0.95 (3.2 $\times 10^{-49}$) & -1.41 $\pm$ 0.11 & {\bf 1.02 $\pm$ 0.03}& 0.052 \\
 "~~~~~~~~~~~~~~~~~~~~"~~~~~~~~~~~~" & Type~1 & 57 &  0.94 (4.8 $\times 10^{-27}$) & -1.12 $\pm$ 0.16 & {\bf 0.93 $\pm$ 0.04} & 0.048 \\
 "~~~~~~~~~~~~~~~~~~~~"~~~~~~~~~~~~" & Type~2 & 18 &  0.97 (4.8  $\times 10^{-11}$) & -1.54 $\pm$ 0.19 & {\bf 1.10 $\pm$ 0.07} & 0.019 \\
$L_{\rm 14-195 keV}$\ vs. $L_{\rm bol}$            & all   & 68 &  0.95 (1.6 $\times 10^{-34}$) & -0.99 $\pm$ 0.14 & {\bf 0.98 $\pm$ 0.04} & 0.057 \\
 "~~~~~~~~~~~~~~~~~~~~"~~~~~~~~~~~~" & Type~1 & 47 &  0.95 (8.8 $\times 10^{-25}$) & -0.63 $\pm$ 0.15 & {\bf 0.90 $\pm$ 0.04} & 0.037 \\
 "~~~~~~~~~~~~~~~~~~~~"~~~~~~~~~~~~" & Type~2 & 11 &  0.98 (3.1  $\times 10^{-7}$) & -1.20 $\pm$ 0.24 & {\bf 1.10 $\pm$ 0.08} & 0.027 \\
 $L_{\rm NUV}$\ vs. $L_{\rm bol}$              & all & 79 &  0.81 (2.0 $\times 10^{-19}$) & -2.63 $\pm$ 0.30 & {\bf 1.21 $\pm$ 0.09} & 0.245 \\
 "~~~~~~~~~~~~~~~~~~~~"~~~~~~~~~~~~" & Type~1 & 45 &  0.79 (9.7 $\times 10^{-11}$) & -3.36 $\pm$ 0.55 & {\bf 1.44 $\pm$ 0.15} & 0.221 \\
 "~~~~~~~~~~~~~~~~~~~~"~~~~~~~~~~~~" & Type~2 & 18 &  0.71 (8.5  $\times 10^{-4}$) & -2.51 $\pm$ 0.68 & {\bf 1.17 $\pm$ 0.24} & 0.221 \\
  $L_{\rm FUV}$\ vs. $L_{\rm bol}$             & all & 67 &  0.75 (3.0 $\times 10^{-13}$) & -3.19 $\pm$ 0.44 & {\bf 1.37 $\pm$ 0.13} & 0.354 \\
 "~~~~~~~~~~~~~~~~~~~~"~~~~~~~~~~~~" & Type~1 & 39 &  0.71 (4.0 $\times 10^{-7}$) & -4.12 $\pm$ 0.85 & {\bf 1.65 $\pm$ 0.23} & 0.311 \\
 "~~~~~~~~~~~~~~~~~~~~"~~~~~~~~~~~~" & Type~2 & 12 &  0.24 (4.4  $\times 10^{-1}$) & undefined & {\bf undefined} & 0.334 \\
$L^{\rm nuc}_{\rm 8.4 GHz}$\ vs. $L_{\rm bol}$  & all & 80 &  0.76  (1.5 $\times 10^{-16}$) &  -7.13  $\pm$ 0.40 & {\bf 1.41 $\pm$ 0.12} & 0.324 \\
 "~~~~~~~~~~~~~~~~~~~~"~~~~~~~~~~~~" & Type~1 & 48 &  0.81 (1.9 $\times 10^{-12}$) &  -8.39 $\pm$ 0.60 & {\bf 1.70 $\pm$ 0.16} & 0.227 \\
 "~~~~~~~~~~~~~~~~~~~~"~~~~~~~~~~~~" & Type~2 & 14 &  0.77 (1.3  $\times 10^{-3}$) & -7.34 $\pm$ 0.92 & {\bf 1.55 $\pm$ 0.33} & 0.181 \\
  $L^{\rm nuc}_{K}$\ vs. $L_{\rm bol}$   & All   & 88 &  0.90 (5.8 $\times 10^{-33}$) &  -1.86 $\pm$ 0.18 & {\bf 1.06 $\pm$ 0.05} & 0.110 \\
 "~~~~~~~~~~~~~~~~~~~~"~~~~~~~~~~~~" & Type~1 & 54 &  0.90 (3.1 $\times 10^{-20}$) &  -1.81 $\pm$ 0.25 & {\bf 1.06 $\pm$ 0.07} & 0.089 \\
 "~~~~~~~~~~~~~~~~~~~~"~~~~~~~~~~~~" & Type~2 & 18 &  0.82 (2.4  $\times 10^{-5}$) & -1.65 $\pm$ 0.43 & {\bf 0.99 $\pm$ 0.15} & 0.106 \\
$L^{\rm nuc}_{\rm 12 \mu m}$\ vs. $L_{\rm bol}$    & All   & 83  &  0.94 (1.7 $\times 10^{-39}$) & -1.24 $\pm$ 0.14 & {\bf 1.09 $\pm$ 0.04 } & 0.076 \\
 "~~~~~~~~~~~~~~~~~~~~"~~~~~~~~~~~~" & Type~1 & 48 &  0.94 (9.7 $\times 10^{-23}$) &  -0.91 $\pm$ 0.20 & {\bf 1.01 $\pm$ 0.05}  & 0.057 \\
 "~~~~~~~~~~~~~~~~~~~~"~~~~~~~~~~~~" & Type~2 & 17 &  0.87  (7.1  $\times 10^{-6}$) & -1.57 $\pm$ 0.46 & {\bf 1.19 $\pm$ 0.16} &  0.089 \\
\hline
$L_{\rm [OIV]26 \mu m}$\ vs. $L_{\rm bol}$     & all & 94 &  0.87  (8.3$\times 10^{-30}$) &  -3.15  $\pm$ 0.18 & {\bf 0.99 $\pm$ 0.05} & 0.140 \\
 "~~~~~~~~~~~~~~~~~~~~"~~~~~~~~~~~~" & Type~1 & 57 &  0.83 (1.4 $\times 10^{-15}$) &  -2.40 $\pm$ 0.23 & {\bf 0.78 $\pm$ 0.06}  & 0.112 \\
 "~~~~~~~~~~~~~~~~~~~~"~~~~~~~~~~~~" & Type~2 & 18 &  0.87 (3.2  $\times 10^{-6}$) & -3.23 $\pm$ 0.41 & {\bf 1.08 $\pm$ 0.14} & 0.099 \\
 $L_{\rm [NeV]14.3 \mu m}$\ vs. $L_{\rm bol}$     & all & 84 &  0.85  (2.1 $\times 10^{-24}$) &  -3.53  $\pm$ 0.20 & {\bf 0.95 $\pm$ 0.06} & 0.110 \\
 "~~~~~~~~~~~~~~~~~~~~"~~~~~~~~~~~~" & Type~1 & 54 &  0.85 (7.8 $\times 10^{-16}$) &  -3.33 $\pm$ 0.25 & {\bf 0.88 $\pm$ 0.07} & 0.114 \\
 "~~~~~~~~~~~~~~~~~~~~"~~~~~~~~~~~~" & Type~2 & 16 &  0.89 (3.8  $\times 10^{-6}$) & -3.65 $\pm$ 0.38 & {\bf 1.01 $\pm$ 0.13} & 0.078 \\
 $L_{\rm [NeV]24.3 \mu m}$\ vs. $L_{\rm bol}$     & all & 73 &  0.79  (1.1 $\times 10^{-16}$) &  -3.42  $\pm$ 0.26 & {\bf 0.93 $\pm$ 0.08} & 0.119\\
 "~~~~~~~~~~~~~~~~~~~~"~~~~~~~~~~~~" & Type~1 & 46 &  0.83 (8.2 $\times 10^{-13}$) &  -2.90 $\pm$ 0.26 & {\bf 0.78 $\pm$ 0.07} & 0.097 \\
 "~~~~~~~~~~~~~~~~~~~~"~~~~~~~~~~~~" & Type~2 & 15 &  0.80 (3.2  $\times 10^{-4}$) & -4.03 $\pm$ 0.65 & {\bf 1.17 $\pm$ 0.21} & 0.087 \\
$L_{\rm [OIII]5007 \mathring{A}}$\ vs. $L_{\rm bol}$  & all & 87 &  0.83  (1.4$\times 10^{-23}$) &  -3.76  $\pm$ 0.25 & {\bf 1.14 $\pm$ 0.07} & 0.142\\
 "~~~~~~~~~~~~~~~~~~~~"~~~~~~~~~~~~" & Type~1 & 53 &  0.83 (1.1 $\times 10^{-14}$) &  -3.27 $\pm$ 0.30 & {\bf 1.00 $\pm$ 0.08} & 0.116 \\
 "~~~~~~~~~~~~~~~~~~~~"~~~~~~~~~~~~" & Type~2 & 18 &  0.65 (3.3  $\times 10^{-3}$) & -4.46 $\pm$ 0.89 & {\bf 1.32 $\pm$ 0.31} & 0.260 \\
\hline
$7.0 \times L^{\rm nuc}_{\rm 12 \mu m} + 389 \times L_{\rm [OIV]26 \mu m}$ vs. $L_{\rm bol}$ & all & 81 &  0.94 (2.0 $\times 10^{-39}$) & -0.05 $\pm$ 0.14 & {\bf 1.04 $\pm$ 0.04} & 0.070 \\
 "~~~~~~~~~~~~~~~~~~~~"~~~~~~~~~~~~" & Type~1 & 48 &  0.94 (8.4 $\times 10^{-24}$) & 0.37 $\pm$ 0.17 & {\bf 0.93 $\pm$ 0.05} & 0.047 \\
 "~~~~~~~~~~~~~~~~~~~~"~~~~~~~~~~~~" & Type~2 & 16 &  0.88 (7.9 $\times 10^{-6}$) & -0.06 $\pm$ 0.40 & {\bf 1.04 $\pm$ 0.14} & 0.078 \\
$5.4 \times L^{\rm nuc}_{\rm 12 \mu m} + 2132 \times L_{\rm [NeV]14.3 \mu m}$ vs. $L_{\rm bol}$ & all & 63 &  0.87 (2.7 $\times 10^{-20}$) & 0.14 $\pm$ 0.23 & {\bf 0.98 $\pm$ 0.07} & 0.080 \\
 "~~~~~~~~~~~~~~~~~~~~"~~~~~~~~~~~~" & Type~1 & 39 &  0.92 (2.8 $\times 10^{-16}$) & 0.67 $\pm$ 0.21 & {\bf 0.84 $\pm$ 0.06} & 0.051 \\
 "~~~~~~~~~~~~~~~~~~~~"~~~~~~~~~~~~" & Type~2 & 14 &  0.70 (5.1 $\times 10^{-3}$) & 0.08 $\pm$ 0.71 & {\bf 0.99 $\pm$ 0.24} & 0.085 \\
$17.3 \times L^{\rm nuc}_{K} + 881 \times L_{\rm [OIII]5007 \mathring{A}}$ vs. $L_{\rm bol}$ & all & 79 &  0.94 (9.3 $\times 10^{-37}$) & 0.00 $\pm$ 0.14 & {\bf 1.01 $\pm$ 0.04} & 0.047 \\
 "~~~~~~~~~~~~~~~~~~~~"~~~~~~~~~~~~" & Type~1 & 50 &  0.93 (5.0 $\times 10^{-23}$) & 0.11 $\pm$ 0.19 & {\bf 0.99 $\pm$ 0.05} & 0.044 \\
 "~~~~~~~~~~~~~~~~~~~~"~~~~~~~~~~~~" & Type~2 & 17 &  0.92 (1.5 $\times 10^{-7}$) & 0.41 $\pm$ 0.25 & {\bf 0.83 $\pm$ 0.09} & 0.039 \\
\hline
\end{tabular}\\[0.2cm] 
\begin{tablenotes}
\footnotesize
\item \textbf{Notes.} Fit results. The columns give for each correlation: (1) variables; (2) Subset of the sample on which the fit was computed: ``all'' indicates the entire sample and Type~1 and Type~2 the Seyfert type subsets; (3) Number of sources; (4) Pearson  correlation coefficient $\rho$ (1: completely correlated variables, 0: uncorrelated variables) with the relative null hypothesis (zero correlation) probability; (5) and (6): Parameters of the linear regression fit using the equation: 
${\rm log(L_y) = \beta + \alpha \times log(L_x)}$; (7): residual variance of the fit $\sigma$. 
\end{tablenotes}
\end{table*}

\begin{figure*}[ht!]
\includegraphics[width = 0.5\textwidth]{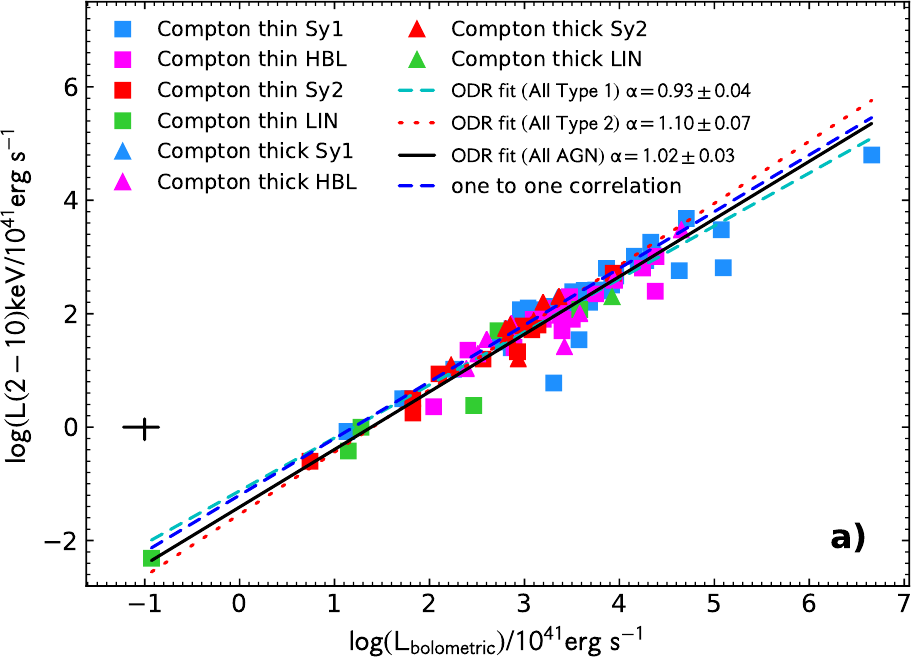}~
\includegraphics[width = 0.5\textwidth]{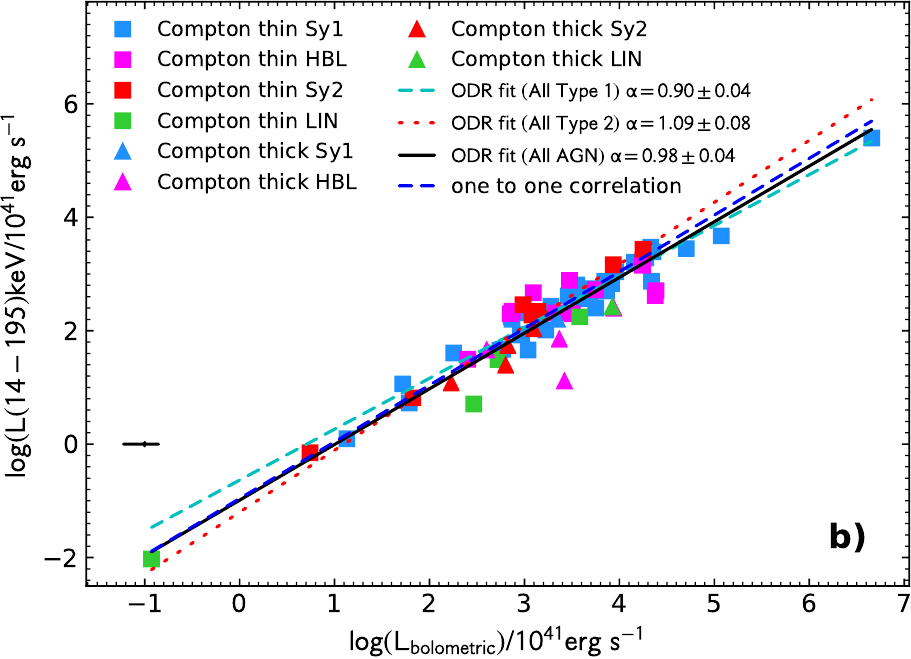}
\caption{{\bf (a:)} Corrected 2-10\,keV 
luminosity versus 
bolometric luminosity. {\bf (b:)} 14-195\,keV 
luminosity versus 
bolometric luminosity.
\label{fig:x-rays-bol}}
\end{figure*}

\begin{figure*}[ht!]
\includegraphics[width = 0.5\textwidth]{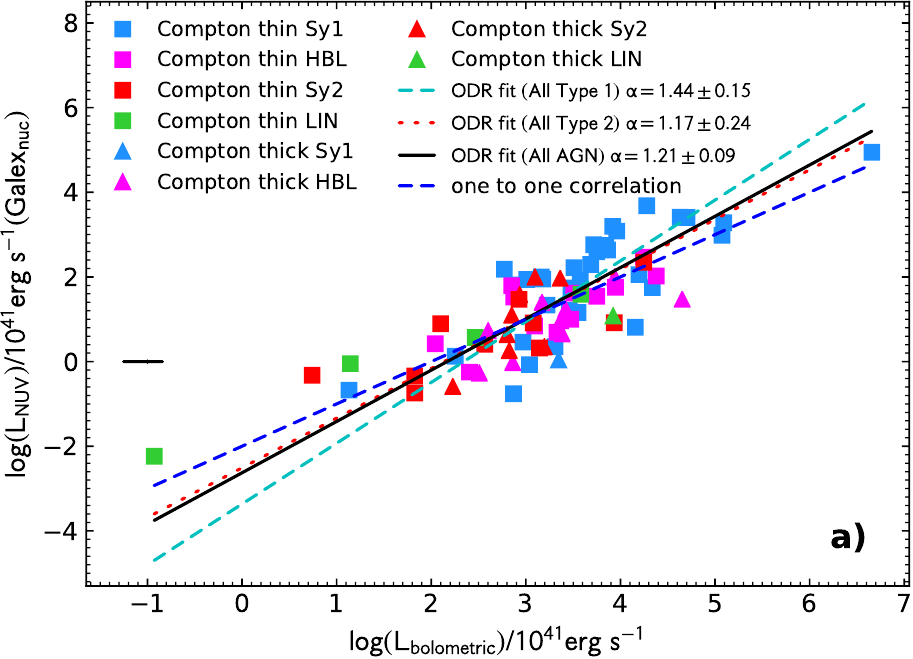}~
\includegraphics[width = 0.5\textwidth]{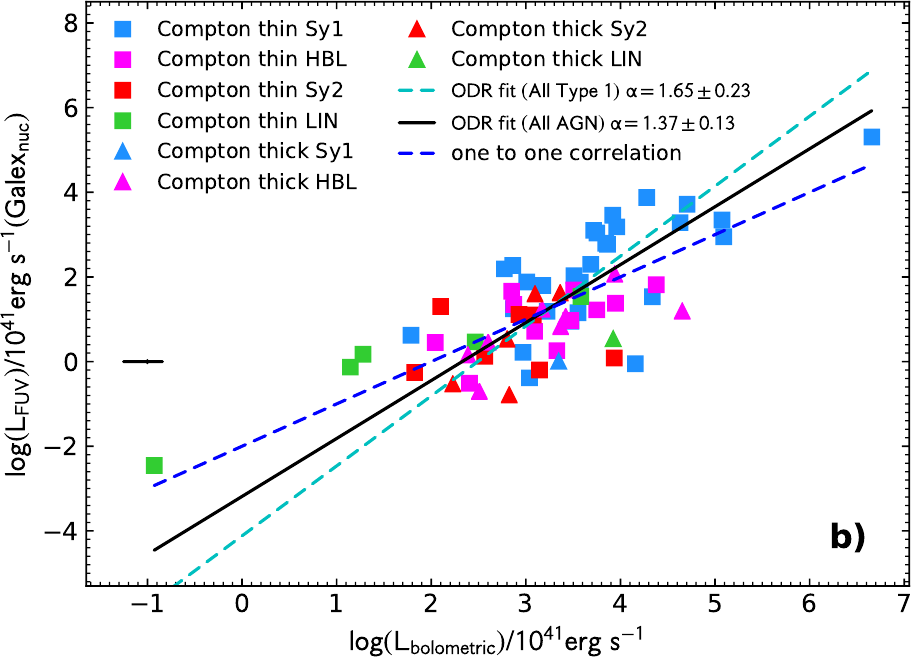}
\caption{{\bf (a:)} GALEX NUV luminosity versus 
bolometric luminosity. {\bf (b:)} GALEX FUV  luminosity versus 
bolometric luminosity.
\label{fig:NUV_FUV_bol}}
\end{figure*}

\begin{figure*}[ht!]
\includegraphics[width = 0.5\textwidth]{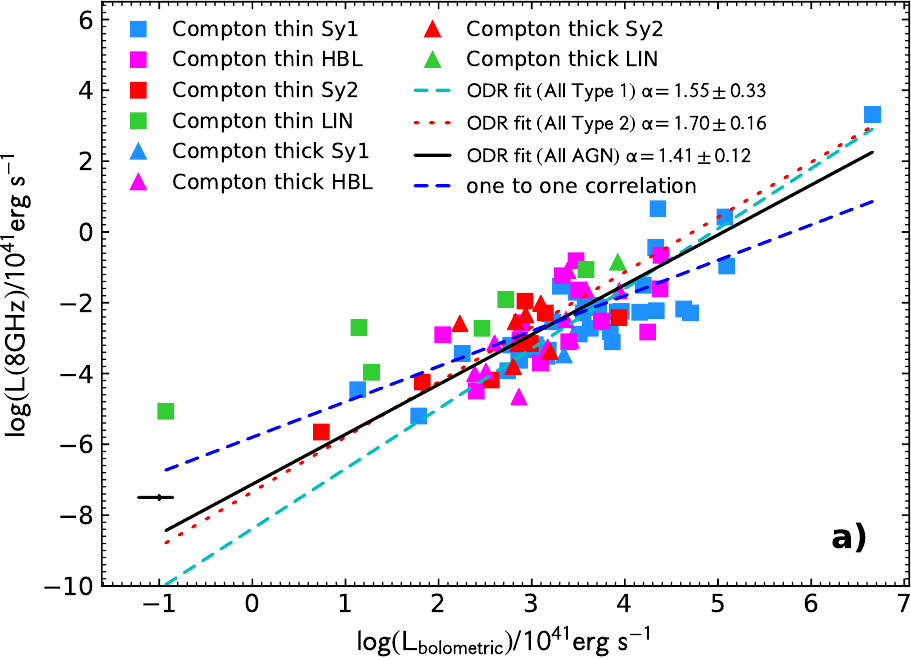}~
\includegraphics[width = 0.5\textwidth]{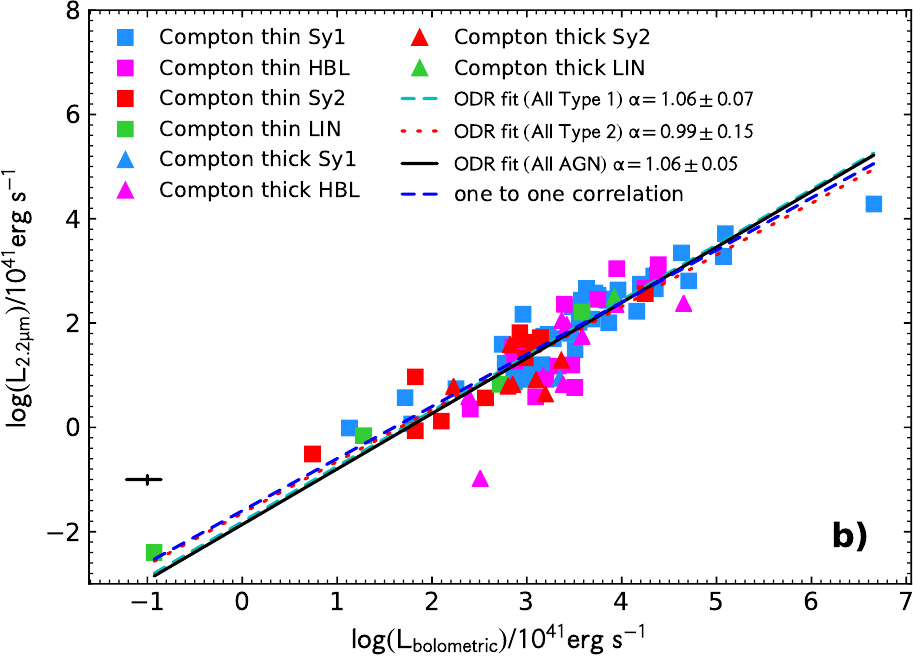}
\caption{{\bf (a:)} 8.4\,GHz luminosity versus 
bolometric luminosity. {\bf (b:)} Nuclear K-band luminosity versus 
bolometric luminosity.
\label{fig:L8GHz_LK_bol}}
\end{figure*}

\begin{figure*}[ht!]
\includegraphics[width = 0.5\textwidth]{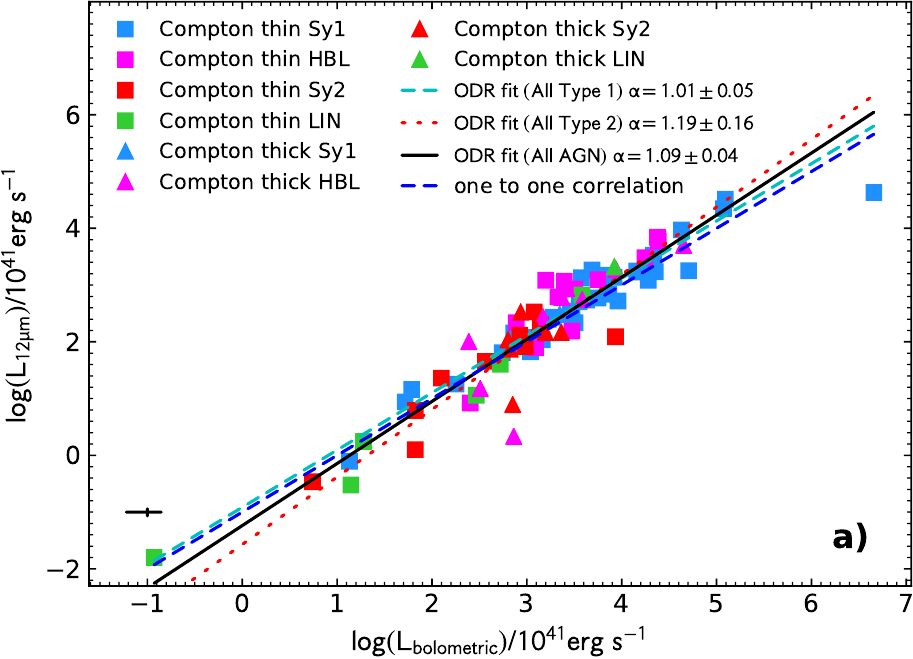}~
\includegraphics[width = 0.5\textwidth]{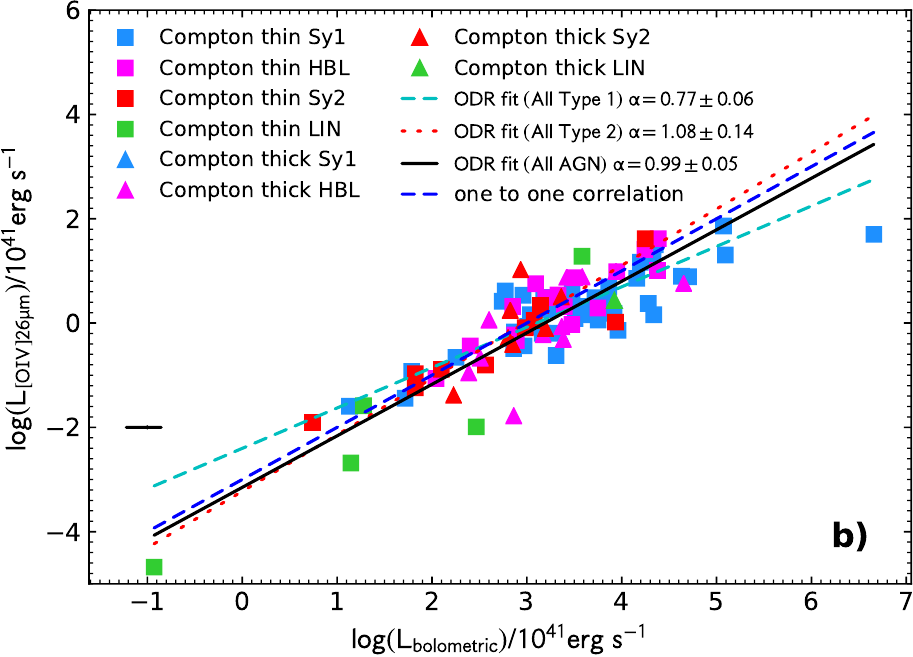}
\caption{{\bf (a:)} Nuclear 12$\mu$m luminosity versus 
bolometric luminosity. {\bf (b:)} [OIV]26$\mu$m luminosity versus 
bolometric luminosity.
\label{fig:L12_oiv_bol}}
\end{figure*}

\begin{figure*}[ht!]
\includegraphics[width = 0.5\textwidth]{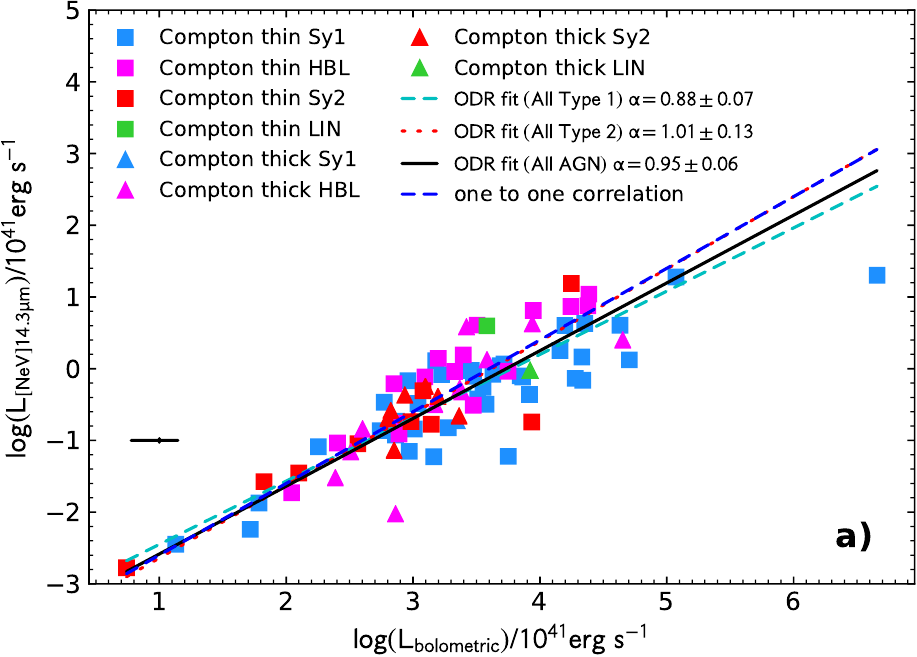}~
\includegraphics[width = 0.5\textwidth]{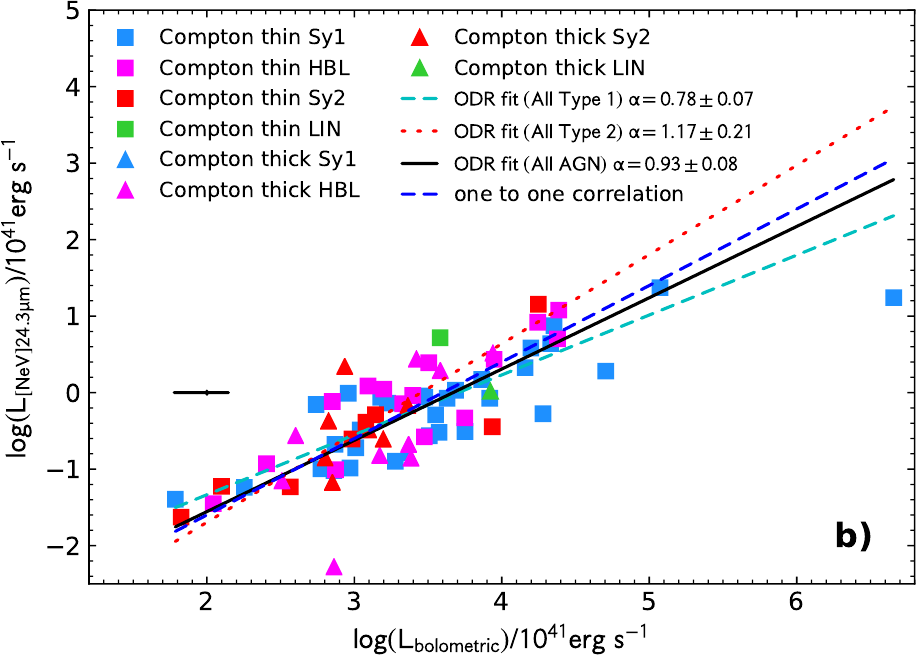}
\caption{{\bf (a:)} [NeV]14.3$\mu$m luminosity versus 
bolometric luminosity. {\bf (b:)} [NeV]24.3$\mu$m luminosity versus 
bolometric luminosity.
\label{fig:Lnev1_nev2_bol}}
\end{figure*}

\begin{figure*}[ht!]
\includegraphics[width = 0.5\textwidth]{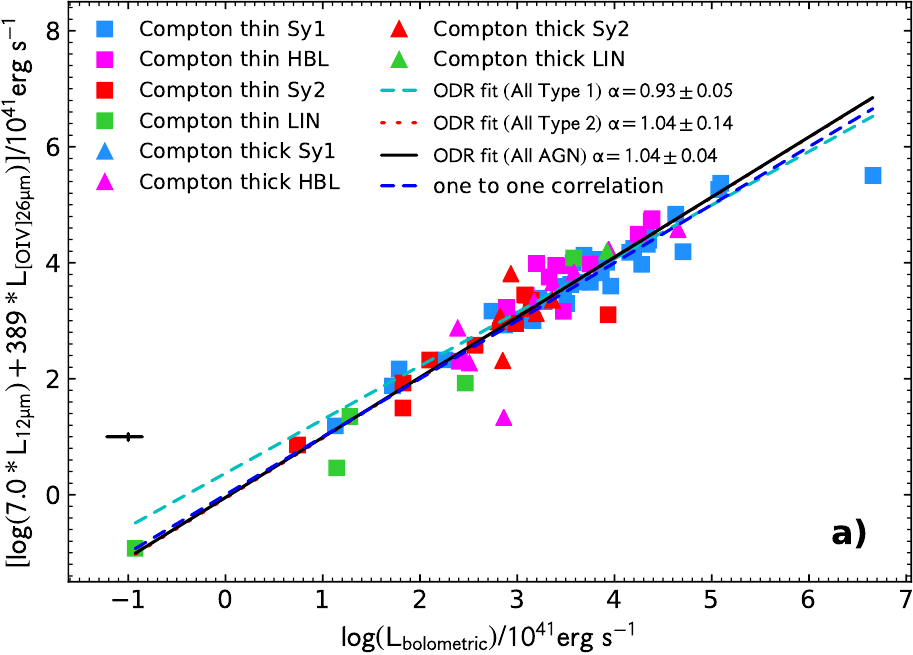}
\includegraphics[width = 0.5\textwidth]{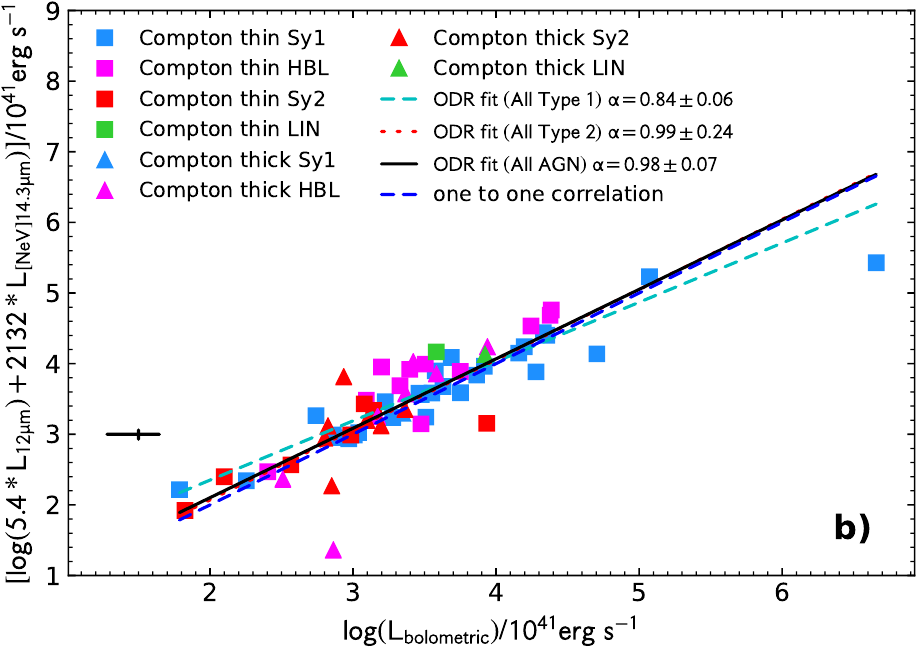}
\caption{{\bf (a:)} Composite 12$\mu$m and [OIV]26$\mu$m luminosity versus 
bolometric luminosity. {\bf (b:)}  Composite 12$\mu$m and [NeV]14.3$\mu$m luminosity versus 
bolometric luminosity.
\label{fig:mix12_oiv_bol}}
\end{figure*}

\begin{figure*}[ht!]
\includegraphics[width = 0.49\textwidth]{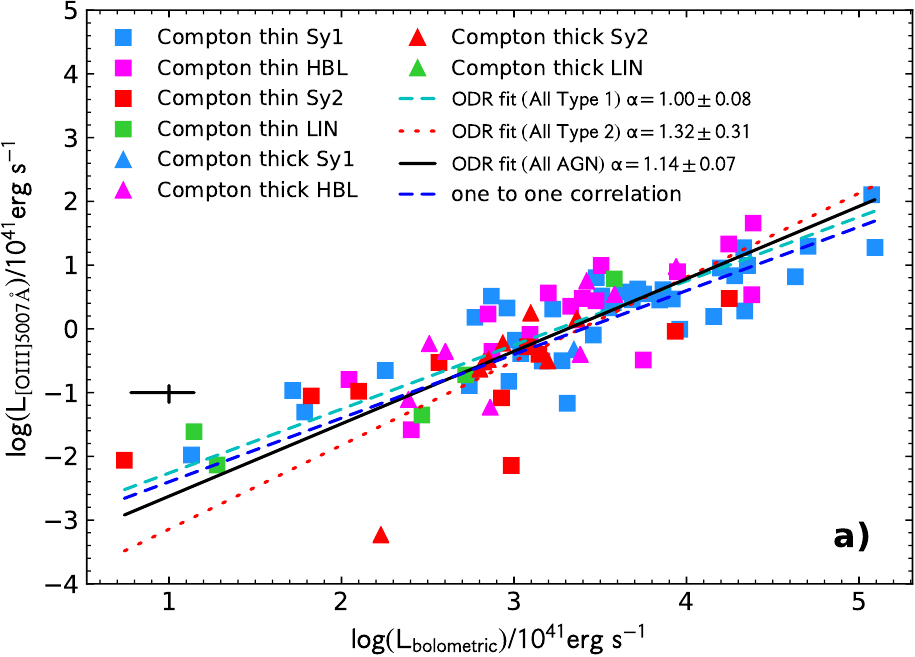}
\includegraphics[width = 0.51\textwidth]{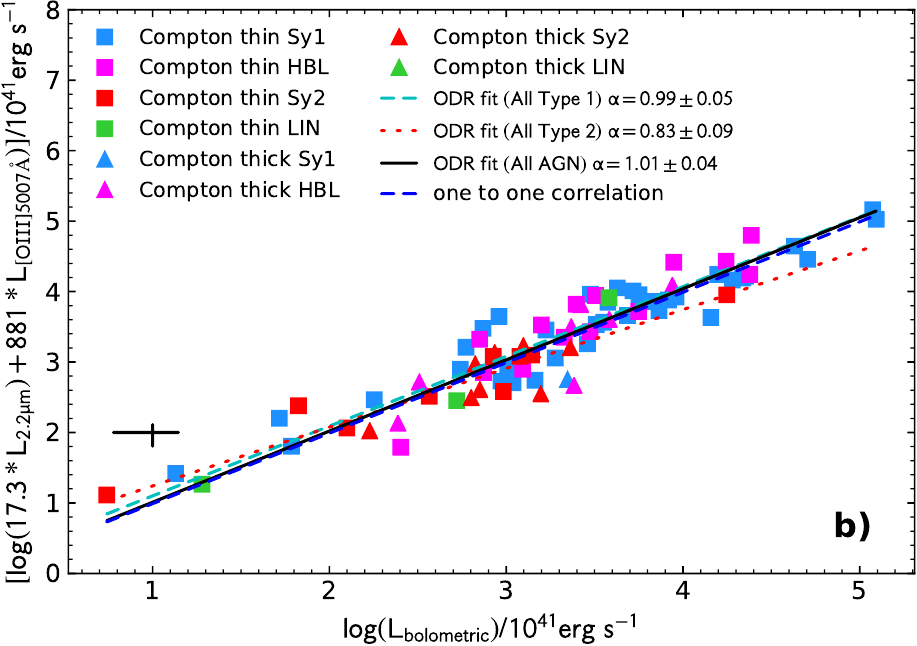}
\caption{{\bf (a:)} [OIII]5007\AA \ luminosity versus 
bolometric luminosity. {\bf (b:)}   Composite nuclear K-band and [OIII]5007\AA \ luminosity versus 
bolometric luminosity.
\label{fig:mixK_o3_bol}}
\end{figure*}

Having obtained the bolometric luminosities for most (82.9\%) of our AGN sample (excluding the lower limits and the two objects with no determination),  which are given in the last data column of Table \ref{tab:sample2}, we computed the correlations between 12$\mu$m nuclear luminosity, K-band nuclear luminosity, NUV corrected luminosity, FUV corrected luminosity, 8.4GHz luminosity, 2-10\,keV luminosity and 15-145\,keV luminosity and the computed bolometric luminosity. We have also computed the correlations of the luminosities of the three mid-IR fine structure lines, which are mainly originated in the AGN narrow line regions, namely [OIV]26$\mu$m, [NeV]14.3$\mu$m and [NeV]24.3$\mu$m with the bolometric luminosities. All correlations are given in  Table \ref{tab:cortot1} and are shown in Figs. \ref{fig:x-rays-bol} -- \ref{fig:mix12_oiv_bol}.

In Fig. \ref{fig:x-rays-bol}a the well known correlation between the 2-10\,keV {\it absorption-corrected} luminosity and the computed bolometric luminosity is shown: including all classes of AGN the correlation is linear (slope 1.00$\pm$0.04) and becomes slightly flatter for the type 1 AGN (slope $\sim$0.90) and slightly steeper for the type 2 AGN (slope $\sim$1.04), while remaining linear within 2 $\sigma$.
The harder X-ray 14-195\,keV (Fig.\ref{fig:x-rays-bol}b) observed luminosity has also a good and almost linear (slope 0.94$\pm$0.04) correlation with the bolometric luminosity if all classes are included, while it becomes significantly steeper for type 2 objects only.

In Fig. \ref{fig:NUV_FUV_bol}a and b are shown the correlations between the NUV and FUV corrected luminosities with the bolometric luminosity, respectively. The NUV luminosity for all classes together correlates almost linearly (within 2\,$\sigma$) with the bolometric luminosity, while the correlation becomes steeper for the type 1 class ($\alpha\,\sim\,1.40$), showing that type 1 objects have a stronger UV-excess at higher luminosities. For the same reason, the FUV luminosity has a correlation steeper than linear with the bolometric luminosity. The slope is $\sim$1.3 for the whole sample and $\sim$1.6 for type 1 AGN only, while no correlation is apparent for the type 2 AGN. This is partly due to the increased contribution of the accretion disk energy in the UV for the more luminous AGN \citep{Sun89}. Moreover, in these two plots, most AGN at bolometric luminosities greater than $\sim$10$^{45}$ erg s$^{-1}$ are Seyfert type 1, known to have more powerful directly observed accretion disks.

Figure \ref{fig:L8GHz_LK_bol}a shows the correlation between the radio 8.4\,GHz luminosity and the bolometric luminosity, respectively. The 8.4\,GHz luminosity has always a steeper than linear correlation with the bolometric luminosity, and its slope increases from the whole sample through type 1 AGN to type 2 objects, showing that at higher luminosities radio loudness is increasing. 

The nuclear K-band luminosity (Fig. \ref{fig:L8GHz_LK_bol}b) has a very good linear correlation with the bolometric luminosity, for all types of objects (slope $\sim\,1.04\,\pm\,0.05$). Therefore the nuclear K-band flux, once corrected for starlight emission, is a good indicator for the bolometric flux.

Fig. \ref{fig:L12_oiv_bol}a and b show the correlations between the nuclear 12\,$\mu$m luminosity and the [OIV]26$\mu$m luminosity and the bolometric luminosity, respectively. The nuclear 12\,$\mu$m luminosity has a strong and linear correlation with the bolometric luminosity. This linearity is known since the pioneering work of \citet{spinoglio1989}, who selected galaxies in the IRAS 12\,$\mu$m band to produce an unbiased sample of active galaxies. The IRAS large-aperture flux densities were indeed contaminated in galaxies by stellar emission, however for the Seyfert galaxies the emission from the AGN tends to dominate the power in this spectral region. Also the [OIV]26$\mu$m luminosity, originating mostly in the NLR of the AGN, has an almost linear correlation for the whole AGN sample ($\alpha\,=\,0.96\,\pm\,0.05$) , with the AGN bolometric luminosity, albeit with a larger scatter with slopes ranging from 0.7 for the type 1 objects to 1.1 for the type 2's. We note that the higher bolometric luminosity type 1 AGN have a weaker emission in the NLR mid-IR fine structure lines, which flattens the slope 
of the correlation. 


Figures \ref{fig:Lnev1_nev2_bol}a and b show the correlations between the [NeV]14.3$\mu$m luminosity and the [NeV]24.3$\mu$m luminosity with the bolometric luminosity, respectively. These correlations are shallower than linear for the whole sample and especially for the brighter type 1 objects, showing again, as for the case of [OIV]26$\mu$m, that at high bolometric luminosity the NLR becomes relatively weaker than expected from a linear correlation. This indicates that the relation between the NLR fine structure mid-IR lines and the bolometric luminosity may not extend linearly up to the regime of quasars, i.e. AGN with bolometric luminosities above $10^{46}\, \rm{erg\,s^{-1}}$. Studies based on brighter AGN samples suggest that at these high luminosities, the forbidden line luminosities can no longer ``keep up" with the non-stellar continuum (see discussion of [OIII] in the Appendix of \citealt{malkan17}). A possible explanation for this ``forbidden line saturation" may be that the narrow line region is becoming ``matter-bounded" \citep{pronik1972,gaskell2021}. There may not be enough gas around luminous quasars to intercept as many of their ionizing photons.


Figures \ref{fig:mix12_oiv_bol}a and b, respectively, show the correlations between a combination of the nuclear 12$\mu$m luminosity and the NLR [OIV]26$\mu$m and [NeV]14.3$\mu$m line luminosities with the bolometric luminosity. The addition of a particular percentage of the luminosity of one of the mid-IR high-ionization fine-structure lines of [OIV]26$\mu$m or [NeV]14.3$\mu$m to the nuclear 12$\mu$m luminosity has the effect of linearizing the correlation with the bolometric luminosity.

\begin{table}[hb!!!]
\centering
\caption{Single and composite continuum- and line-based bolometric corrections in AGN.}\label{tab:lboltracers}
\normalsize
\begin{tabular}{c}
 \bf   Continuum \\
 \hline
$L_{\rm bol}$ = $24.4 \times (L_{\rm 2-10keV})^{0.98 \pm 0.03}$\\
$L_{\rm bol}$ = $10.2 \times (L_{\rm 14-195keV})^{1.02 \pm 0.04}$ \\
$L_{\rm bol}$ = $13.5 \times (L^{\rm nuc}_{\rm 12 \mu m})^{0.91 \pm 0.04}$\\
$L_{\rm bol}$ = $56.2 \times (L^{\rm nuc}_{\rm K})^{0.94 \pm 0.05}$ \\

 \hline \\[-0.25cm]
  \bf   Line      \\
 \hline
$L_{\rm bol}$ =  $1542 \times (L_{\rm [OIV]26})^{1.01 \pm 0.06}$ \\
$L_{\rm bol}$ =  $5420 \times (L_{\rm [NeV]14.3})^{1.06 \pm 0.07}$ \\
$L_{\rm bol}$ =  $4667 \times (L_{\rm [NeV]24.3})^{1.07 \pm 0.09}$ \\
$L_{\rm bol}$ =  $2042 \times (L_{\rm [OIII]5007})^{0.88 \pm 0.06}$\\
 \hline \\[-0.25cm]
    \bf Composite   \\
 \hline
$L_{\rm bol}$ = $7.0 \times L^{\rm nuc}_{\rm 12 \mu m} + 389 \times L_{\rm [OIV]26}$ \\
$L_{\rm bol}$ = $5.4 \times L^{\rm nuc}_{\rm 12 \mu m} + 2132 \times L_{\rm [NeV]14.3}$\\
$L_{\rm bol}$ = $17.3 \times L^{\rm nuc}_{\rm K} + 881 \times L_{\rm [OIII]5007}$\\
  \hline
 \end{tabular}
 \end{table}
 
 
Finally, Figures \ref{fig:mixK_o3_bol}a and b show, respectively, the [OIII]5007\AA \ luminosity as a function of the bolometric luminosity and a combination of the nuclear K-band luminosity and the [OIII]5007\AA \ line luminosity with the bolometric luminosity. The K-band nuclear luminosity was chosen in this case because: (i) it has a good linear correlation with the bolometric luminosity (see Table \ref{tab:cortot1}) and it is at the closest wavelength (about four times longer) to the [OIII]5007\AA \ line, which has already been used to estimate the bolometric luminosity of AGN.

\begin{figure*}[ht!!!]
\includegraphics[width = 0.5\textwidth]{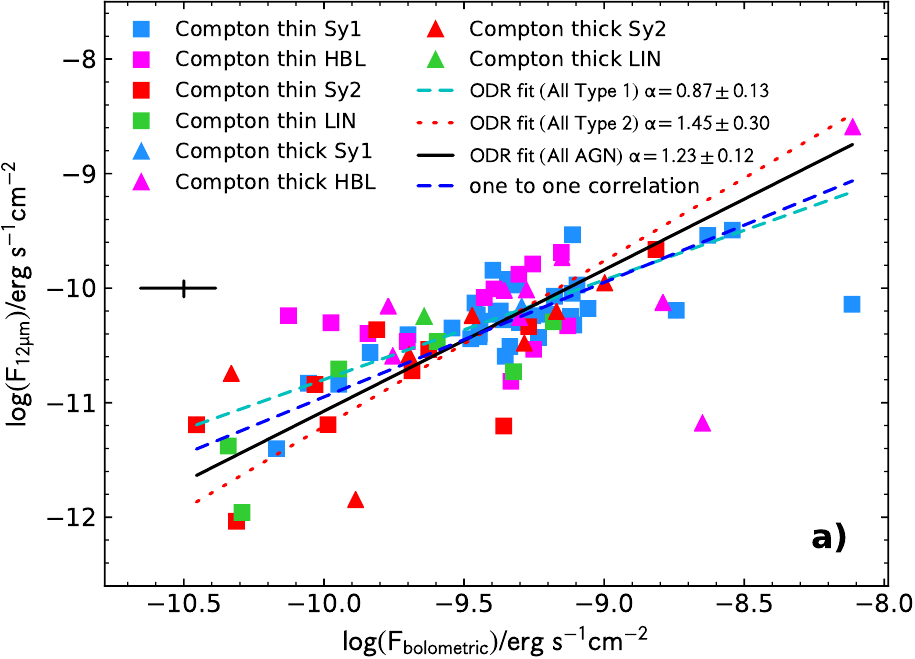}~
\includegraphics[width = 0.5\textwidth]{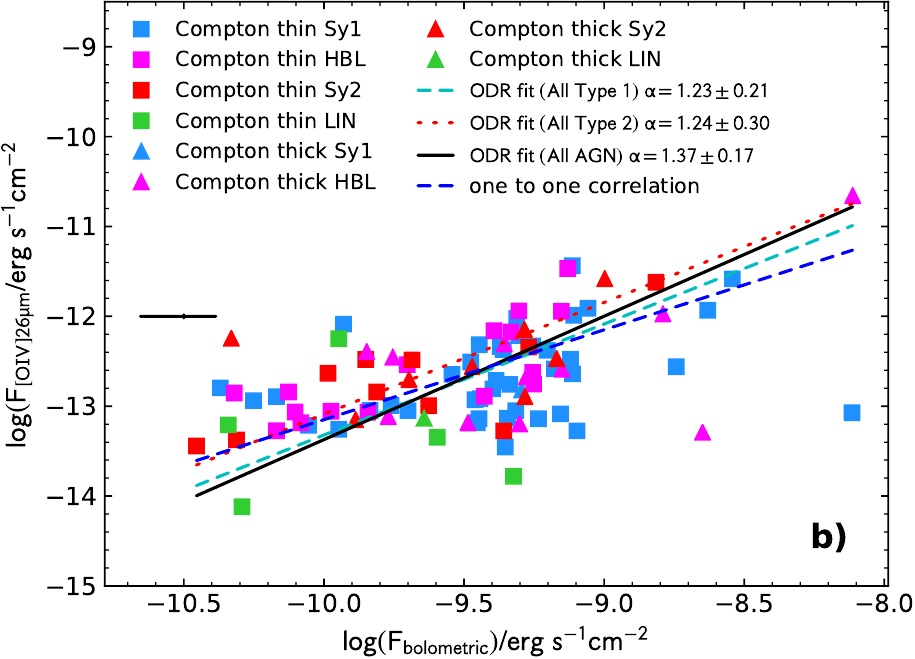}
\caption{{\bf (a:)} 12$\mu$m nuclear flux as a function of the  estimated bolometric flux. {\bf (b:)} [OIV]26$\mu$m line flux as a function of the estimated bolometric flux.
\label{fig:F12_OIV_bolflux}}
\end{figure*}

The combination of a continuum band nuclear luminosity in the IR (both mid-IR and near-IR) with a line luminosity -- when this line is mainly powered by the AGN -- gives a better correlation with the bolometric luminosity, compared to the correlations between the line luminosities and the bolometric luminosity. This can be clearly seen comparing Fig. \ref{fig:L12_oiv_bol}b with Fig. \ref{fig:mix12_oiv_bol}a, Fig. \ref{fig:Lnev1_nev2_bol}a with Fig. \ref{fig:mix12_oiv_bol}b and finally Fig. \ref{fig:mixK_o3_bol}a with Fig. \ref{fig:mixK_o3_bol}b. 
Moreover, in analogy with the measure of the star formation rate (SFR) in galaxies using both the 24\,$\mu$m continuum and the optical H\,$\alpha$ line by \citet{kennicutt2012}, the logic behind the composite tracers is that lines are complementary to continuum, because they account for photons that are not captured by the dust, so the combination of the two is expected to reduce the scatter, as we show in Figs.\,\ref{fig:L12_oiv_bol}, \ref{fig:Lnev1_nev2_bol}, \ref{fig:mix12_oiv_bol} and \ref{fig:mixK_o3_bol}.

In Table \ref{tab:lboltracers} we give our derived bolometric corrections for the full AGN population. The bolometric corrections are given for the continuum X-ray (both 2-10\,keV and 14-195\,keV) and IR luminosities (nuclear 12$\mu$m and K-band), for the luminosities of the brightest lines emitted by the NLR ([OIV]26$\mu$m, [NeV]14.3$\mu$m, [NeV]24.3$\mu$m and [OIII]5007\AA) and for a combination of the mid-IR and near-IR luminosities with the mid-IR fine structure and the [OIII]5007\AA \ line luminosities, respectively. 

We note here that our derived bolometric correction for the [OIII]5007\AA \ line luminosity is lower by about a factor two, compared to the derivation of \citet{Heckman2004}, who give a linear correlation ($L_{\rm bol} \sim 3500\,L_{\rm [OIII]5007 \mathring{A}}$), while we find a shallower slope of $\alpha$ $\sim$ 0.90.

Similar correlations were found for the AGN monocromatic vs. bolometric luminosities when the luminosities are expressed in Eddington units, as shown in Figs.\,\ref{fig:L12_Loiv_lboledd}, \ref{fig:x2-10_x14-195_lboledd}, \ref{fig:L12+oiv_L12+nev1_lboledd} and Table\,\ref{tab:cortot2} in Appendix\,\ref{corr_edd}.

\subsection{Flux-flux correlations}

We show in Fig.\ref{fig:F12_OIV_bolflux}a and b the correlations between the observed 12$\mu$m flux and the [OIV]26$\mu$m line flux, respectively, with the derived bolometric flux of our sample galaxies. The statistics of these correlations are given in Table \ref{tab:cortot}. Although the scatter is large, these correlations are consistent with linearity, within 2$\sigma$. We show this example to demonstrate that our luminosity-luminosity correlations, even if they are boosted by the distance-squared factors, are indeed real, because the corresponding correlations in flux are almost linear. Additionally, we also explored the flux-flux correlations for the 2-10\,keV and 2.2$\mu$m bands. Fig.\,\ref{fig:F2-10_FK_total} and Table\,\ref{tab:cortot} in Appendix\,\ref{corflux} show that the correlation with the bolometric flux is also significant without the distance term.

\begin{table*}[ht!!!]
\centering
\setlength{\tabcolsep}{3.pt}
\caption{Percentiles of the normalized SED distribution as a function of frequency for the four AGN classes. }\label{tab:medsed}
\footnotesize
\vspace{-0.3cm}
\begin{tabular}{cccccccccc}
\multicolumn{10}{c}{\normalsize \bf Seyfert 1} \\
\hline
 $\log \nu$ & $\log (\nu F_\nu)_{10}$ & $\log (\nu F_\nu)_{30}$ & $\log (\nu F_\nu)_{\rm med}$ & $\log (\nu F_\nu)_{70}$ & $\log (\nu F_\nu)_{90}$ & $\log (\nu F_\nu)^{\rm <EW113}_{\rm med}$ & $\log (\nu F_\nu)^{\rm >EW113}_{\rm med}$ & $\log (\nu F_\nu)^{\rm -Ssil}_{\rm med}$ & $\log (\nu F_\nu)^{\rm +Ssil}_{\rm med}$ \\
 {[Hz]} & \multicolumn{9}{c}{[normalized to median $\nu F_\nu$]} \\
\hline
  9.937  &  -5.657  &  -5.249  &   -4.821  &  -4.495  &   -2.82   &   -4.952    &   -4.849   &  -5.027   &  -4.824 \\
 13.411  &   -0.01  &   0.097  &    0.217  &   0.309  &     0.7   &    0.208    &    0.227   &     0.2   &   0.249 \\
 13.645  &  -0.021  &   0.011  &    0.037  &   0.137  &   0.297   &     0.04    &    0.016   &   0.051   &   0.014 \\
 13.726  &   -0.01  &   0.019  &    0.075  &   0.115  &   0.268   &    0.044    &    0.087   &   0.084   &   0.018 \\
 13.837  &  -0.015  &   0.009  &    0.051  &   0.154  &   0.254   &    0.026    &    0.069   &   0.044   &   0.069 \\
 14.147  &   -0.76  &  -0.584  &   -0.373  &  -0.073  &   0.029   &   -0.433    &   -0.316   &  -0.456   &  -0.269 \\
 15.134  &  -1.592  &  -0.606  &   -0.091  &   0.131  &    0.29   &   -0.093    &   -0.046   &   -0.19   &   0.047 \\
 15.306  &  -1.739  &  -0.266  &   -0.039  &   0.133  &   0.522   &   -0.043    &   -0.171   &  -0.179   &  -0.032 \\
 18.049  &  -0.476  &  -0.253  &   -0.128  &     0.0  &   0.218   &   -0.169    &    -0.04   &   -0.04   &  -0.209 \\
 19.112  &  -0.027  &    0.04  &    0.108  &   0.211  &   0.462   &    0.152    &    0.071   &   0.053   &   0.195 \\
\hline \\[-0.25cm]
\multicolumn{10}{c}{\normalsize \bf HBL} \\
\hline
 $\log \nu$ & $\log (\nu F_\nu)_{10}$ & $\log (\nu F_\nu)_{30}$ & $\log (\nu F_\nu)_{\rm med}$ & $\log (\nu F_\nu)_{70}$ & $\log (\nu F_\nu)_{90}$ & $\log (\nu F_\nu)^{\rm <EW113}_{\rm med}$ & $\log (\nu F_\nu)^{\rm >EW113}_{\rm med}$ & $\log (\nu F_\nu)^{\rm -Ssil}_{\rm med}$ & $\log (\nu F_\nu)^{\rm +Ssil}_{\rm med}$ \\
 {[Hz]} & \multicolumn{9}{c}{[normalized to median $\nu F_\nu$]} \\
\hline
  9.936  &  -5.309  &  -4.936  &   -4.293  &  -3.907  &  -3.298   &  -4.011  &  -4.478  &  -4.368  &  -3.335  \\
  13.41  &   0.182  &   0.462  &    0.676  &   0.766  &   1.322   &   0.644  &   0.569  &   0.611  &   0.709  \\
 13.644  &   0.006  &   0.131  &    0.197  &   0.431  &   0.499   &   0.328  &   0.168  &   0.197  &     ---  \\
 13.725  &   0.008  &   0.045  &    0.096  &    0.28  &   0.445   &   0.047  &   0.118  &   0.096  &     ---  \\
 13.836  &  -0.056  &  -0.002  &     0.02  &   0.156  &    0.38   &   0.087  &     0.0  &    0.02  &     ---  \\
 14.146  &  -1.023  &  -0.674  &   -0.189  &   0.003  &   0.407   &  -0.266  &  -0.546  &  -0.402  &  -0.168  \\
 15.133  &  -1.052  &  -0.866  &   -0.405  &  -0.085  &   0.038   &  -1.009  &   -0.18  &  -0.866  &  -0.077  \\
 15.305  &  -1.416  &  -1.112  &   -0.783  &  -0.364  &     0.0   &  -1.128  &  -0.387  &  -1.064  &     0.0  \\
 18.049  &   -0.25  &  -0.033  &    0.165  &   0.253  &   0.919   &  -0.032  &   0.236  &     0.0  &   0.269  \\
 19.112  &  -0.195  &   0.108  &     0.51  &   0.796  &   0.927   &   0.142  &   0.747  &   0.231  &   0.747  \\

\hline \\[-0.25cm]
\multicolumn{10}{c}{\normalsize \bf Seyfert 2} \\
\hline
 $\log \nu$ & $\log (\nu F_\nu)_{10}$ & $\log (\nu F_\nu)_{30}$ & $\log (\nu F_\nu)_{\rm med}$ & $\log (\nu F_\nu)_{70}$ & $\log (\nu F_\nu)_{90}$ & $\log (\nu F_\nu)^{\rm <EW113}_{\rm med}$ & $\log (\nu F_\nu)^{\rm >EW113}_{\rm med}$ & $\log (\nu F_\nu)^{\rm -Ssil}_{\rm med}$ & $\log (\nu F_\nu)^{\rm +Ssil}_{\rm med}$ \\
 {[Hz]} & \multicolumn{9}{c}{[normalized to median $\nu F_\nu$]} \\
\hline
  9.937  &  -5.163  &  -4.825  &    -4.26  &  -3.924  &  -2.734  & -4.825  &  -3.989  &   -4.26  &  -4.825  \\
 13.411  &   0.007  &   0.236  &    0.306  &    0.51  &   1.417  &  0.261  &   0.393  &    0.27  &   0.024  \\
 13.645  &  -0.011  &   0.036  &    0.115  &   0.155  &   0.289  &  0.115  &   0.073  &   0.115  &   0.298  \\
 13.726  &  -0.028  &    0.05  &    0.097  &   0.239  &   0.313  &  0.105  &   0.097  &   0.079  &   0.158  \\
 13.837  &  -0.151  &  -0.005  &    0.063  &   0.148  &   0.231  &  0.097  &  -0.142  &   0.081  &     0.0  \\
 14.147  &  -0.967  &  -0.577  &   -0.275  &     0.0  &   0.733  & -0.622  &  -0.024  &  -0.466  &   0.112  \\
 15.134  &  -1.375  &  -0.733  &    -0.25  &   0.008  &   0.192  & -0.951  &   0.038  &  -0.335  &   0.165  \\
 15.306  &  -1.582  &  -0.919  &   -0.282  &  -0.167  &   0.031  & -0.898  &  -0.271  &   -0.47  &  -0.271  \\
  18.05  &  -0.599  &  -0.172  &   -0.019  &   0.438  &   0.908  & -0.069  &    0.52  &  -0.069  &  -0.038  \\
 19.113  &   0.021  &   0.117  &    0.293  &   0.625  &   1.085  &  0.123  &   0.881  &   0.124  &   0.293  \\

\hline \\[-0.25cm]
\multicolumn{10}{c}{\normalsize \bf LINERs} \\
\hline
 $\log \nu$ & $\log (\nu F_\nu)_{10}$ & $\log (\nu F_\nu)_{30}$ & $\log (\nu F_\nu)_{\rm med}$ & $\log (\nu F_\nu)_{70}$ & $\log (\nu F_\nu)_{90}$ & $\log (\nu F_\nu)^{\rm <EW113}_{\rm med}$ & $\log (\nu F_\nu)^{\rm >EW113}_{\rm med}$ & $\log (\nu F_\nu)^{\rm -Ssil}_{\rm med}$ & $\log (\nu F_\nu)^{\rm +Ssil}_{\rm med}$ \\
 {[Hz]} & \multicolumn{9}{c}{[normalized to median $\nu F_\nu$]} \\
\hline
  9.927  &  -3.537  &  -3.372  &   -3.225  &  -3.084  &   -2.64  &  -3.289  &  -3.196  &  -3.098  &  -3.437 \\
 13.401  &   0.046  &   0.198  &    0.357  &   0.646  &   1.394  &   0.357  &   0.495  &   0.627  &    0.27 \\
 13.635  &  -0.054  &   0.053  &    0.159  &   0.266  &   0.373  &   0.159  &     ---  &     ---  &   0.159 \\
 13.716  &   0.119  &   0.229  &    0.339  &   0.449  &   0.559  &   0.339  &     ---  &     ---  &   0.339 \\
 13.827  &   0.237  &   0.357  &    0.476  &   0.596  &   0.715  &   0.476  &     ---  &     ---  &   0.476 \\
 14.137  &   -0.36  &  -0.079  &    0.106  &   0.321  &   0.542  &  -0.066  &   0.214  &   0.321  &  -0.416 \\
 15.124  &  -0.747  &   -0.21  &      0.0  &   0.019  &   0.209  &   0.048  &  -0.515  &     0.0  &  -0.134 \\
 15.296  &  -0.995  &  -0.304  &   -0.205  &  -0.058  &   0.237  &  -0.163  &  -0.567  &  -0.205  &  -0.019 \\
 18.039  &  -0.353  &     0.0  &      0.0  &     0.0  &   0.049  &     0.0  &     0.0  &     0.0  &     0.0 \\
 19.102  &  -0.039  &   0.027  &    0.135  &    0.25  &   0.293  &     0.0  &   0.207  &   0.278  &  -0.032 \\
\hline\\[-0.25cm]
 \end{tabular}
\begin{tablenotes}
\footnotesize
\item \textbf{Notes.} For each observed frequency (Hz) (col.1), are given the percentiles 10, 30, 50 (median), 70, and 90 (cols. 2-6) calculated at each frequency for the $\nu F_\nu$ distribution of individual sources. The latter have been previously normalized by their median $\nu F_\nu$ over the frequency range. The last four columns (cols. 7-10) correspond to the median $\nu F_\nu$ distribution for sources below ($<$EW113) or above ($>$EW113) the median EW(PAH11.3 $\mu m$) value of their class (0.095 for Sy1, 0.119 for HBL, 0.539 for Sy2), and sources with negative (-Ssil) or positive (+Ssil) silicate strength values. 
\end{tablenotes}
\end{table*}

\subsection{Median spectral energy distributions}

\begin{figure*}[ht!]
\includegraphics[width = 0.5\textwidth]{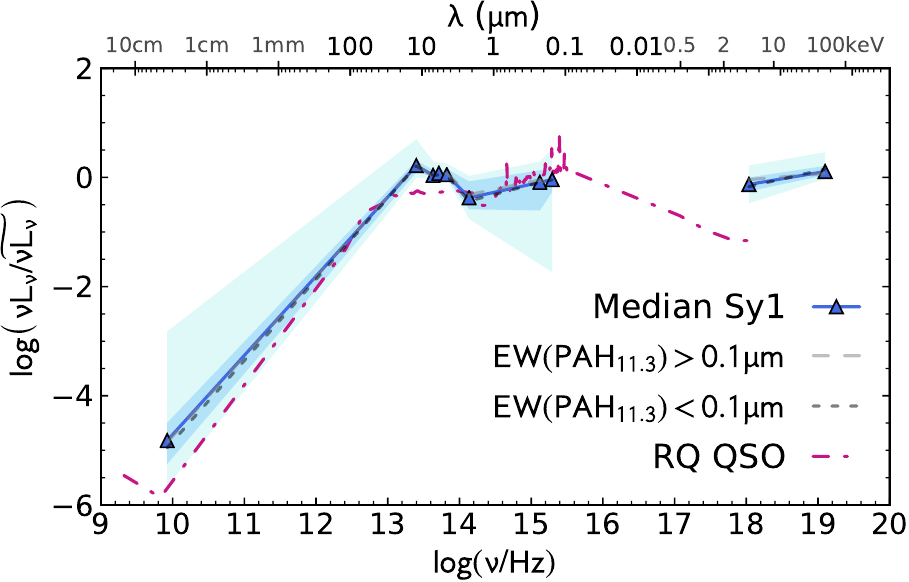}~
\includegraphics[width = 0.5\textwidth]{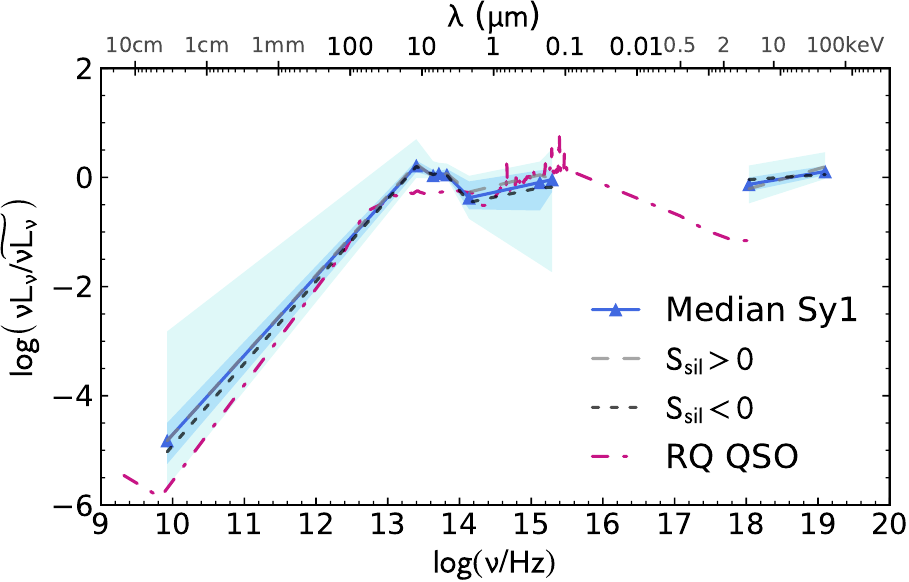}
\caption{Median rest-frame SED for the Seyfert type 1 galaxies, compared to the SED of the radio-quiet quasar population from \citet{Shang2011}, shown as a red dot-dashed line. {\bf (a:)} The population has been divided between the galaxies with high and low value of the EW of the PAH11.3$\mu$m feature {\bf (b:)} The population has been divided between the galaxies with the silicates 9.7$\mu$m feature in emission and in absorption.
\label{fig:Sy1_medianSED}}
\end{figure*}

\begin{figure*}[ht!]
\includegraphics[width = 0.5\textwidth]{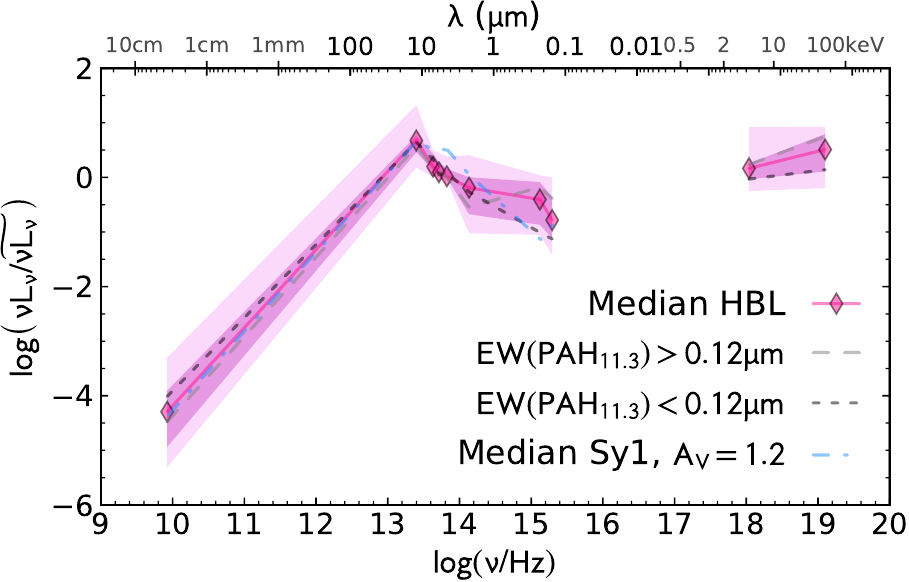}~
\includegraphics[width = 0.5\textwidth]{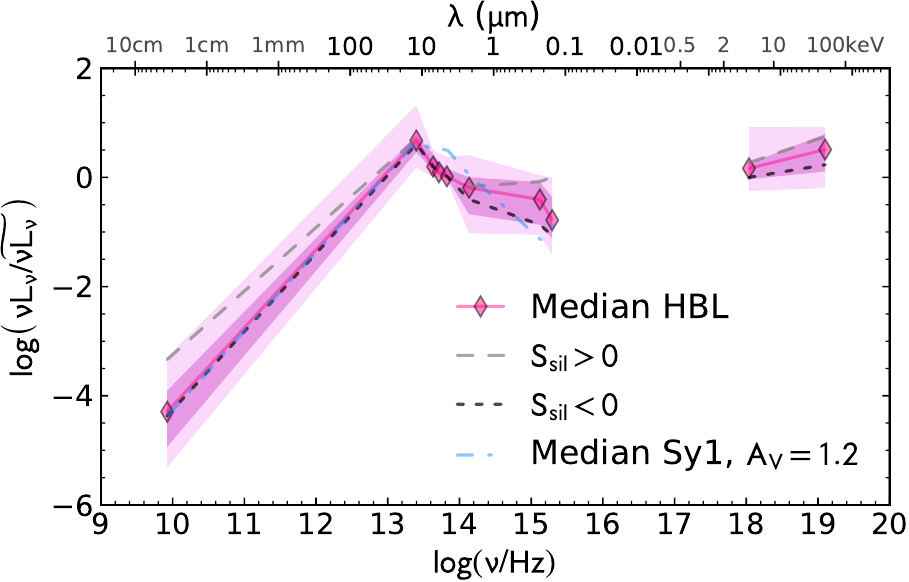}
\caption{ Median rest-frame SED for the HBL galaxies. {\bf (a:)} Same as in Fig. \ref{fig:Sy1_medianSED}a {\bf (b:)} Same as in Fig. \ref{fig:Sy1_medianSED}b. 
\label{fig:hbl_medianSED}}
\end{figure*}

\begin{figure*}[ht!]
\includegraphics[width = 0.5\textwidth]{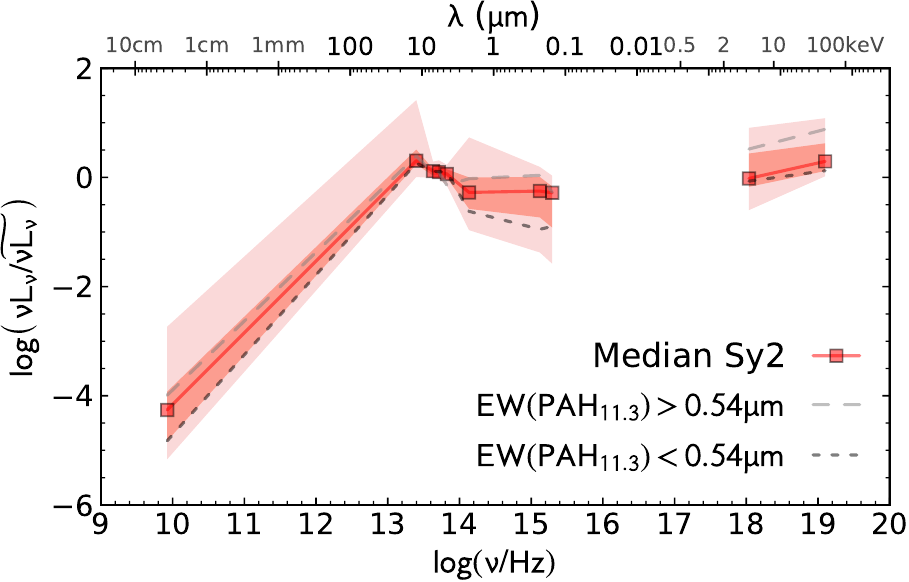}~
\includegraphics[width = 0.5\textwidth]{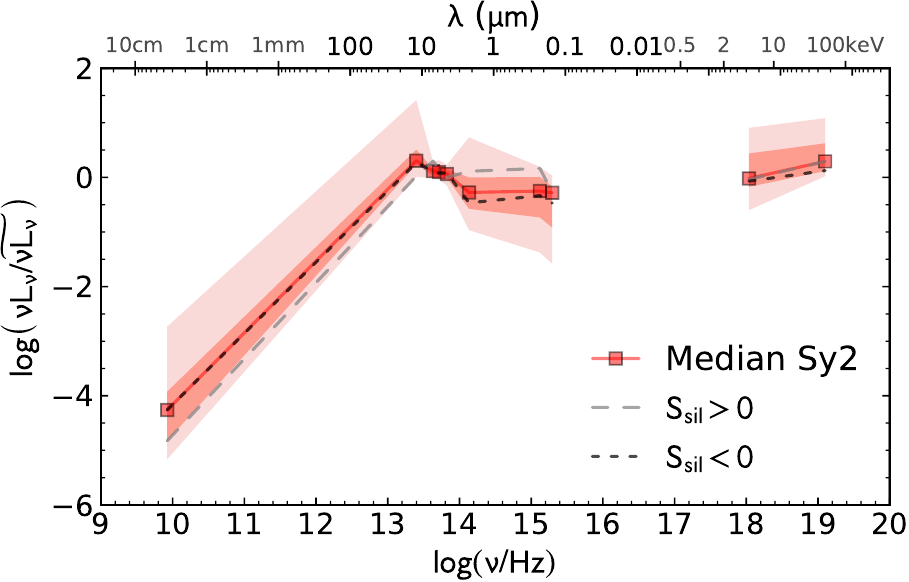}
\caption{Median rest-frame SED for the Seyfert type 2 galaxies. {\bf (a:)} Same as in Fig. \ref{fig:Sy1_medianSED}a {\bf (b:)} Same as in Fig. \ref{fig:Sy1_medianSED}b.
\label{fig:Sy2_medianSED}}
\end{figure*}

\begin{figure*}[ht!]
\includegraphics[width = 0.5\textwidth]{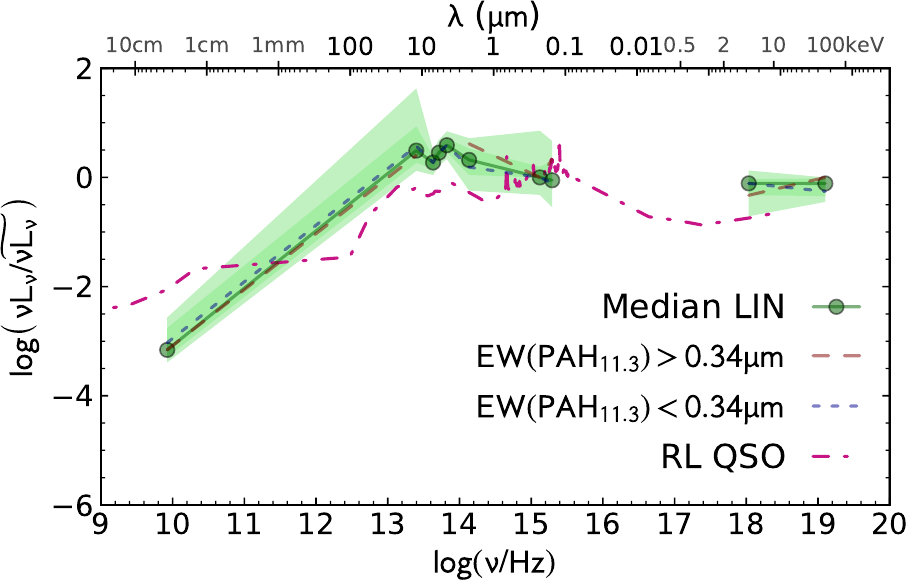}~
\includegraphics[width = 0.5\textwidth]{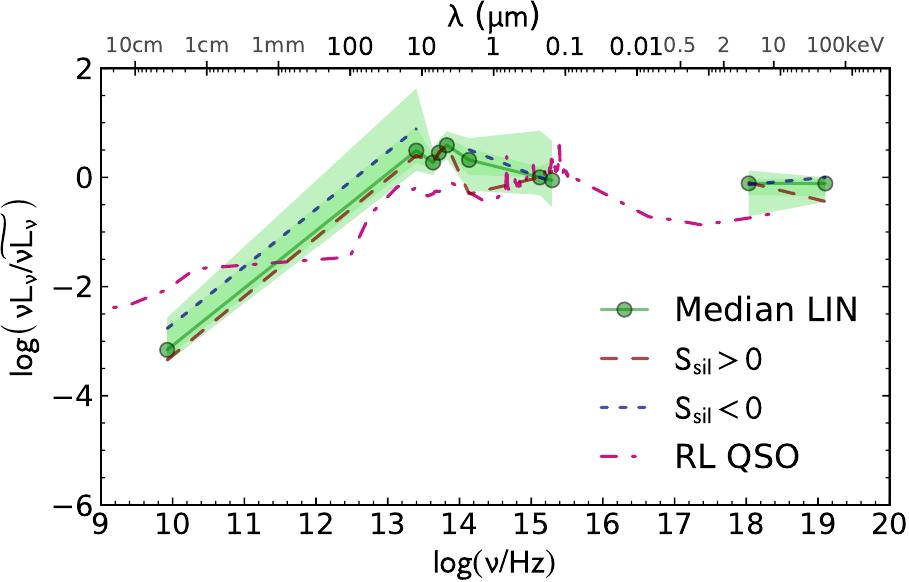}
\caption{Median rest-frame SED for the LINER galaxies, compared to the SED of the radio-loud quasar population from \citet{Shang2011}, shown as a red dot-dashed line. {\bf (a:)} Same as in Fig. \ref{fig:Sy1_medianSED}a {\bf (b:)} Same as in Fig. \ref{fig:Sy1_medianSED}b.
\label{fig:lin_medianSED}}
\end{figure*}

We have computed the median SEDs of the four classes of AGN in our sample: Seyfert type 1, HBL, Seyfert type 2, and LINER galaxies, which are shown in Figs. \ref{fig:Sy1_medianSED}--\ref{fig:lin_medianSED}, respectively, and the relevant values, for each of the 10 photometric bands, are reported in Table\,\ref{tab:medsed}. The individual SEDs of all galaxies of our sample, normalized to their bolometric luminosity, are shown in the Appendix \ref{sed_ind}. The Seyfert type 1 galaxies are shown in Figs. \ref{fig:plot_sy1_1}--\ref{fig:plot_sy1_2}, the Hidden Broad Line Galaxies (HBL) galaxies in Fig. \ref{fig:plot_hbl}, the Seyfert type 2 in Fig. \ref{fig:plot_sy2}, while in Fig. \ref{fig:plot_nsy} are shown the normalized SED of the non-Seyfert galaxies, including LINERs, a galaxy classified as a non-Seyfert (NGC\,1056) and another one classified as a Starburst (NGC\,6810).

Our purpose is twofold. First, we define the median distributions of the four classes of AGN, because they can be used as local templates, to be compared with SED of high-redshift AGN. Secondly, we want to explore if and how the four sub-samples of AGN may still include, even after the aperture corrections we have applied to subtract the galaxy contribution, a significant contribution from starlight emission, or may be affected by dust obscuration (see Section\,\ref{sedcont}).

For each AGN class, the median SED has been derived from the median value of the individual SED distribution evaluated at each frequency sampled (solid line and symbols in Figs.\,\ref{fig:Sy1_medianSED}--\ref{fig:lin_medianSED}). Prior to the median computation, the individual SEDs were normalized to their median $\widetilde{\nu L_\nu}$ value over the frequency range. This approach avoids the arbitrary selection of a reference wavelength for the normalization that would produce a zero dispersion band in the median SED. To account for the dispersion around the median SED we also compute, at each wavelength, the 10, 30, 70, and 90  percentiles (see Table\,\ref{tab:medsed}). The dark-shaded areas in Figs.\,\ref{fig:Sy1_medianSED}--\ref{fig:lin_medianSED} indicate the SED distribution between percentiles 30 and 70, whereas the light-shaded areas correspond to the interval between percentiles 10 and 90.

\section{Discussion}\label{discuss}

\subsection{Median SEDs and host galaxy contribution}\label{sedcont}
Overall, the median SEDs show two maximum values, one in the mid-IR and another in the X-ray range. Sy1 galaxies clearly show a rising optical/UV continuum with increasing frequency, whereas HBL, Sy2 and LINER galaxies tend to show a rather flat or a depressed optical/UV continuum, possibly contaminated by the host galaxy light (see discussion below). For comparison, we have included in Fig.\,\ref{fig:Sy1_medianSED} the median SED of the radio-quiet quasars of the sample of \citet{Shang2011}. The two determinations agree very well in the spectral region from the radio to the UV, while at X-ray frequencies the derivation of \citet{Shang2011} is about one order of magnitude fainter.
The blue optical/UV continuum in Sy1 nuclei is similar to the big blue bump associated with the accretion disk emission in bright quasars \citep{malkan1982}, although the median quasar SEDs show a more pronounced blue bump, a relatively fainter IR bump and fainter X-ray emission  \citep{elvis1994,Krawczyk2013,Saccheo2023}. This is in agreement with the median SED changes reported by \citet{Krawczyk2013} for the less luminous quasars in their sample ($\sim 10^{44.9}\, \rm{erg\,s^{-1}}$), which show a less pronounced blue bump component. The median Sy1 SED in Fig.\,\ref{fig:Sy1_medianSED} extends this trend to lower luminosities, 
with a prominent IR bump brighter than the blue bump component in agreement with the median Seyfert SEDs derived from sub-arcsecond resolution observations by \citet{Prieto2010}.  These changes in the SED are probably due to the decreasing contribution of the accretion disk to the total energy output, or to the increasing fraction of the reprocessed emission in AGN with decreasing luminosity.

When compared to Seyfert galaxies, the median SED of LINERs is brighter at radio wavelengths, in agreement with the results obtained by \citet{Ho2008}. For comparison, we have included in Fig.\,\ref{fig:lin_medianSED} the median SED of the radio-loud quasars of the sample of \citet{Shang2011}. These SEDs do agree to first order, showing a certain radio loudness in the LINER SED, in spite of the fact that our LINER sample is small and not fully representative.
Obscured nuclei, namely HBL and Sy2, shown in Fig.\,\ref{fig:hbl_medianSED} and in Fig.\,\ref{fig:Sy2_medianSED}, respectively, 
have a depressed optical/UV continuum when compared to Seyfert 1 nuclei and a larger scatter in both optical/UV and X-ray ranges, suggesting that obscuration by dust and gas is still a source of scatter even after the absorption corrections have been applied. On the other hand, LINERs show a relatively flat SED in the optical/UV range, possibly dominated by the starlight contribution from the host galaxy. 

Irrespective of the type of AGN considered, the four median SEDs have their minimum scatter in the mid-IR, namely in the four photometric bands at 4.5, 5.8, 7.0\,$\mu$m and, to a lesser extent, at 12\,$\mu$m. This result confirms that the mid-IR continuum contains a constant fraction of the bolometric flux for all types of active galaxies \citep{spinoglio1989}, a discovery that was initially based on large-aperture flux densities from IRAS \citep{Neugebauer1984}. In this work we come to the same conclusion using nuclear 12\,$\mu$m flux measurements at sub-arcsecond resolution, avoiding most of the contribution by the host galaxy emission, and the corresponding corrected mid-IR fluxes from {\it Spitzer}.

To evaluate the possible contamination in the optical/UV range of the median AGN SEDs due to current star formation in the host galaxies, we use the equivalent width of the PAH feature at 11.25$\mu$m as a proxy for the star formation rate \citep[see, e.g.][]{Forster-Schreiber2004, mordini2021}, which is not affected by AGN contamination \citep[e.g.][]{lai2022}. Additionally, dust obscuration can also affect the shape of the optical/UV continuum, either due to nuclear dust (i.e. the torus) or due to host galaxy dust. To investigate this scenario we use the strength of the silicate feature at 9.7\,$\mu$m ($\rm S_{sil}$; \citealt{Spoon2007}), which becomes negative for obscured sources but can also be positive for unobscured sources if dust irradiated by the active nucleus is seen along the line of sight. We have therefore divided the objects in each of the four classes on the basis of these two observed quantities, using as threshold (between low and high values) the median value of the PAH at 11.25$\mu m$, and negative or positive values of $\rm S_{sil}$. The spectroscopic data have been taken from \citet{wu2009}, who measured most of the 12$\mu$m AGN with the {\it Spitzer}-IRS spectrograph in the low resolution mode.

Fig. \ref{fig:Sy1_medianSED}a shows the median SEDs obtained from the Seyfert 1 sample with EW(PAH\,11.25$\mu$m) values below (dotted-green line) and above (dashed-red line) 0.1\,$\mu$m. 
Following the SFR calibration based on the luminosity of the PAH\,11.25$\mu$m derived by \citet{Xie2019}, we obtain a median SFR$\sim 0.7$--$0.8\, \rm{M_\odot\,yr^{-1}}$ for both sub-samples, consistent with moderate star formation in the host galaxies of Seyfert 1s. Analogously, in Fig. \ref{fig:Sy1_medianSED}b, the population of type 1 Seyferts has been divided in those with the silicates in emission (dashed-red line) and in absorption (dotted-green line). The differences in the median SEDs between the Seyfert 1 sub-samples separated by EW(PAH\,11.25$\mu$m) or $\rm S_{sil}$ are negligible for all the wavelengths sampled, suggesting that the median Seyfert type 1 SED is not contaminated by stellar light and does not suffer from heavy dust absorption. 

Figs.\,\ref{fig:hbl_medianSED}a and b show the same analysis for HBL galaxies, which present a similar median EW(PAH\,11.25$\mu$m) value of 0.12\,$\mu$m as Seyfert galaxies. Nevertheless, the HBL populations with EW(PAH\,11.25$\mu$m) $>$ 0.12\,$\mu$m or positive $\rm S_{sil}$ values show a significantly higher continuum in both NUV and FUV bands (dashed-green lines in Figs.\,\ref{fig:hbl_medianSED}a and b) when compared with those below EW(PAH\,11.25$\mu$m) $<$ 0.12\,$\mu$m or negative $\rm S_{sil}$ (dotted-blue lines). 
When the PAH\,11.25 luminosities are translated into SFRs, the sub-sample with EW(PAH\,11.25$\mu$m) $>$ 0.12\,$\mu$m has a median SFR$\sim 0.9\, \rm{M_\odot\,yr^{-1}}$, while the sub-sample with EW(PAH\,11.25$\mu$m) $<$ 0.12\,$\mu$m shows a median SFR$\sim0.6\, \rm{M_\odot\,yr^{-1}}$. As in the case of Seyfert 1 nuclei, the contribution from star formation is moderate, however HBL nuclei are dust obscured and therefore, at similar SFR values, the host galaxy dominates in the UV range. 
Sources with lower star formation rates tend to show a decreasing IR-to-UV continuum and the silicate feature in absorption, consistent with a reddened nuclear continuum, as discussed in Section\,\ref{dust}.

Seyfert 2 galaxies have significantly larger EW(PAH\,11.25$\mu$m) with a median value of 0.54\,$\mu$m, i.e. about five times that of Seyfert 1 and HBL galaxies, meaning that their host galaxies are relatively more active at forming stars in agreement with previous studies \citep{emr87, maiolino1995,buchanan2006}. As in the case of HBL, Figs.\,\ref{fig:Sy2_medianSED}a and b show that the median SED of the Seyfert 2 population with large EW(PAH\,11.25$\mu$m) or positive $\rm S_{sil}$ values present a brighter continuum in the UV but also the \textit{K}-band 
 (dashed-green line in Figs.\,\ref{fig:Sy2_medianSED}), 
in agreement with their higher median SFR of $\sim 4\, \rm{M_\odot\,yr^{-1}}$. However, the population with lower EW(PAH\,11.25$\mu$m) or negative $\rm S_{sil}$ values may still be affected by this contribution, since they show a relatively flat IR-to-UV continuum (dotted-blue line), 
although their median SFR$\sim 0.6\, \rm{M_\odot\,yr^{-1}}$ is comparable to that of Seyfert 1 and HBL host galaxies. 

\begin{figure}[htb!!!]
\includegraphics[width = \columnwidth]{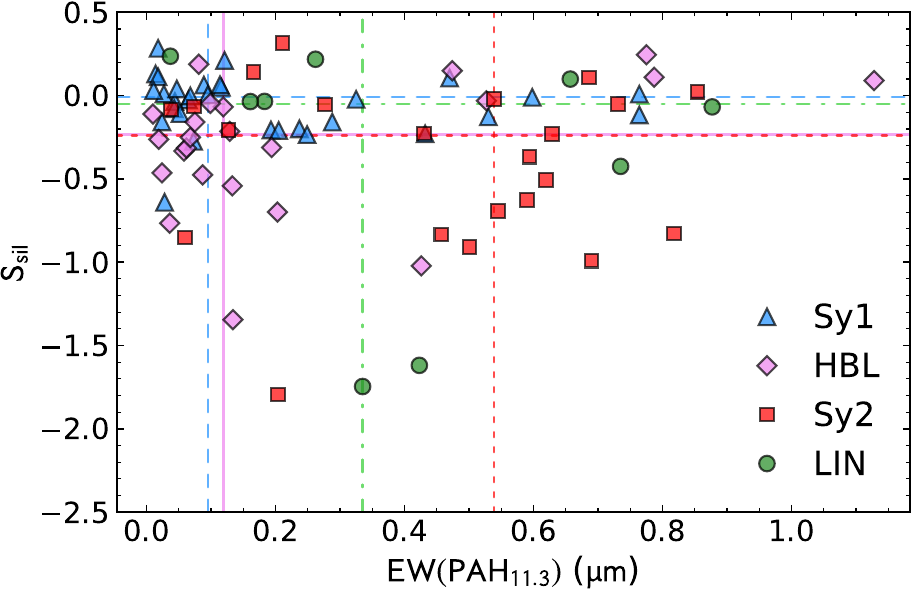} 
\caption{Silicate strength as a function of the EW of the PAH11.3$\mu$m feature for the four AGN populations of Sy1 (blue triangles), HBL (pink diamonds), Sy2 (red squares) and LINERs (green circles). The median EW(PAH11.3$\mu m$) and silicate strength values for Sy1 (dashed blue), HBL (solid pink), Sy2 (dotted red) and LINERs (dot-dashed green) are indicated by the horizontal and vertical lines, respectively.}\label{fig:EW113_Ssil}
\end{figure}

The differences between Seyfert 1, HBL, and Seyfert 2 galaxies are clearly shown in the diagram of the silicate strength at 9.7$\mu$m as a function of the PAH\,11.25$\mu$m feature 
-- $\rm S_{sil}$ vs. EW(PAH\,11.25$\mu$m) -- reported in Fig.\,\ref{fig:EW113_Ssil}. Both HBL and Seyfert 2 galaxies are obscured by a similar amount according to their median $\rm S_{sil} \sim -0.23$, which is larger than that in Seyfert 1s (median $\rm S_{sil} \sim 0.0$). However, the host galaxies of HBLs and Seyfert 1s form stars at a similar rate ($0.6$--$0.9\, \rm{M_\odot\,yr^{-1}}$), about 4--6 times lower than that of Seyfert 2 hosts with EW(PAH\,11.25$\mu$m) $>$ 0.54\,$\mu$m.

The LINER population in our sample is less numerous and more heterogeneous with a large dispersion in EW(PAH\,11.25$\mu$m) and $\rm S_{sil}$ (Fig.\,\ref{fig:EW113_Ssil}). The median EW(PAH\,11.25$\mu$m) of 0.34$\mu$m is not representative of the population, with about half of the sources showing values similar to the Seyfert 1 nuclei 
(median SFR$\sim 0.1\, \rm{M_\odot\,yr^{-1}}$) and the others showing high values indicative of 
more intense star formation activity (median SFR$\sim 1.7\, \rm{M_\odot\,yr^{-1}}$). On the other hand, only two sources show strong silicate absorption, with a median $\rm S_{sil} \sim -0.05$. This dual behavior in the IR spectroscopic properties of LINERs is in agreement with previous studies \citep{dudik2009,jafo2021}. The differences between the median SEDs for the sub-populations with large/small EW(PAH\,11.25$\mu$m) or positive/negative $\rm S_{sil}$ are relatively small (Figs.\,\ref{fig:lin_medianSED}a and b) 
and may be attributed to the small number statistics and the large heterogeneity in the LINER class.

In summary, as can be seen in Fig.\,\ref{fig:EW113_Ssil}, most of the Seyfert type 1 AGN have a low 11.25$\mu$m PAH EW (only 5 Sy1 have EW(PAH\,11.25$\mu$m) $>$ 0.4\,$\mu$m) and have silicate strength around zero. Hidden Broad Line (HBL) galaxies have a larger spread in the silicate strength, while keeping a low number of high PAH EW (6 HBL have EW(PAH\,11.25$\mu$m) $>$ 0.4\,$\mu$m). In contrast, the majority of the Seyfert type 2 galaxies have a large PAH EW (14 Sy2 have EW(PAH\,11.25$\mu$m) $>$ 0.4\,$\mu$m, while only 7 have EW(PAH\,11.25$\mu$m) $<$ 0.4\,$\mu$m).

\subsection{Dust properties in obscured Seyferts}\label{dust}
Figures\,\ref{fig:hbl_medianSED}a and b show that the continuum shape of the median HBL SED with low star formation rates (low EW(PAH\,11.25$\mu$m) or negative $\rm S_{sil}$ (dashed-green lines) can be reproduced by the median Sy1 template after applying a moderate amount of dust reddening. Assuming a foreground dust screen and the extinction curve from \citet{Cardelli1989}, a total extinction of $A_V \sim 1.2\,\rm{mag}$ (dashed-blue line in Figs.\,\ref{fig:hbl_medianSED}a and b) is required. 
This moderate amount of dust obscuration is enough to hide the continuum emission of a Sy1 and thus reproduce the observed SED shape of an HBL nucleus. 

An additional estimate of the dust extinction in these nuclei is provided by the strength of the silicate feature. For a foreground dust screen, the silicate strength corresponds to the optical depth ($\rm S_{sil} = -\Delta \tau_{9.7}$). Comparing the differential extinction in the optical between Seyfert 1 and HBL nuclei ($A_V \sim 1.2\,\rm{mag}$), and the silicate optical depth ($\rm \Delta \tau_{9.7} \sim 0.23$) provides some information about the dust properties. We obtain a ratio of $A_V / \Delta \tau_{9.7} \sim 5.2$, in agreement with typical values in AGN \citep[e.g.][]{lyu2014}, as opposed to the much larger $A_V / \Delta \tau_{9.7} \sim 18$ derived for the local ISM dust in the Milky Way \citep{roche1984}. This suggests that the obscuration may be dominated by large grains that lead to a flatter and featureless extinction curve \citep{maiolino2001a,maiolino2001b,shao2017}. Nevertheless, the effect of a more complex geometry in the dust distribution should be explored in the future to confirm this scenario.

\subsection{Comparison with other bolometric corrections}\label{sec:bolcor}

\begin{figure}[t!!!]
\includegraphics[width = \columnwidth]{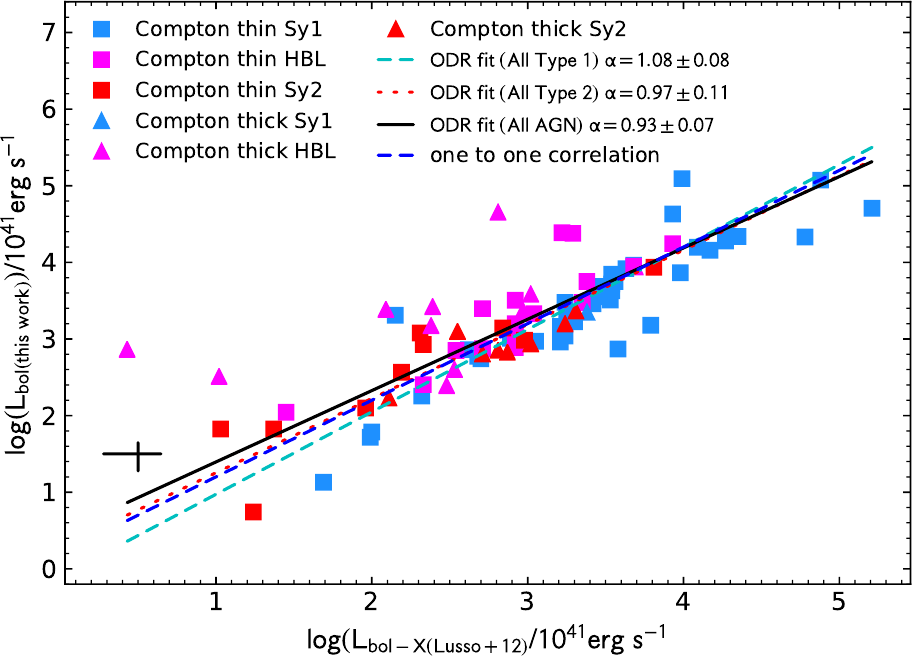}
\caption{Correlation between the bolometric luminosity computed in this work and the bolometric luminosity computed from the 2-10\,keV corrected flux using the bolometric correction of \citet{lusso2012}.}
\label{fig:ourlbol_lusso}
\end{figure}

We compare here the bolometric luminosities derived for type 1 and type 2 AGN in our sample with thosefrom other methods. First, in Fig. \ref{fig:ourlbol_lusso} we compare them with the bolometric luminosities computed from the 2-10\,keV intrinsic flux using the bolometric correction of \citet{lusso2012}. We adopted the polynomial formula for the bolometric luminosity $y = a_1\times x + a_2 \times x^2 + a_3\times x^3 + b$, where $x = {\rm \log(L) - 12}$, L is the bolometric luminosity in units of {\rm L$_{\odot}$}, $y = {\rm \log [L/L_{\rm 2-10 keV}]}$, and the coefficients depend on the AGN type (type 1 AGN: $a_1$ = 0.288, $a_2$ = 0.111, $a_3$ = -0.007 and $b$ = 1.308; type 2 AGN: $a_1$ = 0.230, $a_2$ = 0.050, $a_3$ = 0.001 and $b$ = 1.256; see table\,2 in \citealt{lusso2012}). It appears from Fig. \ref{fig:ourlbol_lusso} that the two determinations of the bolometric luminosities do agree very well, as shown by the linear slope of the correlation ($\alpha$ = 0.98$\pm$ 0.06).

\begin{figure}[t!!!]
\includegraphics[width = \columnwidth]{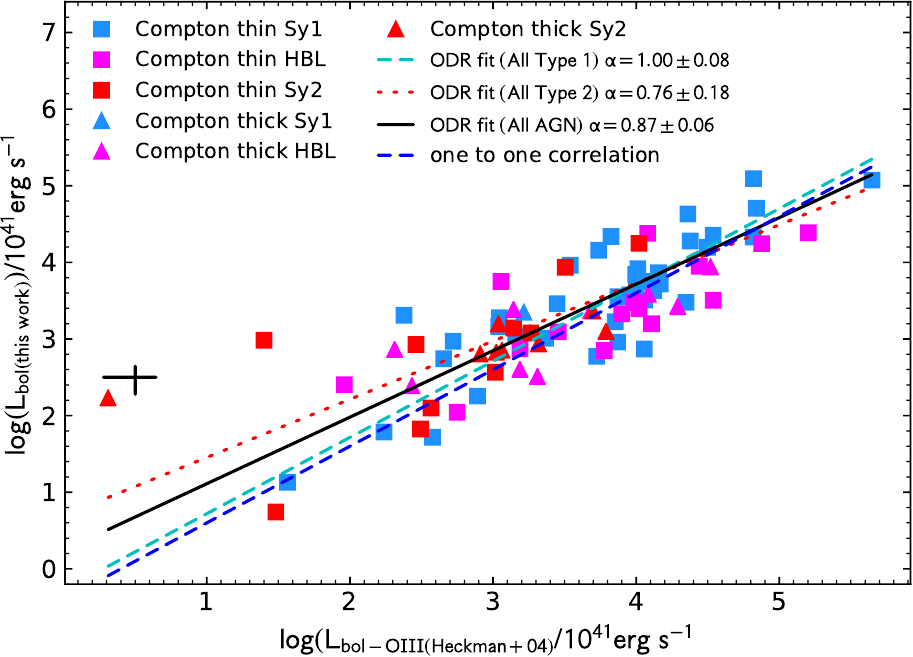}
\caption{Correlation between the bolometric luminosity computed in this work and the bolometric luminosity computed from the [OIII]5007\AA \ line compiled in \citet{spinoglio2022} using the bolometric correction of $L_{\rm bol} \sim 3500\,L_{\rm [OIII]5007 \mathring{A}}$ \citep{Heckman2004}.}
\label{fig:ourlbol_o3}
\end{figure}

In Fig. \ref{fig:ourlbol_o3} we show the comparison of our determination of the bolometric luminosity with the one derived using the [OIII]5007\AA~ line compiled in \citet{spinoglio2022} using the bolometric correction of $L_{\rm bol} \sim 3500\,L_{\rm [OIII]5007 \mathring{A}}$ \citep{Heckman2004}. 
The correlation here is flatter than linear, especially for type 2 objects ($\alpha$ = 0.66$\pm$ 0.16), indicating a deficit of optical forbidden line emission from type 2 AGN, possibly due to obscuration of the emitting NLR.


A recent computation of the hard X-ray 2-10\,keV bolometric correction has been provided by \citet{Duras2020}, who emphasize the dependence of their bolometric corrections on the bolometric luminosity and separately analyze type 1 and type 2 AGN. In the luminosity range 
of $10^{40}$ $\lesssim$ $L_{\rm bol}$ $\lesssim$ $10^{45.5}\,\rm{erg\,s^{-1}}$, we have a similar result when we compare our constant of 15.3, with no dependence on luminosity (see the first line in Table \ref{tab:lboltracers}), with their result presented in the lower panel of their Fig. 4, relative to the average values for type 1 and type 2 sources. Only at higher bolometric luminosities, i.e. for $L_{\rm bol} \gtrsim 10^{46}\,\rm{erg\,s^{-1}}$ the predictions of \citet{Duras2020} show higher bolometric corrections by factors of 3--4, with a strong dependence on luminosity. This discrepancy can be due to the fact that our study is based on a local sample of AGN with moderate luminosities, with $L_{\rm bol} \lesssim 10^{46}\,\rm{erg\,s^{-1}}$, with the only exception of 3C\,273 with $L_{\rm bol} \sim 4\times 10^{47}\,\rm{erg\,s^{-1}}$ (see Table \ref{tab:sample2}). In contrast,  \citet{Duras2020} include in their study various samples of AGN, X-ray selected, covering different redshift ranges. In particular they include 41 sources of the \textsc{WISE}-\textsc{SDSS} Selected Hyper-luminous sample (WISSH) with $L_{\rm bol} > 10^{47}\,\rm{erg\,s^{-1}}$ in the redshift range $2 < z < 4$, observed in the X-rays \citep{Martocchia2017}, as well as 31 high luminosity type 1 AGN with $L_{\rm bol} > 10^{46.5}\,\rm{erg\,s^{-1}}$ in the redshift range $0.9 < z < 5$ from the XXL sample \citep[see Table 2 of][]{Liu2016}. The inclusion of these high luminosity type 1 AGN has the effect of strongly increasing the 2-10\,keV bolometric correction, as can be seen from the upper panel of Fig. 4 of \citet{Duras2020}.

\begin{figure}[ht!!!]
\includegraphics[width = \columnwidth]{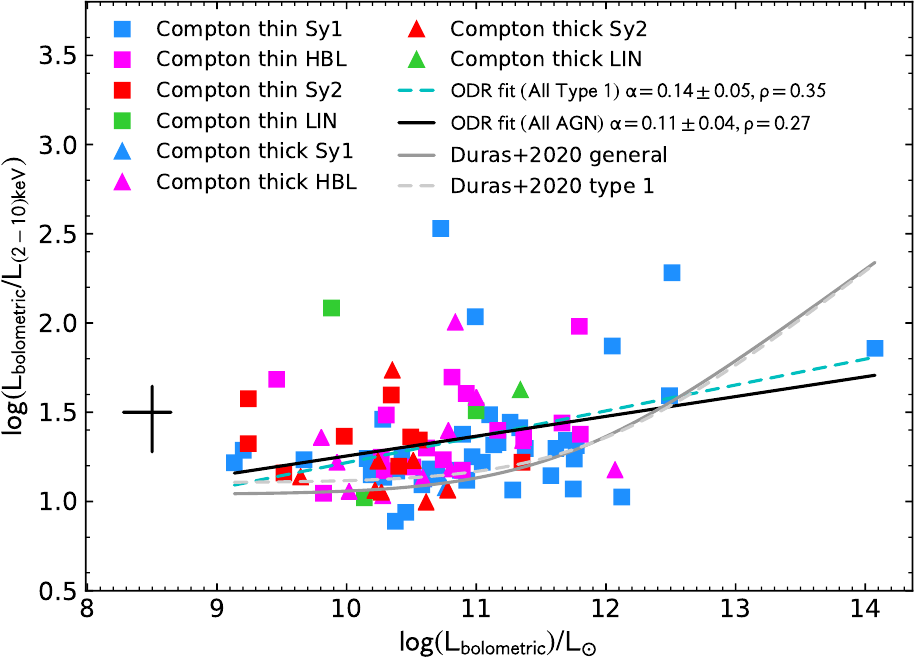}
\caption{The 2-10\,keV bolometric correction as a function of the bolometric luminosity. The ODR fits to the total population and to the type 1 AGN samples have a low Pearson correlation coefficient ($\rho$=0.46 and $\rho$=0.36 for 57 and 90 objects, with P(null)= 3$\times$10$^{-4}$ and  4$\times$10$^{-4}$, respectively) indicating that the correlations are not strong. For comparison the hard X-ray bolometric correction from \citet{Duras2020} has been included for both the general case and for type 1 AGN. } 
\label{fig:2-10bolcor}
\end{figure}

\begin{figure}[ht!!!]
\includegraphics[width = \columnwidth]{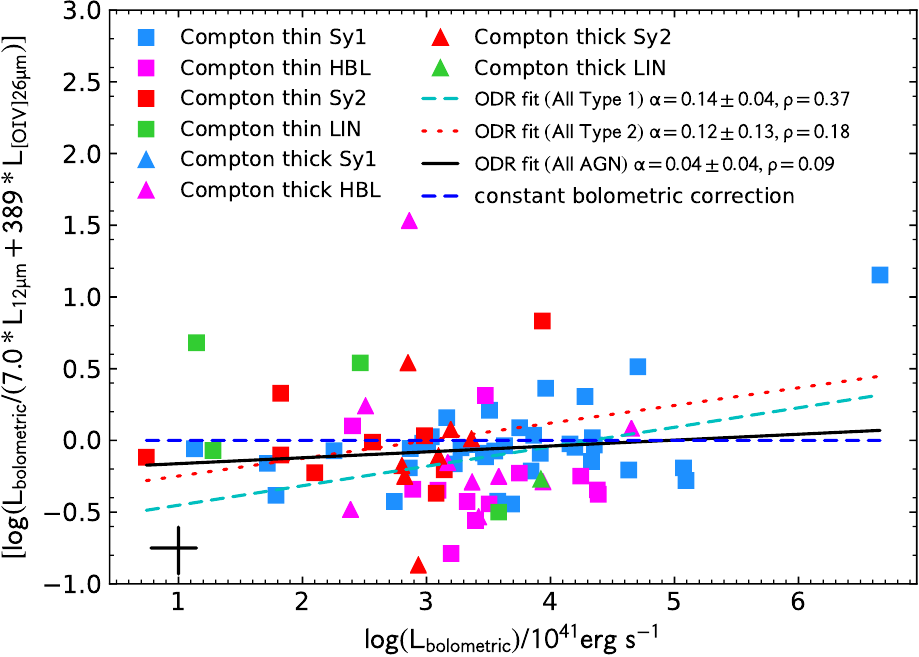}
\caption{ The composite 12$\mu$m and [OIV]26$\mu$m bolometric correction as a function of the bolometric luminosity. The ODR fits to all populations show very low Pearson correlation coefficients, indicating a constant bolometric correction. }
\label{fig:compositebolcor}
\end{figure}

We show in Fig.\,\ref{fig:2-10bolcor} our 2-10\,keV bolometric correction as a function of the bolometric luminosity. 
For comparison, we have included in the plot the fit of the hard X-ray bolometric correction derived from \citet{Duras2020} (see their quation (2) and Fig. 4) for both the general case and for type 1 AGN. In the luminosity range shown in the figure, our results are in a reasonably good agreement with their result. Nevertheless, it can be seen that the derived statistics shown in Fig.\,\ref{fig:2-10bolcor} indicates a poor correlation, that is consistent with a flat distribution, i.e. a constant bolometric correction as a function of luminosity.
The same conclusion of a constant bolometric correction is also valid for the combination of the 12$\mu$m luminosity and the [OIV]26$\mu$m line luminosity, as is shown in Fig. \ref{fig:compositebolcor}.




\section{Summary and conclusions}

In this work we have used the 12$\mu$m sample of local AGN to derive the nuclear SED and the AGN bolometric luminosities, as much as possible free from galactic contamination.

For each galaxy we used a 10-band spectral energy distribution, from the radio to the hard X-rays. To isolate 
genuine AGN continuum, we included sub-arcsecond photometric data where available, and corrected the bands contaminated by stellar light from the host galaxy. Both the radio observations at 8.4\,GHz and most of the 12$\mu$m photometric data are taken with sub-arcsecond apertures, while the mid-IR bands at 7.0, 5.8, and 4.5\,$\mu$m have been corrected using the 12$\mu$m data. The nuclear K-band photometry was taken from \citet{spinoglio2022} and the UV data from GALEX have been corrected for the telescope PSF in the two NUV and FUV bands. Finally, the absorption-corrected 2-10\,keV and observed 14-195\,keV X-ray fluxes intrinsically originate in the active nuclei.

Using the 10-band {\it nuclear} photometric data, we derived the median Spectral Energy Distributions (SED) for each AGN type, namely Seyfert type 1, Seyfert nuclei with hidden broad-lines (HBL), Seyfert type 2 and LINERs. The median Seyfert 1 SED shows the characteristic big blue bump feature in the UV, nevertheless the largest contribution to the bolometric luminosity comes from the IR peak and the X-ray continuum. The median SEDs of both HBL and type 2 nuclei are affected by starlight contamination in the optical/UV range. The median nuclear SED of obscured type 1 nuclei can be reproduced by applying a moderate foreground dust extinction of $A_V \sim 1.2\,\rm{mag}$ to the median Seyfert 1 SED.

We find that the 12$\mu$m and the \textit{K}-band nuclear luminosities have good linear correlations with the bolometric luminosity, similar to those in the X-rays. We derive bolometric corrections for either continuum bands (K-band, 12$\mu$m, 2-10\,keV and 14-195\,keV) and narrow emission lines (mid-IR high ionization lines of [OIV] and [NeV] and optical [OIII]5007\AA) as well as for combinations of IR continuum and line emission. We find that a combination of continuum plus line emission accurately predicts the bolometric luminosity up to quasar luminosities ($10^{46}\,\rm{erg\, s^{-1}}$). This is the case of the 12$\mu$m continuum plus the [OIV]26$\mu$m or the [NeV]14.3$\mu$m mid-IR lines, and the 2.2$\mu$m continuum plus the [OIII]5007\AA \ optical line. This result reflects the fact that the IR continuum includes a large fraction of (and is proportional to) the bolometric luminosity of the AGN, through both the nuclear continuum emission from hot dust/torus and the gas emission from the narrow-line region. %

The \textit{James Webb Space Telescope} \citep[][]{Gardner2006} will be able to measure this bolometric luminosity in virtually any AGN in the local Universe. For AGN at redshifts z$<$1 the [NeV]14.3$\mu$m line can also be used.

\begin{acknowledgments}
This research is based on ultraviolet observations made with the Galaxy Evolution Explorer, obtained from the MAST data archive at the Space Telescope Science Institute, which is operated by the Association of Universities for Research in Astronomy, Inc., under NASA contract NAS 5–26555. 
This research has made use of the NASA/IPAC Extragalactic Database (NED), which is operated by the Jet Propulsion Laboratory, California Institute of Technology, under contract with the National Aeronautics and Space Administration.
JAFO acknowledges financial support by the Spanish Ministry of Science and Innovation (MCIN/AEI/10.13039/501100011033) and ``ERDF A way of making Europe'' though the grant PID2021-124918NB-C44; MCIN and the European Union -- NextGenerationEU through the Recovery and Resilience Facility project ICTS-MRR-2021-03-CEFCA. We thank the anonymous referee for his/her constructive report that helped to improve the paper and correct an error in the X-ray fluxes conversion.
\end{acknowledgments}

\bibliography{bolometric_luminosities_local_agn}{}
\bibliographystyle{aasjournal}


\clearpage
\begin{ThreePartTable}
\setlength{\tabcolsep}{2.pt}
\setlength{\LTcapwidth}{\textwidth}
\scriptsize
\begin{longtable}{rlccccc}
\caption{The AGN sample: coordinates, redshift, types {\bf and radio loudness (flagged as RL)}.{(\bf Note: This table is the full version of Table \ref{tab:sample0} and will be published in electronic form).} }\label{tab:sample0long}
\\ \hline\\[-0.3cm]
\bf n. & \bf Name         & \bf R.A. (J2000.0)  & \bf Dec. (J2000.0) & \bf $z$     & \bf Type & Radio \\
       & (1)             & (2)               & (3)               & (4)         & (5)      & (6) \\
\hline\\[-0.4cm]
\endfirsthead
\caption{continued.}\\
\hline\\[-0.3cm]
\bf n. &\bf Name        & \bf R.A. (J2000.0)  & \bf Dec. (J2000.0) & \bf $z$     & \bf Type & Radio \\
       & (1)            & (2)               & (3)               & (4)         & (5)       & (6) \\
\hline\\[-0.4cm]
\endhead
\endfoot
\hline\\
\endlastfoot
  1 &  MRK0335          &  00:06:19.5  & +20:12:10  &  0.0258  & Sy1     &   \\  
  2 &  NGC34=Mrk938=N17 &  00:11:06.5  & -12:06:26  &  0.0196  & Sy2     &   \\  
  3 &  IRASF00198-7926  &  00:21:57.0  & -79:10:14  &  0.0728  & Sy2     &   \\  
  4 &  ESO012-G021      &  00:40:47.8  & -79:14:27  &  0.0300  & Sy1     &   \\  
  5 &  NGC0262=MRK348   &  00:48:47.1  & +31:57:25  &  0.0150  & HBL     & RL \\  
  6 &  Izw001=UGC00545  &  00:53:34.9  & +12:41:36  &  0.0611  & Sy1     &   \\  
  7 &  IRASF00521-7054  &  00:53:56.2  & -70:38:03  &  0.0689  & HBL     &   \\  
  8 &  ESO541-IG012     &  01:02:17.5  & -19:40:09  &  0.0566  & Sy2     &   \\  
  9 &  NGC0424          &  01:11:27.5  & -38:05:01  &  0.0118  & HBL CT  &   \\ 
 10 &  NGC0526A         &  01:23:54.2  & -35:03:56  &  0.0191  & Sy1     &   \\  
 11 &  NGC0513          &  01:24:26.8  & +33:47:58  &  0.0195  & HBL     &   \\  
 12 &  IRASF01475-0740  &  01:50:02.7  & -07:25:48  &  0.0177  & HBL     & RL  \\  
 13 &  MRK1034NED02     &  02:23:22.0  & +32:11:50  &  0.0338  & Sy1     &   \\  
 14 &  ESO545-G013      &  02:24:40.2  & -19:08:27  &  0.0337  & Sy1     &   \\  
 15 &  NGC0931=Mrk1040  &  02:28:14.5  & +31:18:42  &  0.0167  & Sy1     &   \\  
 16 &  NGC1068          &  02:42:40.7  & -00:00:48  &  0.0038  & HBL CT  &   \\ 
 17 &  NGC1056          &  02:42:48.5  & +28:34:29  &  0.0052  & Sy2     &   \\  
 18 &  NGC1097          &  02:46:19.1  & -30:16:28  &  0.0042  & LIN     &   \\  
 19 &  NGC1125          &  02:51:40.4  & -16:39:02  &  0.0109  & HBL CT  &   \\ 
 20 &  NGC1144          &  02:55:12.2  & -00:11:01  &  0.0288  & Sy2     &   \\  
 21 &  MCG-02-08-039    &  03:00:29.8  & -11:24:59  &  0.0299  & HBL     &   \\  
 22 &  NGC1194          &  03:03:49.2  & -01:06:12  &  0.0136  & Sy1 CT  &   \\ 
 23 &  NGC1241          &  03:11:14.7  & -08:55:20  &  0.0135  & Sy2     &   \\  
 24 &  NGC1320          &  03:24:48.7  & -03:02:33  &  0.0089  & Sy2     &   \\  
 25 &  NGC1365          &  03:33:36.4  & -36:08:25  &  0.0055  & Sy1     &   \\  
 26 &  NGC1386          &  03:36:45.4  & -35:59:57  &  0.0029  & HBL CT  &   \\ 
 27 &  IRASF03362-1642  &  03:38:34.5  & -16:32:16  &  0.0369  & Sy2     &   \\  
 28 &  IRASF03450+0055  &  03:47:40.2  & +01:05:14  &  0.0310  & Sy1     & RL  \\  
 29 &  NGC1566          &  04:20:00.6  & -54:56:17  &  0.0050  & Sy1     &   \\  
 30 &  3C120            &  04:33:11.1  & +05:21:16  &  0.0330  & Sy1     & RL  \\  
 31 &  MRK0618          &  04:36:22.2  & -10:22:34  &  0.0356  & Sy1     &   \\  
 32 &  IRASF04385-0828  &  04:40:54.9  & -08:22:22  &  0.0151  & HBL     &   \\  
 33 &  NGC1667          &  04:48:37.1  & -06:19:12  &  0.0152  & Sy2     &   \\  
 34 &  ESO033-G002      &  04:55:59.6  & -75:32:27  &  0.0181  & Sy2     &   \\  
 35 &  ESO362-G018      &  05:19:35.5  & -32:39:30  &  0.0124  & Sy1     &   \\  
 36 &  IRASF05189-2524  &  05:21:01.4  & -25:21:45  &  0.0426  & HBL     &   \\  
 37 &  ESO253-G003      &  05:25:18.3  & -46:00:20  &  0.0425  & Sy2     &   \\  
 38 &  IRASF05563-3820  &  05:58:02.0  & -38:20:05  &  0.0339  & Sy1     &   \\  
 39 &  MRK0006          &  06:52:12.2  & +74:25:37  &  0.0188  & Sy1     &   \\  
 40 &  MRK0009          &  07:36:57.0  & +58:46:13  &  0.0399  & Sy1     &   \\  
 41 &  MRK0079          &  07:42:32.8  & +49:48:35  &  0.0222  & Sy1     &   \\  
 42 &  IRASF07599+6508  &  08:04:30.5  & +64:59:53  &  0.1483  & Sy1     &   \\  
 43 &  NGC2639          &  08:43:38.0  & +50:12:20  &  0.0111  & Sy1     &   \\  
 44 &  IRASF08572+3915  &  09:00:25.3  & +39:03:54  &  0.0583 & Sy2/LIN  &   \\
 45 &  MRK0704          &  09:18:26.0  & +16:18:19  &  0.0292  & Sy1     &   \\  
 46 &  UGC05101         &  09:35:51.6  & +61:21:11  &  0.0394  & LIN CT  & RL  \\ 
 47 &  NGC2992          &  09:45:42.0  & -14:19:35  &  0.0077  & Sy1     &   \\  
 48 &  MRK1239          &  09:52:19.1  & -01:36:43  &  0.0199  & Sy1     &   \\  
 49 &  M81              &  09:55:33.2  & +69:03:55  &  -0.00013 & LIN    &   \\  
 50 &  3C234            &  10:01:49.5  & +28:47:09  &  0.1849  & Sy1     & RL  \\  
 51 &  NGC3079          &  10:01:57.8  & +55:40:47  &  0.0037  & Sy2 CT  &   \\ 
 52 &  NGC3227          &  10:23:30.6  & +19:51:54  &  0.0039  & Sy1     &   \\  
 53 &  NGC3511          &  11:03:23.7  & -23:05:11  &  0.0037  & Sy1     &   \\  
 54 &  NGC3516          &  11:06:47.5  & +72:34:07  &  0.0088  & Sy1     &   \\  
 55 &  MCG+00-29-023    &  11:21:12.2  & -02:59:03  &  0.0249  & Sy2     &   \\  
 56 &  NGC3660          &  11:23:32.2  & -08:39:30  &  0.0123  & Sy2     &   \\  
 57 &  NGC3982          &  11:56:28.1  & +55:07:31  &  0.0037  & HBL     &   \\  
 58 &  NGC4051          &  12:03:09.6  & +44:31:53  &  0.0023  & Sy1     &   \\  
 59 &  UGC07064         &  12:04:43.3  & +31:10:38  &  0.0250  & HBL     &   \\  
 60 &  NGC4151          &  12:10:32.6  & +39:24:21  &  0.0033  & Sy1     &   \\  
 61 &  NGC4253=MRK766   &  12:18:26.5  & +29:48:46  &  0.0129  & Sy1     &   \\  
 62 &  NGC4388          &  12:25:46.7  & +12:39:41  &  0.0084  & HBL     &   \\  
 63 &  3C273            &  12:29:06.7  & +02:03:09  &  0.1583  & Sy1     & RL  \\  
 64 &  NGC4501=M88      &  12:31:59.0  & +14:25:10  &  0.0076  & Sy2     &   \\  
 65 &  NGC4579=M58      &  12:37:43.5  & +11:49:05  &  0.0051  & LIN     &   \\  
 66 &  NGC4593          &  12:39:39.4  & -05:20:39  &  0.0090  & Sy1     &   \\  
 67 &  NGC4594=M104     &  12:39:58.8  & -11:37:28  &  0.0034  & LIN     &   \\  
 68 &  NGC4602          &  12:40:36.5  & -05:07:55  &  0.0085  & Sy1     &   \\  
 69 &  TO1238-36=I3639  &  12:40:52.9  & -36:45:22  &  0.0109  & HBL CT  &   \\ 
 70 &  M-2-33-34=N4748  &  12:52:12.4  & -13:24:54  &  0.0146  & Sy1     &   \\  
 71 &  MRK0231          &  12:56:14.2  & +56:52:25  &  0.0422  & Sy1     & RL  \\  
 72 &  NGC4922          &  13:01:24.5  & +29:18:30  &  0.0236  & Sy2     &   \\  
 73 &  NGC4941          &  13:04:13.1  & -05:33:06  &  0.0037  & Sy2     &   \\  
 74 &  NGC4968          &  13:07:06.0  & -23:40:43  &  0.0099  & Sy2     &   \\  
 75 &  NGC5005          &  13:10:56.2  & +37:03:33  &  0.0032  & LIN     &   \\  
 76 &  NGC5033          &  13:13:27.5  & +36:35:38  &  0.0029  & Sy1     &   \\  
 77 &  MCG-03-34-064    &  13:22:24.4  & -16:43:43  &  0.0165  & HBL     &   \\  
 78 &  NGC5135          &  13:25:44.0  & -29:50:02  &  0.0137  & Sy2 CT  &   \\ 
 79 &  NGC5194=M51A     &  13:29:52.3  & +47:11:54  &  0.0015  & Sy2     &   \\  
 80 &  MCG-06-30-015    &  13:35:53.7  & -34:17:45  &  0.0077  & Sy1     &   \\  
 81 &  IRASF13349+2438  &  13:37:18.7  & +24:23:03  &  0.1076  & Sy1     &   \\  
 82 &  NGC5256=Mrk266   &  13:38:17.8  & +48:16:35  &  0.0279  & Sy2     &   \\  
 83 &  MRK0273          &  13:44:42.1  & +55:53:13  &  0.0378  & LIN     &   \\  
 84 &  IC4329A          &  13:49:19.3  & -30:18:34  &  0.0161  & Sy1     &   \\  
 85 &  NGC5347          &  13:53:17.8  & +33:29:27  &  0.0078  & HBL CT  &   \\ 
 86 &  MRK0463=UGC8850  &  13:56:02.9  & +18:22:19  &  0.0504  & HBL     &   \\  
 87 &  NGC5506          &  14:13:14.8  & -03:12:27  &  0.0062  & Sy2     &   \\  
 88 &  NGC5548          &  14:17:59.5  & +25:08:12  &  0.0172  & Sy1     &   \\  
 89 &  MRK0817          &  14:36:22.1  & +58:47:39  &  0.0315  & Sy1     &   \\  
 90 &  IRASF15091-2107  &  15:11:59.8  & -21:19:02  &  0.0446  & Sy1     &   \\  
 91 &  NGC5929          &  15:26:06.1  & +41:40:14  &  0.0083  & HBL     &   \\  
 92 &  NGC5953          &  15:34:32.3  & +15:11:42  &  0.0066  &  {\bf nSy}     &   \\  
 93 &  ARP220=UGC9913   &  15:34:57.3  & +23:30:12  &  0.0181  & Sy2     &   \\  
 94 &  N5995=M-2-40-4   &  15:48:24.9  & -13:45:28  &  0.0252  & HBL     &   \\  
 95 &  IRASF15480-0344  &  15:50:41.5  & -03:53:18  &  0.0303  & HBL     &   \\  
 96 &  ESO141-G055      &  19:21:14.3  & -58:40:13  &  0.0371  & Sy1     &   \\  
 97 &  IRASF19254-7245  &  19:31:22.5  & -72:39:20  &  0.0617  & HBL CT  & RL  \\ 
 98 &  NGC6810          &  19:43:34.1  & -58:39:21  &  0.0068  & LIN     &   \\  
 99 &  NGC6860          &  20:08:46.1  & -61:05:56  &  0.0149  & Sy1     &   \\  
100 &  NGC6890          &  20:18:18.1  & -44:48:23  &  0.0081  & Sy2     &   \\  
101 &  MRK0509          &  20:44:09.7  & -10:43:25  &  0.0344  & Sy1     &   \\  
102 &  IC5063           &  20:52:02.0  & -57:04:09  &  0.0113  & HBL     &   \\  
103 &  MRK0897          &  21:07:45.8  & +03:52:40  &  0.0263  & Sy2     &   \\  
104 &  NGC7130          &  21:48:19.5  & -34:57:09  &  0.0162  & Sy2 CT  &   \\ 
105 &  NGC7172          &  22:02:01.7  & -31:52:18  &  0.0087  & Sy2 CT  &   \\ 
106 &  IRASF22017+0319  &  22:04:19.2  & +03:33:50  &  0.0611  & HBL     &   \\  
107 &  NGC7213          &  22:09:16.2  & -47:10:00  &  0.0058  & LIN     &   \\  
108 &  3C445            &  22:23:49.6  & -02:06:12  &  0.0562  & Sy1     & RL  \\  
109 &  NGC7314          &  22:35:46.0  & -26:03:02  &  0.0048  & HBL     &   \\  
110 &  MCG+03-58-007    &  22:49:36.9  & -19:16:24  &  0.0315  & HBL     &   \\  
111 &  NGC7469          &  23:03:15.6  & +08:52:26  &  0.0163  & Sy1     &   \\  
112 &  NGC7496          &  23:09:47.2  & -43:25:40  &  0.0055  & Sy2     &   \\  
113 &  NGC7582          &  23:18:23.5  & -42:22:14  &  0.0053  & Sy2 CT  &   \\ 
114 &  NGC7590          &  23:18:55.0  & -42:14:17  &  0.0053  & Sy2     &   \\  
115 &  NGC7603          &  23:18:56.6  & +00:14:38  &  0.0295  & Sy1     &   \\  
116 &  NGC7674          &  23:27:56.7  & +08:46:45  &  0.0289  & HBL CT  &   \\ 
117 &  CGCG381-051      &  23:48:41.3  & +02:14:21  &  0.0307  & Sy2     &   \\  

\end{longtable}
\end{ThreePartTable}

\clearpage

\begin{ThreePartTable}
\begin{TableNotes}
\scriptsize
\item \textbf{Notes.} The columns give for each AGN in the sample: (1) name; (2) AGN type: Sy1 = Seyfert type 1; Sy2 = Seyfert type 2; HBL = Hidden Broad Line Region AGN; LIN = LINER galaxy; (3) redshift;  
(4) 8.4GHz nuclear flux density \citep{Thean2000} ; (5) VLT-VISIR subarcsecond nuclear 12$\mu$m flux density from \citet{Asmus2014}, $\dag$: for these 27 AGN, no sub-arcsec detection is available at 12$\mu$m, we used instead  the 10.8$\mu$m flux density measured with a 1.5 \arcsec aperture, from \citet{Gorjian2004}, $\ddag$: for these 2 objects, we use the 10.6$\mu$m flux density measured with a 3.9 \arcsec aperture, from \citet{Rieke1978}; (6) {\it Spitzer}-IRS 12$\mu$m continuum flux density as derived from \citet{Gallimore2010}, using the [NeII]12.8$\mu$m line flux and the correspondent equivalent width; (7) R12 is the ratio between the {\it Spitzer}-IRS continuum and the VLT-VISIR subarcsecond flux density at 12$\mu$m;  (8) {\it Spitzer}-IRS 7$\mu$m continuum flux density as derived from \citet{Gallimore2010}, using the [ArII]7.0$\mu$m line flux and the correspondent equivalent width corrected for R12; (9){\it Spitzer}-IRAC 5.8$\mu$m flux density corrected for R12; (10) {\it Spitzer}-IRAC 4.5$\mu$m flux density corrected for R12; (11) 2.2$\mu$m nuclear emission derived from Keck/2MASS \citep{spinoglio2022}.
\end{TableNotes}
\setlength{\tabcolsep}{1.pt}
\setlength{\LTcapwidth}{\textwidth}
\footnotesize
\begin{longtable*}{rlcccccccccc}
\caption{Low frequency nuclear photometry of the AGN sample. {(\bf Note: This table is the full version of Table \ref{tab:sample1} and will be published in electronic form).}}\label{tab:sample3} 
\\ \hline\\[-0.2cm]    
\bf n. & \bf Name & $z$  & \bf Type & \bf F(8.4GHz) & \bf F(12$\mu$m) & \bf F(12$\mu$m) & \bf R12          & \bf F(7.0$\mu$m) & \bf F(5.8$\mu$m) & \bf F(4.5$\mu$m) & \bf F(K)  \\[0.05cm]
    &      &      &      &  VLA          &   VLT     &   IRS   &  IRS/VLT & IRS$_{\rm COR}$   & IRS$_{\rm COR}$        & IRS$_{\rm COR}$  & Keck/2MASS    \\[0.05cm]
    &      &      &      &  \multicolumn{3}{c}{(mJy)}          &          & \multicolumn{4}{c}{(mJy)}                                         \\[0.05cm]
    & (1)  & (2)  & (3)  & (4)  & (5)  &  (6)                  & (7)      & (8)           & (9)                & (10)         & (11)   \\[0.1cm]
\hline\\[-0.2cm]
\endfirsthead
\caption{continued.}\\
\\ \hline\\[-0.2cm]    
 \bf n. & \bf Name & $z$  &  \bf Type & \bf F(8.4GHz) &  \bf F(12$\mu$m) & \bf F(12$\mu$m) & \bf R12          & \bf F(7.0$\mu$m) & \bf F(5.8$\mu$m) & \bf F(4.5$\mu$m) & \bf F(K)  \\[0.05cm]
       &         &      &          &  VLA          &   VLT     &   IRS   &  IRS/VLT & IRS$_{\rm COR}$   & IRS$_{\rm COR}$        & IRS$_{\rm COR}$        & Keck/2MASS   \\[0.05cm]
       &         &         &  &  \multicolumn{3}{c}{(mJy)}  & & \multicolumn{4}{c}{(mJy)}   \\
       [0.05cm]
 & (1)             & (2)            & (3)         & (4)     & (5)      &  (6)    & (7)             & (8)                & (9)        & (10)      & (11)   \\[0.1cm]
\hline\\[-0.2cm]
\endhead
\endfoot
\hline\\
\insertTableNotes
\endlastfoot
   1 & MRK0335            &  0.0258 & Sy1 &     2.1 &    151.$\dag$  &   190.52 &    1.26   &  100.32 &  77.91 &  67.53 &   16.00    \\  
   2 & NGC34=Mrk938=N17   &  0.0196 & Sy2 &    14.5 &    57.8  &   210.38 &    3.64   &    ---  &   ---  &   ---  &    5.25    \\  
   3 & IRASF00198-7926    &  0.0728 & Sy2 &    ---  &    ---   &    ---   &    ---    &    ---  &   ---  &   ---  &    2.15    \\  
   4 & ESO012-G021        &  0.0300 & Sy1 &    ---  &    ---   &   115.67 &    ---    &    ---  &   ---  &   ---  &    5.10    \\  
   5 & NGC0262=MRK348     &  0.0150 & HBL &   346.0 &    117.$\dag$  &   304.83 &    2.61   &   47.89 &  37.73 &  28.48 &    2.17    \\  
   6 & IZw001=UGC00545    &  0.0611 & Sy1 &     0.9 &    425.9 &    ---   &    ---    &    ---  &   ---  &   ---  &   18.40    \\  
   7 & IRASF00521-7054    &  0.0689 & HBL &    ---  &    ---   &    ---   &    ---    &    ---  &   ---  &   ---  &    7.24    \\  
   8 & ESO541-IG012       &  0.0566 & Sy2 &     0.8 &    ---   &    ---   &    ---    &    ---  &   ---  &   ---  &    6.85    \\  
   9 & NGC0424            &  0.0118 & HBL &    11.9 &   736.2  &   979.81 &    1.33   &  415.77 & 310.84 & 226.16 &   24.90    \\  
  10 & NGC0526A           &  0.0191 & Sy1 &     7.1 &   236.4  &   239.51 &    1.01   &  101.73 &  97.91 &  82.61 &    8.79    \\  
  11 & NGC0513            &  0.0195 & HBL &     1.2 &    ---   &    91.37 &    ---    &    ---  &   ---  &   ---  &    1.51    \\  
  12 & IRASF01475-0740    &  0.0177 & HBL &   129.8 &    ---   &   197.32 &    ---    &    ---  &   ---  &   ---  &    0.66    \\  
  13 & MRK1034NED02       &  0.0338 & Sy1 &    11.6 &    ---   &    ---   &    ---    &    ---  &   ---  &   ---  &    ---     \\ 
  14 & ESO545-G013        &  0.0337 & Sy1 &     0.2 &    ---   &    ---   &    ---    &    ---  &   ---  &   ---  &    1.53    \\ 
  15 & NGC0931=Mrk1040    &  0.0167 & Sy1 &  $<$0.4 &   212.$\dag$   &   453.62 &    2.14   &  106.49 &  95.99 &  66.69 &    7.11    \\ 
  16 & NGC1068            &  0.0038 & HBL &    30.3 &  10205.8 &    ---   &    ---    &    ---  &   ---  &   ---  &   196.00   \\ 
  17 & NGC1056            &  0.0052 & NSy &     0.6 &    ---   &   102.82 &    ---    &    ---  &   ---  &   ---  &    ---     \\ 
  18 & NGC1097            &  0.0042 & LIN &     3.1 &   16.8   &   448.86 &   26.72   &    ---  &   ---  &   ---  &    1.22    \\ 
  19 & NGC1125            &  0.0109 & HBL &     3.0 &    ---   &   136.08 &    ---    &    ---  &   ---  &   ---  &    ---     \\ 
  20 & NGC1144            &  0.0288 & Sy2 &     2.3 &   25.1   &    95.73 &    3.81   &    ---  &   ---  &   ---  &    ---     \\ 
  21 & MCG-02-08-039      &  0.0299 & HBL &  $<$0.3 &   230.7  &   199.59 &    0.87   &   60.46 &  26.01 &  16.53 &    0.29    \\ 
  22 & NGC1194            &  0.0136 & Sy1 &     0.9 &   276.6  &   272.17 &    0.98   &  167.06 & 139.23 &  81.20 &    1.46    \\ 
  23 & NGC1241            &  0.0135 & Sy2 &     6.8 &   ---    &    65.32 &    ---    &    ---  &   ---  &   ---  &    ---     \\ 
  24 & NGC1320=MRK607     &  0.0089 & Sy2 &     1.0 &   230.   &   408.25 &    1.77   &   89.86 &  65.97 &  41.41 &    2.40    \\ 
  25 & NGC1365            &  0.0055 & Sy1 &     2.0 &   360.7  &  1406.21 &    3.90   &    ---  &   ---  &   ---  &   40.20    \\
  26 & NGC1386            &  0.0029 & HBL &     6.8 &   299.3  &   522.57 &    1.75   &  121.18 &  90.09 &  56.59 &    0.39    \\  
  27 & IRASF03362-1642    &  0.0369 & SY2 &     1.5 &   148.$\dag$   &    ---   &    ---    &    ---  &   ---  &   ---  &    ---     \\  
  28 & IRASF03450+0055    &  0.0310 & Sy1 &     6.8 &     98.$\dag$  &   250.40 &    2.56   &   42.04 &  36.08 &  29.04 &    9.91    \\  
  29 & NGC1566            &  0.0050 & Sy1 &    ---  &   59.3   &   149.69 &    2.52   &   32.77 &  30.38 &  26.07 &    4.61    \\  
  30 & 3C120              &  0.0330 & Sy1 &  2105.2 &    266.0 &    ---   &    ---    &    ---  &   ---  &   ---  &   12.90    \\  
  31 & MRK0618            &  0.0356 & Sy1 &     2.9 &   ---    &    ---   &    ---    &    ---  &   ---  &   ---  &    9.20    \\  
  32 & IRASF04385-0828    &  0.0151 & HBL &     6.0 &   161.$\dag$   &   498.98 &    3.10   &    ---  &   ---  &   ---  &    4.70    \\  
  33 & NGC1667            &  0.0152 & Sy2 &     1.5 &    5.7   &    74.98 &   13.15   &    ---  &   ---  &   ---  &    0.89    \\  
  34 & ESO033-G002        &  0.0181 & Sy2 &    ---  &  174.2   &   250.40 &    1.44   &   76.43 &  54.82 &  41.81 &    4.05    \\  
  35 & ESO362-G018        &  0.0124 & Sy1 &     1.0 &  157.5   &   157.25 &    1.00   &   53.43 &  57.19 &  31.85 &    1.94    \\  
  36 & IRASF05189-2524    &  0.0426 & HBL &     6.9 &  650.9   &    ---   &    ---    &    ---  &   ---  &   ---  &   14.30    \\  
  37 & ESO253-G003        &  0.0425 & Sy2 &  $<$2.4 &    ---   &    ---   &    ---    &    ---  &   ---  &   ---  &    3.08    \\
  38 & IRASF05563-3820    &  0.0339 & Sy1 &     2.6 &    426.5 &    ---   &    ---    &    ---  &   ---  &   ---  &   21.70    \\ 
  39 & MRK0006            &  0.0188 & Sy1 &    27.9 &   160.$\ddag$   &   244.95 &    1.53   &   75.08 &  68.52 &  66.30 &   18.00    \\ 
  40 & MRK0009            &  0.0399 & Sy1 &     0.6 &     58.$\dag$  &    ---   &    ---    &    ---  &   ---  &   ---  &    8.95    \\ 
  41 & MRK0079            &  0.0222 & Sy1 &     0.8 &    237.$\dag$  &   340.21 &    1.44   &  124.72 & 101.78 &  75.44 &    6.37    \\ 
  42 & IRASF07599+6508    &  0.1483 & Sy1 &     5.8 &    180.$\dag$  &    ---   &    ---    &    ---  &   ---  &   ---  &   13.20    \\ 
  43 & NGC2639            &  0.0111 & Sy1 &   118.0 &    ---   &   41.33  &    ---    &    ---  &   ---  &   ---  &    ---     \\ 
  44 & IRASF08572+3915    &  0.0583 & LIN &     3.8 &    602.1 &    ---   &    ---    &    ---  &   ---  &   ---  &    0.96    \\ 
  45 & MRK0704            &  0.0292 & Sy1 &     0.9 &   297.$\dag$   &   367.43 &    1.24   &  154.57 & 106.70 &  86.73 &    9.91    \\ 
  46 & UGC05101           &  0.0394 & LIN &    45.1 &    227.0 &    ---   &    ---    &    ---  &   ---  &   ---  &    6.25    \\ 
  47 & NGC2992            &  0.0077 & Sy1 &     5.5 &  191.2   &   443.85 &    2.32   &   56.91 &  62.38 &  45.79 &    4.36    \\ 
  48 & MRK1239            &  0.0199 & Sy1 &    10.5 &  574.3   &   843.73 &    1.47   &  282.48 & 234.22 & 188.55 &   21.50    \\ 
  49 & M81=NGC3031        &  0.00014& LIN &   221.0 &   136.9  &    ---   &    ---    &    ---  &   ---  &   ---  &    6.31    \\ 
  50 & 3C234              &  0.1849 & Sy1 &    38.8 &   110.$\dag$   &    ---   &    ---    &    ---  &   ---  &   ---  &    1.72    \\ 
  51 & NGC3079            &  0.0037 & Sy2 &    93.3 &   ---    &   244.95 &    ---    &    ---  &   ---  &   ---  &   13.70    \\ 
  52 & NGC3227            &  0.0039 & Sy1 &    12.2 &  200.5   &   468.97 &    2.34   &   75.77 &  66.01 &  41.81 &   11.30    \\ 
  53 & NGC3511            &  0.0037 & Sy1 &  $<$0.3 &  ---     &    54.43 &    ---    &    ---  &   ---  &   ---  &    ---     \\ 
  54 & NGC3516            &  0.0088 & Sy1 &     3.1 &  230.$\ddag$    &   381.04 &    1.66   &  117.89 &  85.77 &  70.98 &    6.02    \\ 
  55 & MCG+00-29-023      &  0.0249 & Sy2 &  $<$0.3 &   58.$\dag$    &   160.44 &    2.77   &   18.39 &  19.88 &   8.97 &    1.98    \\ 
  56 & NGC3660            &  0.0123 & Sy2 &  $<$0.3 &   25.7   &   35.31  &    1.37   &    8.73 &   6.04 &   7.13 &    0.27    \\ 
  57 & NGC3982            &  0.0037 & HBL &     0.8 &    26.5  &   89.43  &    3.37   &    ---  &   ---  &   ---  &    ---     \\ 
  58 & NGC4051            &  0.0023 & Sy1 &     0.6 &   464.0  &   508.05 &    1.09   &  218.00 & 151.79 & 109.41 &    6.85    \\ 
  59 & UGC07064           &  0.0250 & HBL &  $<$0.3 &  ---     &   102.06 &    ---    &    ---  &   ---  &   ---  &    1.72    \\ 
  60 & NGC4151            &  0.0033 & Sy1 &    10.9 &  1287.4  &  2041.28 &    1.59   &  411.51 & 328.46 & 243.44 &   21.10    \\ 
  61 & NGC4253=MRK766     &  0.0129 & Sy1 &     8.7 &   253.$\dag$   &   381.04 &    1.51   &   86.45 &  69.65 &  56.04 &   11.30    \\ 
  62 & NGC4388            &  0.0084 & HBL &     1.4 &   187.8  &   520.15 &    2.77   &   75.55 &  59.46 &  29.64 &    1.69    \\ 
  63 & 3C273              &  0.1583 & Sy1 &  41725.0&    289.5 &    ---   &    ---    &    ---  &   ---  &   ---  &   23.80    \\ 
  64 & NGC4501=M88        &  0.0076 & Sy2 &  $<$0.2 &     3.7  &   55.16  &   14.91   &    ---  &   ---  &   ---  &    0.47    \\ 
  65 & NGC4579=M58        &  0.0051 & LIN &    36.5 &    74.6  &   93.43  &    1.25   &   51.09 &  65.31 &  73.22 &    ---     \\ 
  66 & NGC4593            &  0.0090 & Sy1 &     1.9 &  227.4   &   399.18 &    1.76   &  126.09 &  92.97 &  76.22 &    6.08    \\ 
  67 & NGC4594=M104       &  0.0034 & LIN &    86.6 &      4.4 &   103.92 &   23.62   &    ---  &   ---  &   ---  &    ---     \\ 
  68 & NGC4602            &  0.0085 & Sy1 &  $<$0.2 &   ---    &   41.24  &    ---    &    ---  &   ---  &   ---  &    ---     \\ 
  69 & TOL1238-36=I3639   &  0.0109 & HBL &     2.3 &   386.1  &   497.68 &    1.29   &   95.84 &  69.59 &  39.10 &    2.51    \\ 
  70 & M-2-33-34=N4748    &  0.0146 & Sy1 &     1.5 &    ---   &   84.13  &    ---    &    ---  &   ---  &   ---  &    2.44    \\ 
  71 & MRK0231            &  0.0422 & Sy1 &   234.5 &    1235.$\dag$ &    ---   &    ---    &    ---  &   ---  &   ---  &   46.60    \\ 
  72 & NGC4922            &  0.0236 & Sy2 &  $<$0.3 &    162.$\dag$  &    ---   &    ---    &    ---  &   ---  &   ---  &    ---     \\ 
  73 & NGC4941            &  0.0037 & Sy2 &     2.1 &   76.1   &   92.12  &    1.21   &   27.28 &  23.46 &  21.89 &   21.10    \\
  74 & NGC4968            &  0.0099 & Sy2 &     2.1 &   250.$\dag$   &   367.43 &    1.47   &   67.55 &  48.44 &  29.60 &    1.38    \\
  75 & NGC5005            &  0.0032 & LIN &     8.8 &     7.0  &   149.69 &   21.38   &    ---  &   ---  &   ---  &    6.79    \\
  76 & NGC5033            &  0.0029 & Sy1 &     2.1 &    15.9  &   149.25 &    9.39   &    ---  &   ---  &   ---  &    3.60    \\
  77 & MCG-03-34-064      &  0.0165 & HBL &    42.2 &    530.6 &    ---   &    ---    &    ---  &   ---  &   ---  &    0.66    \\
  78 & NGC5135            &  0.0137 & Sy2 &  $<$2.3 &   132.0  &   349.63 &    2.65   &   43.09 &  14.12 &  11.25 &    3.19    \\
  79 & NGC5194=M51A       &  0.0015 & Sy2 &     0.5 &     25.8 &    ---   &    ---    &    ---  &   ---  &   ---  &    4.25    \\
  80 & MCG-06-30-015      &  0.0077 & Sy1 &  $<$0.3 &   340.8  &   344.75 &    1.01   &  180.99 & 155.50 & 126.34 &   12.20    \\
  81 & IRASF13349+2438    &  0.1076 & Sy1 &     4.7 &    476.4 &    ---   &    ---    &    ---  &   ---  &   ---  &   13.90    \\
  82 & NGC5256=Mrk266     &  0.0279 & Sy2 &     2.9 &     72.$\dag$  &   108.87 &    1.51   &   17.64 &  28.64 &  11.77 &    1.91    \\
  83 & MRK0273            &  0.0378 & LIN &    30.5 &      79.$\dag$ &    ---   &    ---    &    ---  &   ---  &   ---  &    3.53    \\
  84 & IC4329A            &  0.0161 & Sy1 &    10.4 &  1157.7  &  1197.55 &    1.03   &  928.28 & 467.70 & 369.68 &   20.10    \\
  85 & NGC5347            &  0.0078 & HBL &     0.8 &   278.0  &   337.49 &    1.21   &   74.48 &  50.66 &  30.64 &    1.96    \\
  86 & MRK0463=UGC8850    &  0.0504 & HBL &    43.6 &   395.$\dag$   &    ---   &    ---    &    ---  &   ---  &   ---  &   16.00    \\
  87 & NGC5506            &  0.0062 & Sy2 &    67.6 &   870.8  &  1290.86 &    1.48   &  519.14 & 494.00 & 383.37 &   42.10    \\
  88 & NGC5548            &  0.0172 & Sy1 &     2.2 &   123.8  &   299.39 &    2.42   &   39.63 &  31.14 &  21.42 &    3.22    \\
  89 & MRK0817            &  0.0315 & Sy1 &     2.8 &    234.$\dag$  &   299.39 &    1.28   &   84.80 &  81.21 &  58.54 &    9.04    \\
  90 & IRASF15091-2107    &  0.0446 & Sy1 &     7.8 &    145.$\dag$  &    ---   &    ---    &    ---  &   ---  &   ---  &    8.63    \\
  91 & NGC5929            &  0.0083 & HBL &     9.1 &   ---    &   24.68  &    ---    &    ---  &   ---  &   ---  &    ---     \\
  92 & NGC5953            &  0.0066 &  {\bf nSy} &     1.1 & $<$29.5  &   174.47 & $>$5.91   &    ---  &   ---  &   ---  &    ---     \\
  93 & ARP220=UGC9913     &  0.0181 & Sy2 &    63.0 &     142.$\dag$ &    ---   &    ---    &    ---  &   ---  &   ---  &    3.25    \\
  94 & N5995=M-2-40-4     &  0.0252 & HBL &     2.4 &   332.4  &   482.13 &    1.45   &  176.72 & 112.52 &  88.94 &   13.90    \\
  95 & IRASF15480-0344    &  0.0303 & HBL &    12.4 &   102.$\dag$   &   217.74 &    2.13   &   30.50 &  17.05 &  11.52 &    1.87    \\
  96 & ESO141-G055        &  0.0371 & Sy1 &     --- &    148.7 &    ---   &    ---    &    ---  &   ---  &   ---  &   11.20    \\
  97 & IRASF19254-7245    &  0.0617 & HBL &     --- &   221.5  &    ---   &    ---    &    ---  &   ---  &   ---  &    1.94    \\
  98 & NGC6810            &  0.0068 & SB  &     --- &    44.4  &   518.94 &   11.69   &    ---  &   ---  &   ---  &    2.96    \\
  99 & NGC6860            &  0.0149 & Sy1 &    ---  &  206.1   &   217.74 &    1.06   &  123.24 & 102.04 &  71.56 &    6.98    \\
 100 & NGC6890            &  0.0081 & Sy2 &    0.5  &  116.6   &   174.19 &    1.49   &   52.79 &  35.21 &  24.90 &    1.74    \\
 101 & MRK0509            &  0.0344 & Sy1 &     2.2 &  256.4   &    ---   &    ---    &    ---  &   ---  &   ---  &   16.90    \\
 102 & IC5063             &  0.0113 & HBL &    230. &   820.6  &  1052.39 &    1.28   &  329.95 & 170.84 &  96.92 &    3.63    \\
 103 & MRK0897            &  0.0263 & Sy2 &     3.5 &    8.2   &   97.98  &   11.95   &    ---  &   ---  &   ---  &    8.55    \\
 104 & NGC7130            &  0.0162 & Sy2 &    18.1 &   104.5  &   319.80 &    3.06   &    ---  &   ---  &   ---  &    0.96    \\
 105 & NGC7172            &  0.0087 & Sy2 &     4.7 &   185.0  &   217.74 &    1.18   &  108.32 & 154.89 & 107.56 &    8.87    \\
 106 & IRASF22017+0319    &  0.0611 & HBL &     0.2 &    137.$\dag$  &    ---   &    ---    &    ---  &   ---  &   ---  &    3.87    \\
 107 & NGC7213            &  0.0058 & LIN &   183.8 &  203.3   &   293.94 &    1.45   &   76.81 &  94.75 &  95.17 &    6.19    \\
 108 & 3C445              &  0.0562 & Sy1 &   58.1  &  180.0   &    ---  &    ---    &    ---  &   ---  &   ---  &    ---      \\
 109 & NGC7314            &  0.0048 & HBL &     0.7 &   61.5   &   122.48 &    1.99   &   15.88 &  16.37 &  13.91 &    3.02    \\
 110 & MCG+03-58-007      &  0.0315 & HBL &     0.4 &    200.$\dag$  &   293.94 &    1.47   &   91.05 &  79.00 &  61.31 &    7.18    \\
 111 & NGC7469            &  0.0163 & Sy1 &    16.0 &  1173.8  &  1001.59 &    0.85   &  308.16 & 317.36 & 176.85 &   13.90    \\
 112 & NGC7496            &  0.0055 & Sy2 &     3.8 &   169.5  &   199.31 &    1.18   &   33.81 &  40.40 &  17.86 &    0.71    \\
 113 & NGC7582            &  0.0053 & Sy2 &    51.8 &   443.2  &   952.60 &    2.15   &  186.13 & 240.12 & 132.50 &   42.50    \\
 114 & NGC7590            &  0.0053 & Sy2 &  $<$0.2 & $<$10.6  &   60.48  & $>$5.71   &    ---  &   ---  &   ---  &   16.30    \\
 115 & NGC7603            &  0.0295 & Sy1 &     3.3 &    102.$\dag$  &   281.24 &    2.76   &   72.01 &  64.88 &  56.76 &   15.40    \\
 116 & NGC7674            &  0.0289 & HBL &    12.8 &  382.2   &   533.45 &    1.40   &  139.15 & 105.68 &  67.56 &    8.39    \\
 117 & CGCG381-051        &  0.0307 & Sy2 &     0.6 &    68.$\dag$   &   113.06 &    1.66   &   15.59 &   9.38 &   5.05 &    0.67    \\
\end{longtable*}
\end{ThreePartTable}

\clearpage

\begin{ThreePartTable}
\begin{TableNotes}
\footnotesize
\item \textbf{Notes.} The columns give for each AGN in the sample: (1) name; (2) AGN type: Sy1 = Seyfert type 1; Sy2 = Seyfert type 2; HBL = Hidden Broad Line Region AGN; LIN = LINER galaxy; (3) redshift; (4) (4) Near-ultraviolet flux (NUV) from the Mikulski Archive for Space Telescopes \citep[MAST, ][]{conti2011} GALEX Catalog Search; (5) Far-ultraviolet flux (FUV) from the Mikulski Archive for Space Telescopes \citep[MAST, ][]{conti2011} GALEX Catalog Search; (6) 2-10\,keV absorption corrected X-ray flux from the references of column (10); (7) 14-195\,keV observed X-ray flux from the references of column (11); (8) Computed bolometric flux (see text); (9) Logarithm of the total (bolometric) luminosity; (10) photon index of the 2-10\,keV observations $\Gamma_1$, from either \citet{Brightman2011} or \citet{ricci2017};  (11) photon index of the 14-195\,keV observations $\Gamma_2$, from \citet{oh2018}; (12) reference of the 2-10keV flux, Ref$_{1}$: (1): derived from the absorption corrected luminosity of \citet{Brightman2011}, (2): derived from the absorption corrected luminosity of \citet{Guainazzi2005}, (3): \citet{Ghosh1992}, (4): \citet{Reeves2000}, (5): derived from the absorption corrected luminosity of \citet{Tan2012}, (6): \citet{ricci2017}, (7): derived from the absorption corrected luminosity of \citet{Marinucci2012}, (8): \citet{Bi2020}, (9): derived from the absorption corrected luminosity of \citet{Iyomoto1996}, (10): derived from the absorption corrected luminosity of \citet{Saade2022}, (11): derived from the absorption corrected luminosity of \citet{Levenson2009}, (12): \citet{Rivers2013}, (13): \citet{Walton2021}, (14): \citet{Bassani1999}, (15): \citet{Tanimoto2022}, (16): \citet{DellaCeca2008}, (17): \citet{Iyomoto2001}, (18): \citet{Boissay2016}, (19): \citet{Brightman2008}, (20): derived from the absorption corrected luminosity of \citet{Akylas2009}, (21): \citet{Chen2022}, (22): derived from the absorption corrected luminosity of \citet{Zhou2010}, (23): \citet{Vasylenko2018}, (24): derived from the absorption corrected luminosity of \citet{Yamada2023}, (25): \citet{Lutz2004}, (26): \citet{Osorio-Clavijo2022}, (27): \citet{Cardamone2007}, (28): \citet{Osorio-Clavijo2023}, (29): \citet{Corral2014}, (30): \citet{Braito2009}, (31): \citet{Zhou2010}; (13) reference of the 14-195keV flux, Ref$_{2}$: (32): \citet{oh2018}, (33): \citet{Deluit2004}, (34): \citet{Cusumano2010}; (14) Bolometric flux/luminosity flag: L = lower limit to the bolometric flux and luminosity; N = data not available for a proper integration of bolometric flux and luminosity.
\end{TableNotes}
\setlength{\tabcolsep}{1.pt}
\setlength{\LTcapwidth}{\textwidth}
\footnotesize
\begin{longtable*}{rlccccccccccccc}
\caption{High frequency nuclear photometry of the AGN sample {(\bf Note: This table is the full version of Table \ref{tab:sample2} and will be published in electronic form).}}\label{tab:sample4} 
\\ \hline\\[-0.2cm]    
\bf n. &\bf Name   & $z$     & \bf Type &  \bf F$_{NUV}$ & \bf F$_{FUV}$ & \bf F$_{2-10keV}$    & \bf F$_{14-195keV}$ & \bf F$_{bol}$   & \bf $L_{\rm bol}$ & $\Gamma_1$ & $\Gamma_2$  & Ref$_{1}$  & Ref$_{2}$ & Flag \\[0.05cm]
         &                  &            &   &          \multicolumn{2}{c}{(mJy)}   & \multicolumn{3}{c}{($10^{-12} \rm{erg s^{-1} cm^{-2}}$)}   &  log($\rm{erg\,s^{-1}}$) &   &    & \\[0.05cm]
         & (1)             & (2)      & (3)   & (4)  & (5)   &  (6)    & (7)             & (8)                & (9)   & (10)  & (11) & (12) & (13) & (14)   \\[0.1cm]
\hline\\[-0.2cm]
\endfirsthead
\caption{continued.}\\
\hline\\[-0.2cm]
\caption{High frequency nuclear photometry of the AGN sample {(\bf Note: This table is the full version of Table \ref{tab:sample2} and will be published in electronic form).}}
\\ \hline\\[-0.2cm]    
\bf n. &\bf Name   & $z$     & \bf Type &  \bf F$_{NUV}$ & \bf F$_{FUV}$ & \bf F$_{2-10keV}$    & \bf F$_{14-195keV}$ & \bf F$_{bol}$   & \bf $L_{\rm bol}$ & $\Gamma_1$ & $\Gamma_2$  & Ref$_{1}$  & Ref$_{2}$ & Flag \\[0.05cm]
         &                  &            &   &          \multicolumn{2}{c}{(mJy)}   & \multicolumn{3}{c}{($10^{-12} \rm{erg s^{-1} cm^{-2}}$)}   &  log($\rm{erg\,s^{-1}}$) &   &    & \\[0.05cm]
         & (1)             & (2)      & (3)   & (4)  & (5)   &  (6)    & (7)             & (8)                & (9)   & (10)  & (11) & (12) & (13) & (14)   \\[0.1cm]
\hline\\[-0.2cm]
\endhead
\endfoot
\hline\\
\insertTableNotes
\endlastfoot
   1 & MRK0335            &  0.0258 & Sy1 &  3.517  & 1.880  &   16.69 &    15.97 &     402.72   &     44.751   &  2.03 & 2.31 &   (1)  & (32) &  \\  
   2 & NGC34=Mrk938=N17   &  0.0196 & Sy2 &  0.072  & 0.246  &    2.35 &      --- &     105.68   &     43.927   &  1.9  &  --- &   (2)  &      &  \\  
   3 & IRASF00198-7926    &  0.0728 & Sy2 &  ---    & 0.133  &    ---  &    21.80 &     148.25   &     45.248   &  ---  &  --- &        & (33) &  \\  
   4 & ESO012-G021        &  0.0300 & Sy1 &  ---    & ---    &    5.51 &     4.00 &      47.86   &     43.959   &  1.21 &  --- &   (3)  & (34) &  \\  
   5 & NGC0262=MRK348     &  0.0150 & HBL &  0.089  & 0.143  &   37.45 &   144.81 &     642.69   &     44.475   &  1.68 & 1.90 &   (1)  & (32) &  \\  
   6 & IZw001=UGC00545    &  0.0611 & Sy1 &  1.108  & 2.196  &    6.51 &      --- &     516.42   &     45.631   &  2.16 & ---  &   (4)  &      &  \\  
   7 & IRASF00521-7054    &  0.0689 & HBL &  0.011  & 0.038  &    3.46 &      --- &      83.37   &     44.948   &  1.80 & ---  &   (5)  &      &  \\  
   8 & ESO541-IG012       &  0.0566 & Sy2 &  0.047  & ---    &    ---  &      --- &$>$   68.23   &$>$  44.682   &  ---  & ---  &        &      & L \\  
   9 & NGC0424            &  0.0118 & HBL &  0.103  & 0.212  &   28.40 &    21.51 &     814.70   &     44.368   &  2.11 & 1.91 &   (6)  & (32) &  \\  
  10 & NGC0526A           &  0.0191 & Sy1 &  0.084  & 0.127  &   23.10 &    73.91 &     468.81   &     44.551   &  1.51 & 1.93 &   (1)  & (32) &  \\  
  11 & NGC0513            &  0.0195 & HBL &  0.126  & 0.280  &    5.57 &    24.47 &      93.97   &     43.872   &  1.69 & 1.74 &   (1)  & (32) &  \\ 
  12 & IRASF01475-0740    &  0.0177 & HBL &  ---    & 0.046  &   21.50 &      --- &     372.39   &     44.384   &  2.19 & ---  &   (7)  &      &  \\ 
  13 & MRK1034NED02       &  0.0338 & Sy1 &  0.031  & 0.087  &    ---  &      --- &$>$   77.62   &$>$  44.275   &  ---  & ---  &        &      & L \\
  14 & ESO545-G013        &  0.0337 & Sy1 &  0.119  & 0.250  &    4.19 &      --- &      62.37   &     44.178   &  ---  & ---  &   (8)  &      &  \\
  15 & NGC0931=Mrk1040  &  0.0167 & Sy1 & $<$0.029 & $<$0.188 &  29.19 &    62.22 &     498.88   &     44.460   &  2.01 & 2.09 &   (6)  & (32) &  \\
  16 & NGC1068            &  0.0038 & HBL &  1.714  & 3.778  &   76.40 &    37.90 &    8994.98   &     44.421   &  1.84 & 1.82 &   (6)  & (32) &  \\
  17 & NGC1056            &  0.0052 & NSy &  ---    & ---    &    ---  &      --- &        ---   &      ---     &  ---  & ---  &        &      & N \\
  18 & NGC1097            &  0.0042 & LIN &  0.182  & ---    &    2.39 &      --- &      53.09   &     42.279   &  1.80 & ---  &   (9)  &      &  \\
  19 & NGC1125            &  0.0109 & HBL &  0.052  & 0.143  &   12.50 &    16.23 &     164.06   &     43.602   &  1.9  & 2.13 &   (6)  & (32) &  \\
  20 & NGC1144            &  0.0288 & Sy2 &  0.003  & 0.032  &   26.40 &    74.31 &     492.04   &     44.935   &  1.69 & 1.73 &   (6)  & (32) &  \\
  21 & MCG-02-08-039      &  0.0299 & HBL &  ---    & ---    &    3.75 &      --- &      83.56   &     44.199   &  1.9  & ---  &   (1)  &      &  \\
  22 & NGC1194            &  0.0136 & Sy1 &  0.012  & 0.019  &   42.50 &    36.22 &     583.45   &     44.347   &  1.9  & 1.86 &   (6)  & (32) &  \\
  23 & NGC1241            &  0.0135 & Sy2 &  ---    & 0.058  &    ---  &    11.64 &$>$  117.22   &$>$  43.644   &  ---  & ---  &        & (32) & L \\
  24 & NGC1320=MRK607     &  0.0089 & Sy2 &  0.092  & 0.173  &   29.30 &    13.13 &     390.84   &     43.802   &  1.9  & ---  &   (1)  & (32) &  \\
  25 & NGC1365            &  0.0055 & Sy1 &  ---    & ---    &   44.15 &    63.52 &     893.31   &     43.741   &  2.81 & 1.99 &   (1)  & (32) &  \\
  26 & NGC1386            &  0.0029 & HBL &  0.051  & 0.202  &   97.90 &      --- &    1892.34   &     43.509   &  3.35 & ---  &   (10) &      &  \\ 
  27 & IRASF03362-1642    &  0.0369 & SY2 &  ---    & ---    &    ---  &      --- &$>$   18.66   &$>$  43.735   &   --- & ---  &        &      & L \\ 
  28 & IRASF03450+0055    &  0.0310 & Sy1 &  3.241  & 4.740  &    ---  &      --- &$>$  159.96   &$>$  44.512   &   --- & ---  &        &      & L \\ 
  29 & NGC1566            &  0.0050 & Sy1 &  ---    & ---    &    5.34 &    19.54 &     102.57   &     42.717   &  1.73 & 1.96 &   (11) & (32) &  \\ 
  30 & 3C120              &  0.0330 & Sy1 &  ---    & ---    &   42.52 &    95.38 &     977.24   &     45.354   &  1.75 & 2.01 &   (1)  & (32) &  \\ 
  31 & MRK0618            &  0.0356 & Sy1 &  2.135  & 1.450  &    8.37 &    18.30 &     192.75   &     44.717   &  2.08 & 2.00 &   (1)  & (32) &  \\ 
  32 & IRASF04385-0828    &  0.0151 & HBL &  ---    & ---    &    4.70 &      --- &     164.44   &     43.889   &  1.60 & ---  &   (12) &      &  \\ 
  33 & NGC1667          &  0.0152 & Sy2 & $<$0.070 &$<$0.171 &   11.53 &      --- &     148.59   &     43.851   &  1.9  & ---  &        &      &  \\ 
  34 & ESO033-G002        &  0.0181 & Sy2 &  0.083  & 0.080  &    6.70 &    24.49 &     176.20   &     44.078   &  1.82 & 2.04 &   (13) & (32) &  \\ 
  35 & ESO362-G018        &  0.0124 & Sy1 &  2.598  & 1.350  &    6.90 &    48.89 &     229.09   &     43.861   &  1.53 & ---  &   (1)  & (32) &  \\ 
  36 & IRASF05189-2524    &  0.0426 & HBL &  0.078  & 0.185  &    5.80 &     9.71 &     610.94   &     45.378   &  2.08 & 2.06 &   (6)  & (32) &  \\ 
  37 & ESO253-G003        &  0.0425 & Sy2 &  0.585  & 0.805  &    ---  &    15.51 &$>$  123.31   &$>$  44.681   &  ---  & ---  &        & (32) & L \\
  38 & IRASF05563-3820    &  0.0339 & Sy1 &  0.064  & 0.155  &   46.26 &    27.45 &     891.25   &     45.338   &  1.67 & 2.47 &   (1)  & (32) &  \\
  39 & MRK0006            &  0.0188 & Sy1 &  0.054  & 0.495  &   15.04 &    56.70 &     408.32   &     44.477   &  1.50 & 1.91 &   (1)  & (32) &  \\
  40 & MRK0009            &  0.0399 & Sy1 &  ---    & ---    &    6.80 &     9.84 &     124.45   &     44.629   &  1.86 & 2.23 &   (6)  & (32) &  \\
  41 & MRK0079            &  0.0222 & Sy1 &  2.572  & 2.819  &   54.07 &    42.72 &     709.58   &     44.864   &  1.57 & 2.13 &   (1)  & (32) &  \\
  42 & IRASF07599+6508    &  0.1483 & Sy1 &  0.644  & 3.100  &    ---  &      --- &$>$  239.33   &$>$  46.119   &  ---  & ---  &        &      & L \\
  43 & NGC2639          &  0.0111 & Sy1 & $<$0.011 &$<$0.058 &    2.06 &      --- &     805.38   &     44.309   &  1.9  & ---  &  (14)  &      &  \\
  44 & IRASF08572+3915    &  0.0583 & LIN &  0.020  & 0.047  &    ---  &      --- &$>$  416.87   &$>$  45.495   &  ---  & ---  &        &      & L \\
  45 & MRK0704            &  0.0292 & Sy1 &  1.584  & 2.083  &   12.44 &    36.84 &     387.26   &     44.844   &  1.73 & 2.02 &   (1)  & (32) &  \\
  46 & UGC05101           &  0.0394 & LIN &  0.005  & 0.025  &    5.40 &     7.13 &     251.19   &     44.923   &  1.00 & 2.11 &  (15)  & (32) &  \\
  47 & NGC2992            &  0.0077 & Sy1 &  0.015  & 0.046  &   89.67 &    32.65 &     903.65   &     44.039   &  1.59 & 1.89 &   (1)  & (32) &  \\
  48 & MRK1239            &  0.0199 & Sy1 &  0.407  & 0.672  &    3.70 &      --- &     456.04   &     44.575   &  ---  & ---  &  (16)  &      &  \\
  49 & M81=NGC3031        &  0.00014& LIN &  0.385  & 0.945  &   10.50 &    20.26 &     297.85   &     40.071   &  1.83 & 1.99 &   (6)  & (32) &  \\
  50 & 3C234              &  0.1849 & Sy1 &  0.139  & 0.090  &    3.73 &     5.82 &     132.13   &     46.073   &  1.31 & 2.28 &   (6)  & (32) &  \\
  51 & NGC3079            &  0.0037 & Sy2 &  0.046  & 0.060  &   38.0  &    36.74 &     609.54   &     43.228   &  2.42 & 1.77 &  (17)  & (32) &  \\
  52 & NGC3227            &  0.0039 & Sy1 &  ---    & 0.280  &   29.07 &   112.47 &     580.76   &     43.254   &  1.54 & 2.08 &  (18)  & (32) &  \\
  53 & NGC3511            &  0.0037 & Sy1 &  ---    & 0.093  &    ---  &      --- &       ---    &       ---    &  ---  & ---  &        &      & N \\
  54 & NGC3516            &  0.0088 & Sy1 &  2.112  & 3.604  &   30.70 &   112.42 &     644.17   &     44.009   &  2.03 & 1.93 &   (6)  & (32) &  \\
  55 & MCG+00-29-023      &  0.0249 & Sy2 &  ---    & ---    &    ---  &      --- &$>$   14.49   &$>$  43.275   &  ---  & ---  &        &      & L \\
  56 & NGC3660            &  0.0123 & Sy2 &  0.284  & 0.166  &    2.43 &      --- &      40.36   &     43.099   &  ---  & ---  &  (19)  &      &  \\
  57 & NGC3982            &  0.0037 & HBL &  ---    & 0.219  &   208.6 &      --- &    2618.18   &     43.862   &  1.83 & ---  &  (10)  &      &  \\
  58 & NGC4051            &  0.0023 & Sy1 &  1.699  & ---    &    25.2 &    42.49 &     570.16   &     42.786   &  1.70 & 2.22 &  (10)  & (32) &  \\
  59 & UGC07064           &  0.0250 & HBL &  0.158  & 0.319  &    2.70 &    13.48 &      53.95   &     43.850   &  1.67 & 1.69 &   (6)  & (32) &  \\
  60 & NGC4151            &  0.0033 & Sy1 &  3.488  & 0.051  &   210.0 &   618.88 &    3349.65   &     43.869   &  1.73 & 1.88 &  (20)  & (32) &  \\
  61 & NGC4253=MRK766     &  0.0129 & Sy1 &  0.199  & 0.414  &    28.0 &    26.17 &     489.78   &     44.225   &  2.02 & 2.26 &  (21)  & (32) &  \\
  62 & NGC4388            &  0.0084 & HBL &  0.158  & 0.313  &   47.54 &   278.91 &     859.01   &     44.093   &  1.60 & 1.77 &   (1)  & (32) &  \\
  63 & 3C273              &  0.1583 & Sy1 &  17.320 & 11.24  &  106.30 &   421.57 &    7177.94   &     47.658   &  1.58 & 1.75 &   (1)  & (32) &  \\
  64 & NGC4501=M88        &  0.0076 & Sy2 &  0.021  & 0.025  &    2.31 &      --- &      56.62   &     42.824   &  4.05 & ---  &  (10)  &      &  \\
  65 & NGC4579=M58        &  0.0051 & LIN &  0.242  & 0.456  &    3.91 &     8.27 &     552.08   &     43.466   &  1.78 & ---  &   (1)  & (32) &  \\
  66 & NGC4593          &  0.0090 & Sy1 & $<$0.129 &$<$3.961 &   61.26 &    88.30 &     879.02   &     44.163   &  2.03 & 1.84 &  (22)  & (32) &  \\
  67 & NGC4594=M104       &  0.0034 & LIN &  0.138  & 0.247  &    1.39 &      --- &      59.57   &     42.145   &  1.95 & ---  &   (1)  &      &  \\
  68 & NGC4602            &  0.0085 & Sy1 &  0.012  & 0.039  &    ---  &      --- &$>$    0.16   &$>$  40.384   &  ---  & ---  &        &      & L \\
  69 & TOL1238-36=I3639   &  0.0109 & HBL &  0.312  & 0.665  &   38.98 &      --- &     609.54   &     44.172   &  ---  & ---  &  (10)  &      &  \\
  70 & M-2-33-34=N4748    &  0.0146 & Sy1 &  1.568  & 2.268  &    8.32 &     9.35 &     134.90   &     43.773   &  1.91 & 2.06 &  (23)  & (32) &  \\
  71 & MRK0231            &  0.0422 & Sy1 &  ---    & ---    &    2.98 &      --- &$>$  928.97   &$>$  45.552   &  1.66 & ---  &   (1)  &      & L \\
  72 & NGC4922            &  0.0236 & Sy2 &  0.004  & 0.010  &    1.48 &      --- &$>$ 1066.60   &$>$  45.095   &  ---  & ---  &   (10) &      & L \\
  73 & NGC4941           &  0.0037 & Sy2 & $<$0.045 &$<$0.042 &   5.48 &    20.16 &     240.44   &     42.825   &  1.95 & 1.64 &   (6)L & (32) &  \\
  74 & NGC4968           &  0.0099 & Sy2 & $<$0.052 &$<$0.072 &   68.3 &      --- &     781.63   &     44.196   &  1.33 & ---  &   (1)  &      &  \\
  75 & NGC5005           &  0.0032 & LIN & $<$0.083 &$<$0.094 &   0.61 &      --- &$>$   57.02   &$>$  42.074   &  1.57 & ---  &   (10) &      & L \\
  76 & NGC5033           &  0.0029 & Sy1 & $<$0.023 &$<$0.081 &    4.2 &     6.26 &      79.07   &     42.130   &  1.72 & 1.63 &   (6)  & (32) &  \\
  77 & MCG-03-34-064      &  0.0165 & HBL &  0.390  & 0.486  &   12.32 &    30.98 &     567.54   &     44.505   &  2.50 & 2.20 &   (1)  & (32) &  \\
  78 & NGC5135            &  0.0137 & Sy2 &  0.480  & 1.578  &   44.90 &      --- &     595.66   &     44.363   &  ---  & ---  &   (24) &      &  \\
  79 & NGC5194=M51A       &  0.0015 & Sy2 &  ---    & 0.676  &    4.7  &    13.26 &     121.06   &     41.741   &  2.61 & ---  &   (1)  & (32) &  \\
  80 & MCG-06-30-015      &  0.0077 & Sy1 &  0.060  & 0.156  &    44.9 &    59.53 &     770.90   &     43.970   &  2.09 & 2.47 &   (1)  & (32) &  \\
  81 & IRASF13349+2438    &  0.1076 & Sy1 &  0.165  & 0.527  &    2.36 &      --- &     451.86   &     46.093   &  2.31 & ---  &   (1)  &      &  \\
  82 & NGC5256=Mrk266     & 0.0279 & Sy2 & $<$0.087 &$<$0.155 &   0.86 &      --- &      52.48   &     43.935   &  2.74 & ---  &   (1)  &      &  \\
  83 & MRK0273            &  0.0378 & LIN &  0.052  & 0.087  &    3.5  &     5.18 &     125.03   &     44.582   &  2.25 & ---  &   (14) & (32) &  \\
  84 & IC4329A            &  0.0161 & Sy1 &  0.007  & 0.080  &   168.  &   263.25 &    2679.17   &     45.158   &  1.89 & 2.05 &   (25) & (32) &  \\
  85 & NGC5347            &  0.0078 & HBL &  0.051  & ---    &   7.44  &      --- &     196.34   &     43.388   &  ---  & ---  &   (26) &      &  \\
  86 & MRK0463=UGC8850    &  0.0504 & HBL &  ---    & ---    &   17.0  &     8.48 &     438.53   &     45.386   &  1.64 & 1.71 &   (10) & (32) &  \\
  87 & NGC5506            &  0.0062 & Sy2 &  0.036  & 0.174  &   69.3  &   239.40 &    1778.28   &     44.144   &  1.80 & 2.11 &   (1)  & (32) &  \\
  88 & NGC5548            &  0.0172 & Sy1 &  0.781  & 1.792  &   35.04 &    86.47 &     526.02   &     44.509   &  1.82 & 1.91 &   (1)  & (32) &  \\
  89 & MRK0817            &  0.0315 & Sy1 &  6.230  & 5.044  &   13.60 &    28.77 &     393.55   &     44.918   &  2.14 & 1.86 &   (6)  & (32) &  \\
  90 & IRASF15091-2107  &  0.0446 & Sy1 & $<$0.064 &$<$0.182 &   16.86 &    32.67 &     365.59   &     45.197   &  1.99 & 1.97 &   (6)  & (32) &  \\
  91 & NGC5929            &  0.0083 & HBL &  0.088  & 0.123  &    1.4  &      --- &      78.34   &     43.043   &  ---  & ---  &   (27) &      &  \\
  92 & NGC5953            &  0.0066 & LIN &  0.120  & 0.484  &    0.01 &      --- &$>$   12.47   &$>$  42.045   &  ---  & ---  &   (28) &      & L \\
  93 & ARP220=UGC9913   &  0.0181 & Sy2 & $<$0.030 &$<$0.034 &    0.11 &      --- &$>$  107.65   &$>$  43.865   &  ---  & ---  &   (29) &      & L \\
  94 & N5995=M-2-40-4     &  0.0252 & HBL &  0.056  & 0.176  &   14.90 &    35.21 &     420.73   &     44.749   &  1.67 & 2.05 &   (6)  & (32) &  \\
  95 & IRASF15480-0344    &  0.0303 & HBL &  ---    & ---    &    4.60 &      --- &     197.24   &     44.583   &  5.66 & ---  &   (1)  &      &  \\
  96 & ESO141-G055        &  0.0371 & Sy1 &  11.82  & 11.21  &   26.5  &    58.77 &     645.65   &     45.279   &  2.05 & 2.06 &   (25) & (32) &  \\
  97 & IRASF19254-7245    &  0.0617 & HBL &  0.009  & 0.025  &   33.28 &      --- &     533.33   &     45.653   &  1.9  & ---  &   (30) &      &  \\
  98 & NGC6810            &  0.0068 & SB  &  0.024  & 0.231  &    0.08 &      --- &      46.99   &     42.647   &  1.9  & ---  &   (1)  &      &  \\
  99 & NGC6860            &  0.0149 & Sy1 &  ---    & ---    &   25.60 &    51.52 &     413.05   &     44.278   &  2.11 & 2.05 &   (6)  & (32) &  \\
 100 & NGC6890            &  0.0081 & Sy2 &  0.044  & 0.125  &   10.2  &      --- &     273.53   &     43.565   &  3.87 & ---  &   (1)  &      &  \\
 101 & MRK0509            &  0.0344 & Sy1 &  9.498  & 6.855  &  170.80 &   100.14 &    2009.09   &     45.705   &  1.67 & 2.08 &   (31) & (32) &  \\
 102 & IC5063             &  0.0113 & HBL &  0.031  & 0.123  &   41.1  &    67.76 &     810.96   &     44.328   &  1.90 & 1.90 &   (6)  & (32) &  \\
 103 & MRK0897            &  0.0263 & Sy2 &  0.335  & 0.873  &   ---   &      --- &$>$   70.47   &$>$  44.011   &  ---  & ---  &        &      & L \\
 104 & NGC7130            &  0.0162 & Sy2 &  0.324  & 1.200  &   11.9  &    17.41 &     230.14   &     44.097   &  2.03 & 1.88 &  (24)  & (32) &  \\
 105 & NGC7172            &  0.0087 & Sy2 &  ---    & ---    &   34.10 &   160.02 &     623.73   &     43.984   &  1.46 & 1.84 &   (6)  & (32) &  \\
 106 & IRASF22017+0319    &  0.0611 & HBL &  ---    & 0.251  &    7.20 &    16.16 &     212.32   &     45.244   &  1.69 & 2.24 &   (6)  & (32) &  \\
 107 & NGC7213            &  0.0058 & LIN &  ---    & ---    &   62.92 &    39.04 &     767.36   &     43.720   &  1.91 & 1.90 &   (1)  & (32) &  \\
 108 & 3C445              &  0.0562 & Sy1 &  ---    & ---    &   24.40 &    39.82 &     308.32   &     45.331   &  1.74 & 2.04 &   (6)  & (32) &  \\
 109 & NGC7314            &  0.0048 & HBL &  0.029  & 0.078  &   41.80 &    57.42 &     540.75   &     43.404   &  2.14 & 1.94 &   (6)  & (32) &  \\
 110 & MCG+03-58-007      &  0.0315 & HBL &  ---    & ---    &    2.13 &      --- &     118.30   &     44.396   &  1.9  & ---  &   (1)  &      &  \\
 111 & NGC7469            &  0.0163 & Sy1 &  1.603  & 2.368  &   25.19 &    70.63 &     883.08   &     44.686   &  2.09 & 2.08 &   (1)  & (32) &  \\
 112 & NGC7496            &  0.0055 & Sy2 &  1.090  & 1.978  &   ---   &      --- &$>$   79.62   &$>$  42.691   &  ---  & ---  &        &      & L \\
 113 & NGC7582            &  0.0053 & Sy2 &  0.012  & 0.205  &   59.86 &    82.28 &    1169.50   &     43.825   &  1.89 & 2.13 &   (1)  & (32) &  \\
 114 & NGC7590            &  0.0053 & Sy2 &  0.048  & 0.133  &    ---  &      --- &$>$   97.95   &$>$  42.748   &  ---  & ---  &        &      & L \\
 115 & NGC7603            &  0.0295 & Sy1 &  3.769  & 4.434  &   22.40 &    52.96 &     496.59   &     44.961   &  1.88 & 1.93 &   (6)  & (32) &  \\
 116 & NGC7674            &  0.0289 & HBL &  0.297  & 0.377  &   20.13 &    12.60 &     493.17   &     44.940   &  3.28 & ---  &   (1)  & (32) &  \\
 117 & CGCG381-051        &  0.0307 & Sy2 &  0.031  & ---    &    ---  &      --- &$>$   26.55   &$>$  43.724   &   --- & ---  &        &      & L \\
 \end{longtable*}
\end{ThreePartTable}

\clearpage

%






\appendix

\section{SED of all galaxies of the sample}\label{sed_ind}

\begin{figure*}[ht!]
\includegraphics[width = 0.5\textwidth]{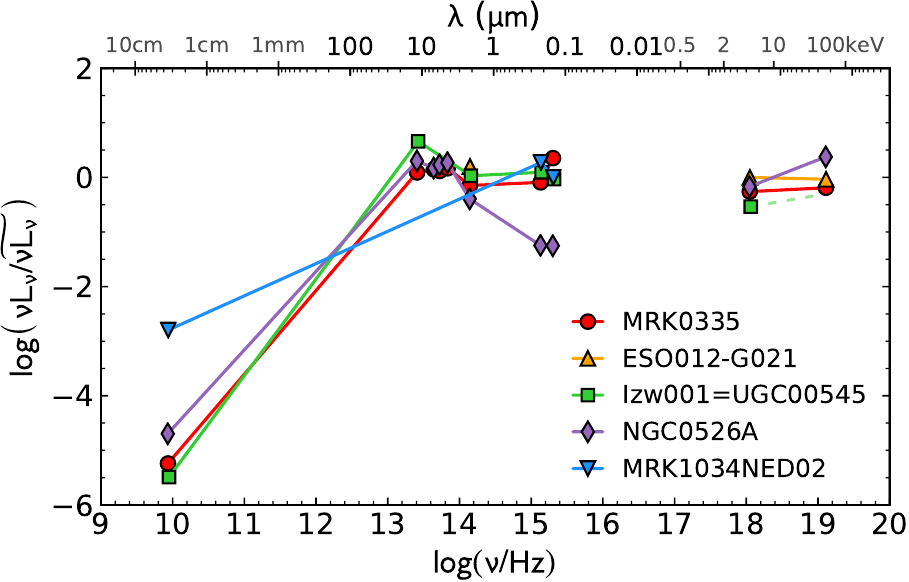}~
\includegraphics[width = 0.5\textwidth]{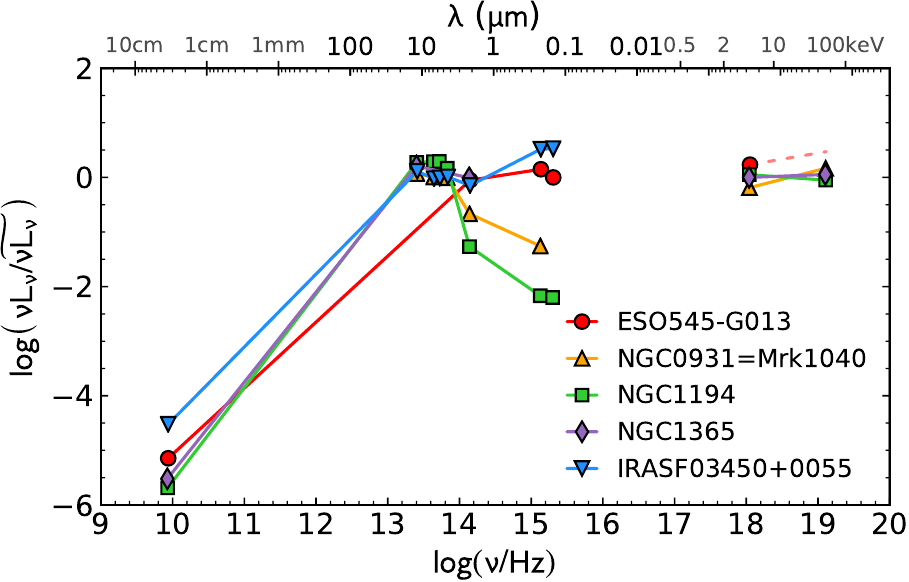}\\
\includegraphics[width = 0.5\textwidth]{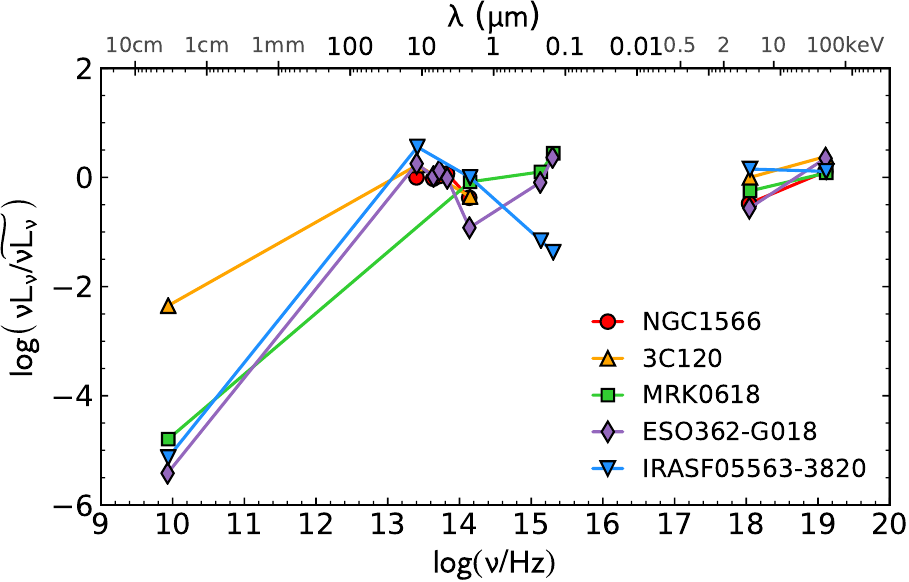}~
\includegraphics[width = 0.5\textwidth]{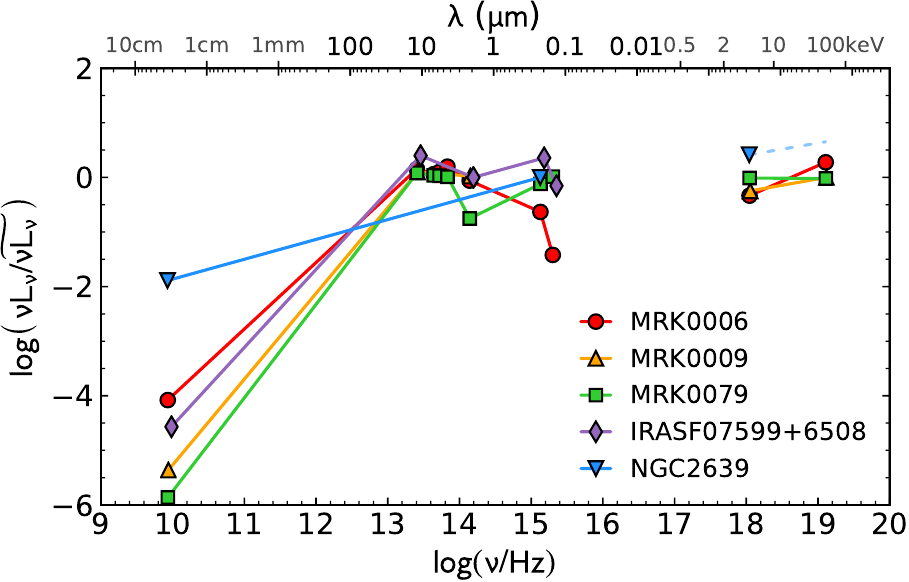}\\
\includegraphics[width = 0.5\textwidth]{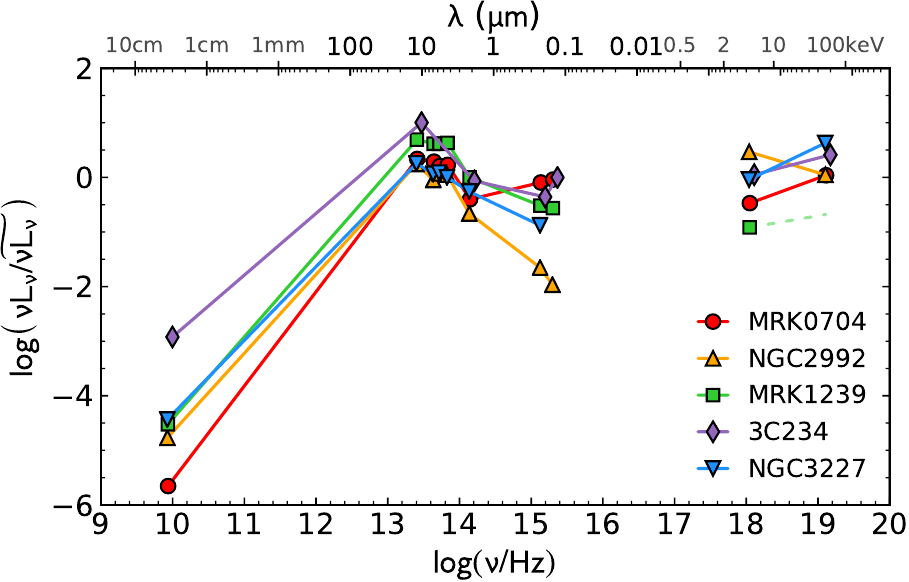}~
\includegraphics[width = 0.5\textwidth]{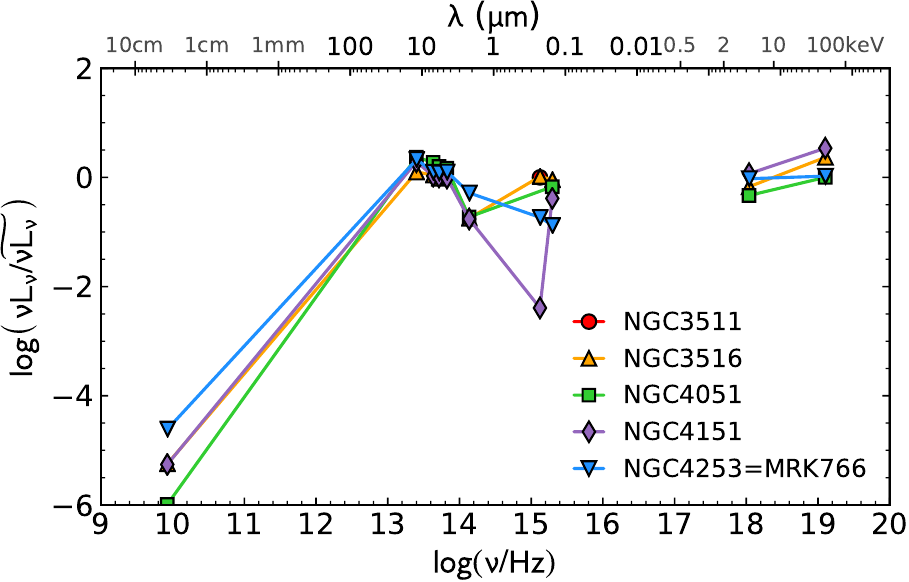}\\
\includegraphics[width = 0.5\textwidth]{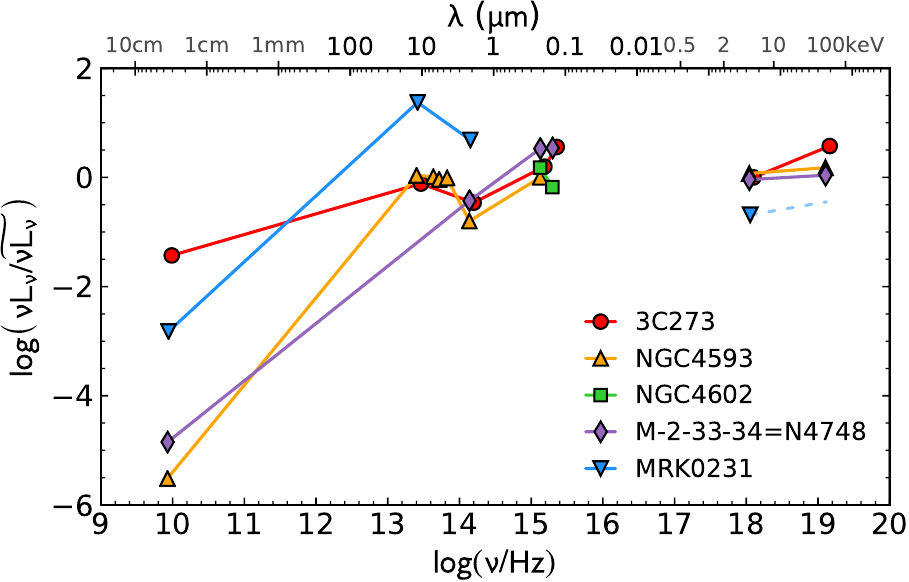}~
\includegraphics[width = 0.5\textwidth]{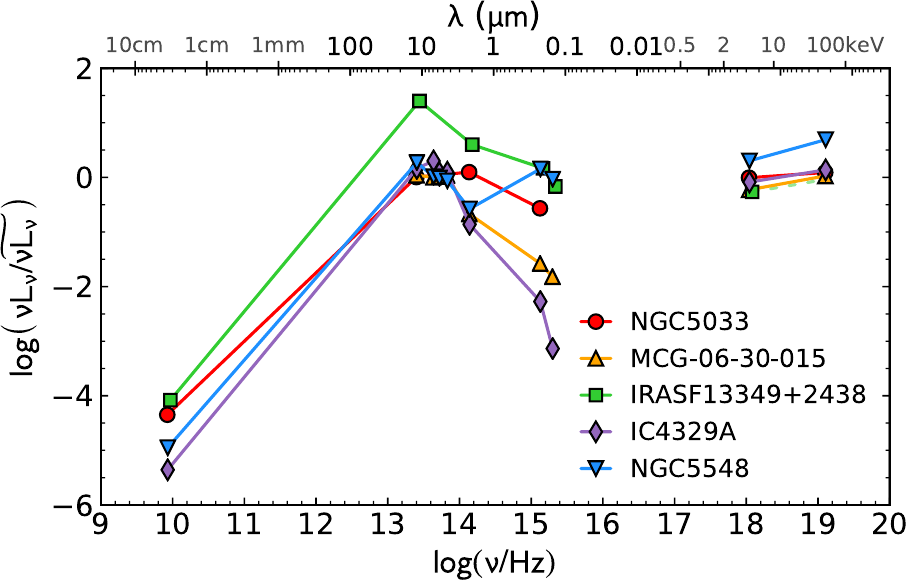}\\
\caption{Normalized {\bf rest-frame} SEDs of Seyfert type 1. 
\label{fig:plot_sy1_1}}
\end{figure*}

\begin{figure*}[ht!]
\includegraphics[width = 0.5\textwidth]{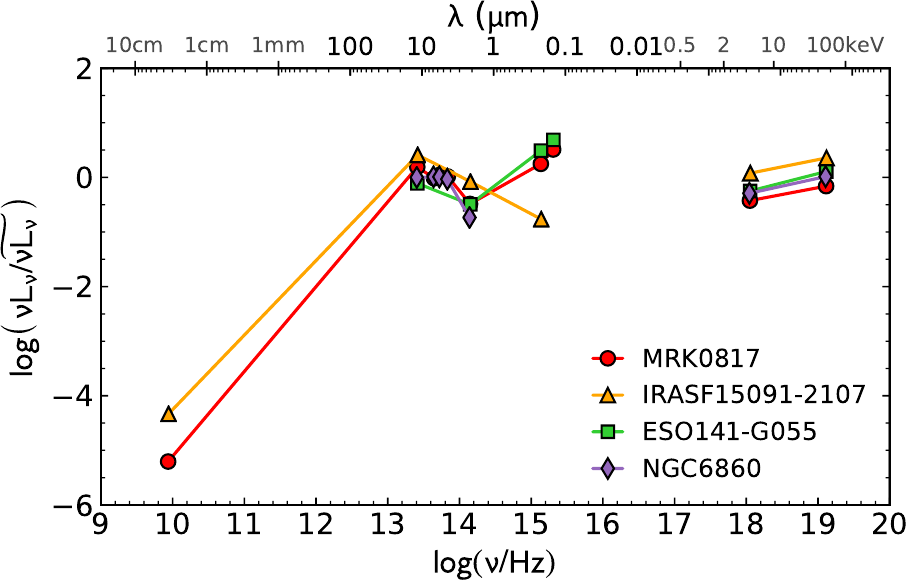}~
\includegraphics[width = 0.5\textwidth]{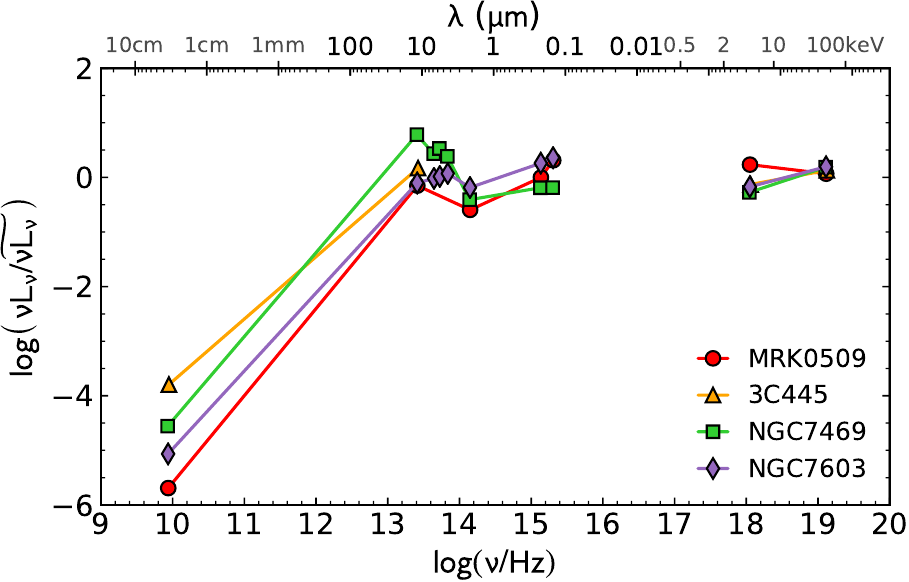}

\caption{Normalized rest-frame SEDs of Seyfert type 1 (continued). 
\label{fig:plot_sy1_2}}
\end{figure*}

\begin{figure*}[ht!]
\includegraphics[width = 0.5\textwidth]{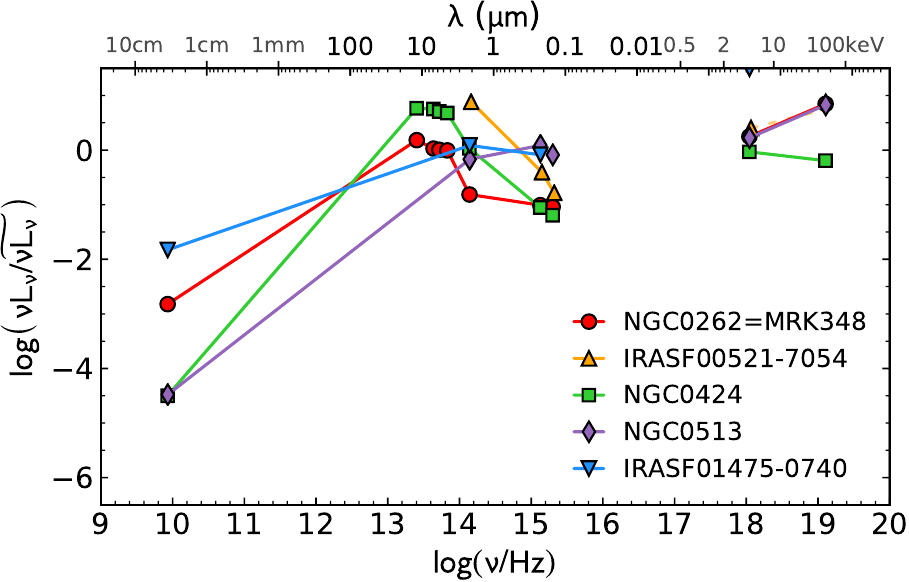}~
\includegraphics[width = 0.5\textwidth]{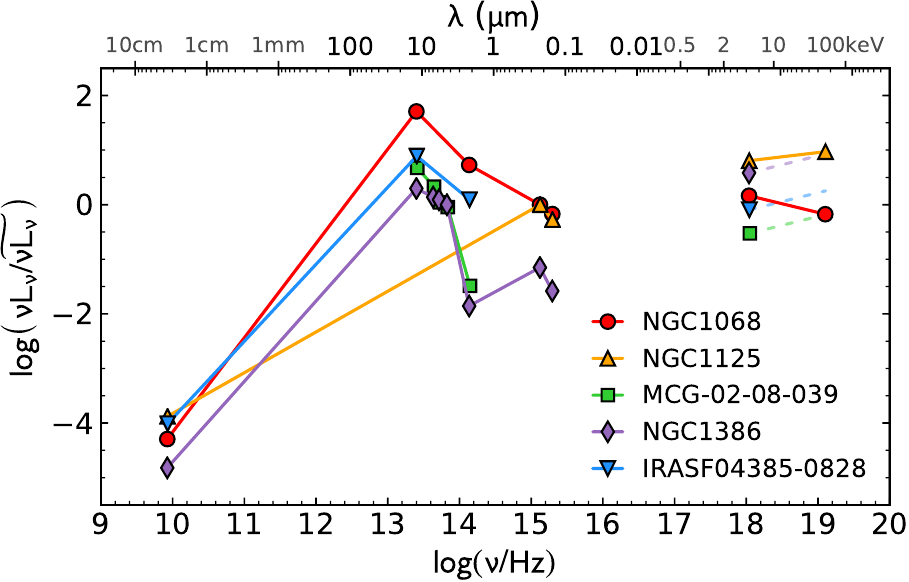}\\
\includegraphics[width = 0.5\textwidth]{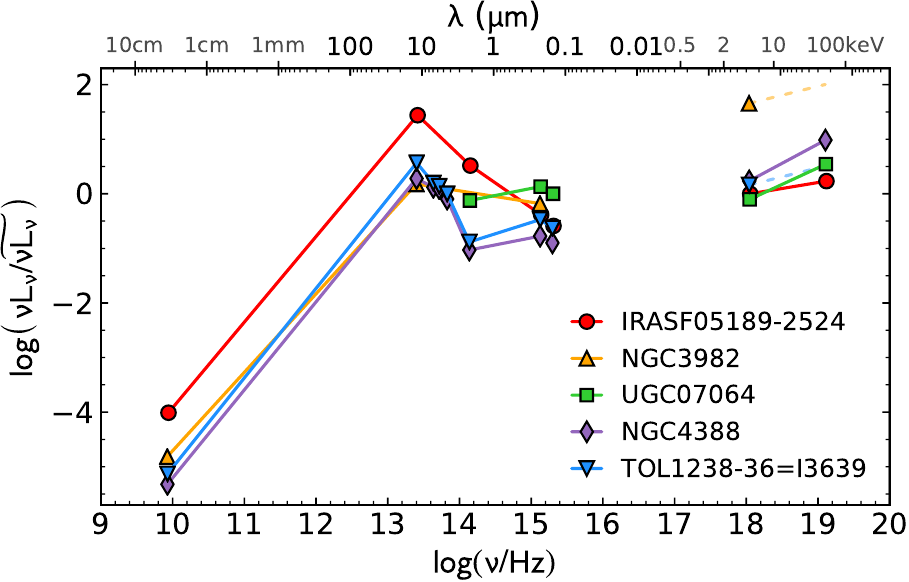}~
\includegraphics[width = 0.5\textwidth]{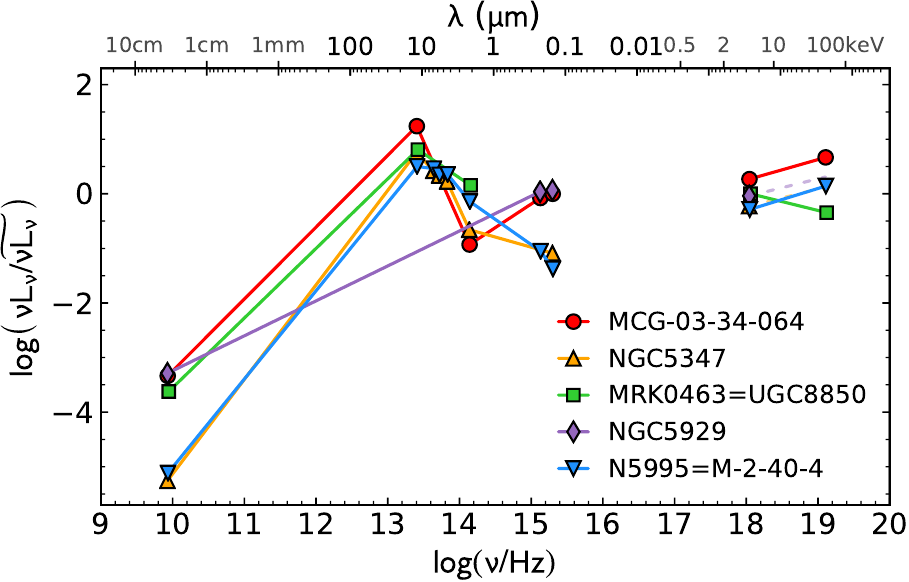}\\
\includegraphics[width = 0.5\textwidth]{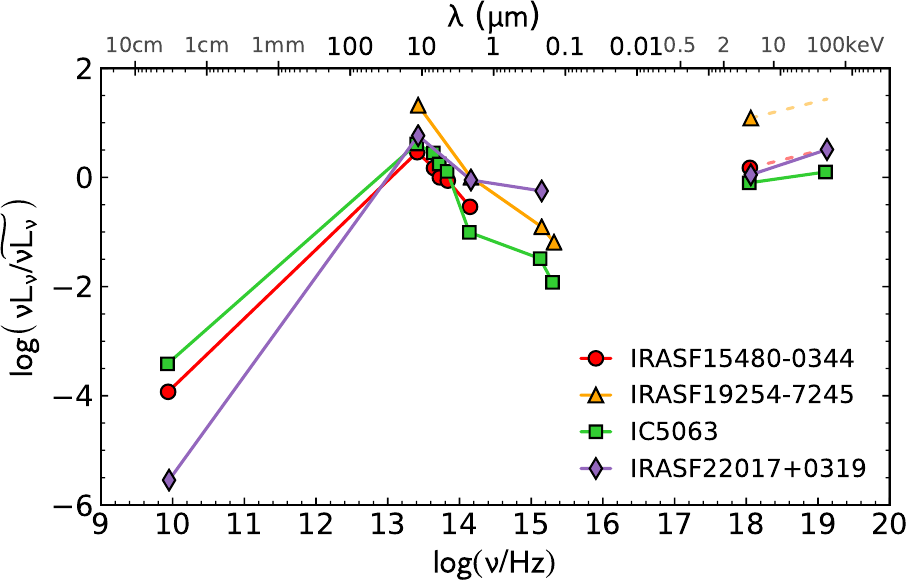}~
\includegraphics[width = 0.5\textwidth]{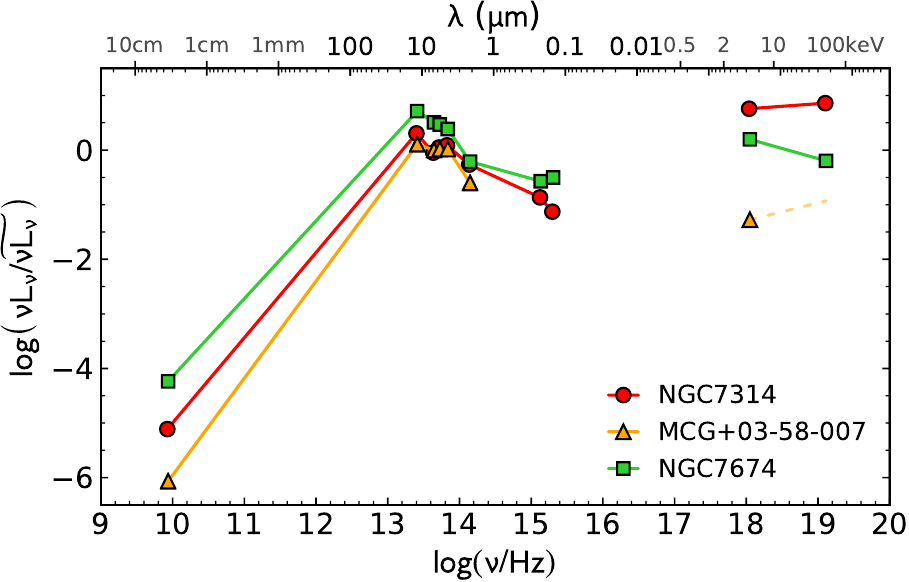}

\caption{Normalized rest-frame SEDs of Hidden Broad Line Region Galaxies (Type 1). 
\label{fig:plot_hbl}}
\end{figure*}

\begin{figure*}[ht!]
\includegraphics[width = 0.5\textwidth]{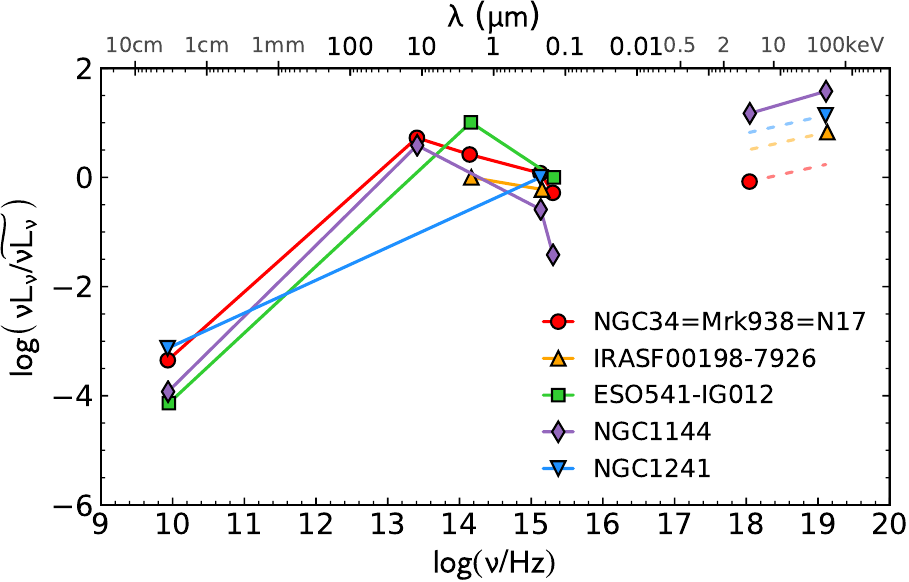}~
\includegraphics[width = 0.5\textwidth]{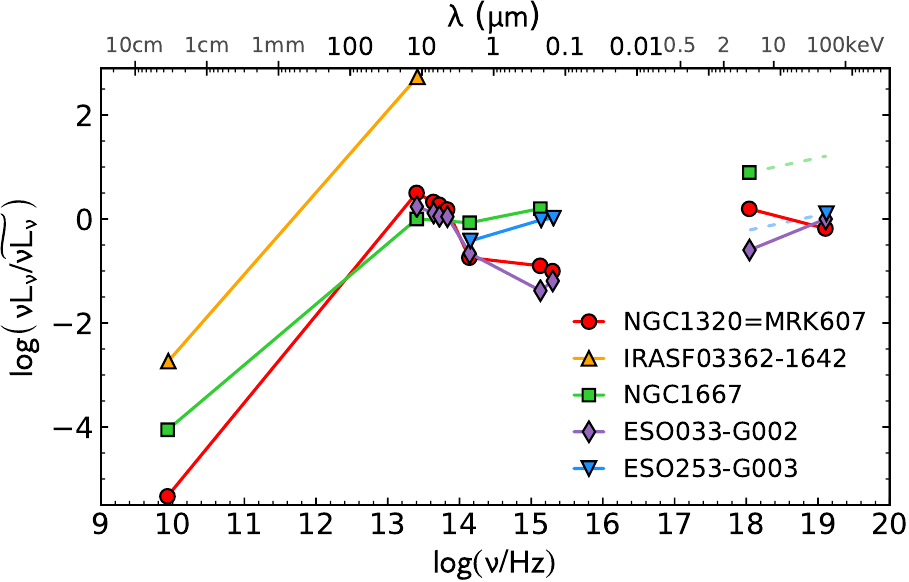}\\
\includegraphics[width = 0.5\textwidth]{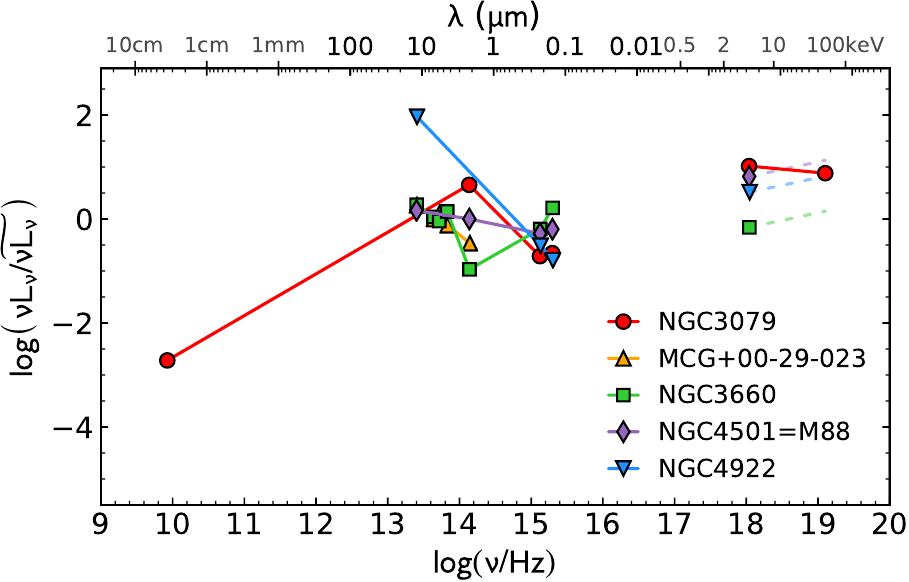}~
\includegraphics[width = 0.5\textwidth]{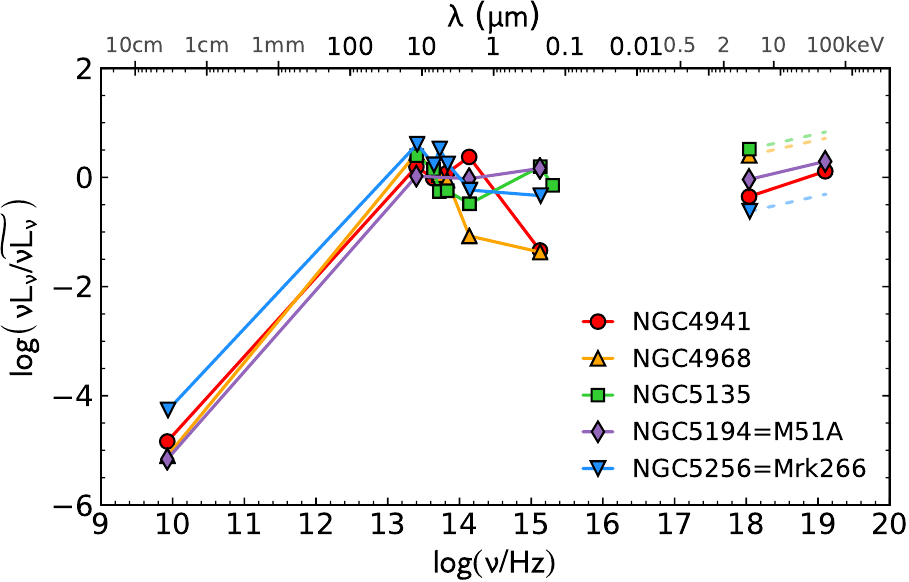}\\
\includegraphics[width = 0.5\textwidth]{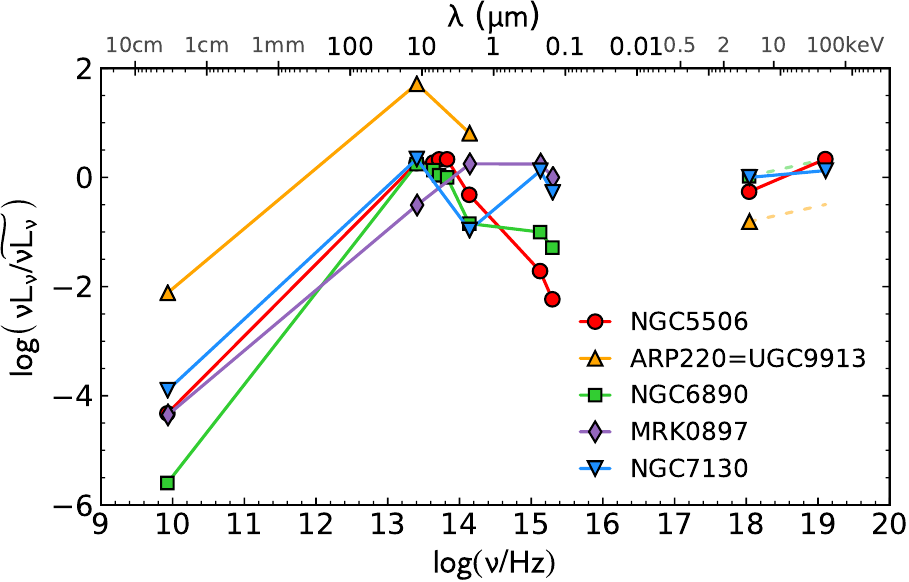}~
\includegraphics[width = 0.5\textwidth]{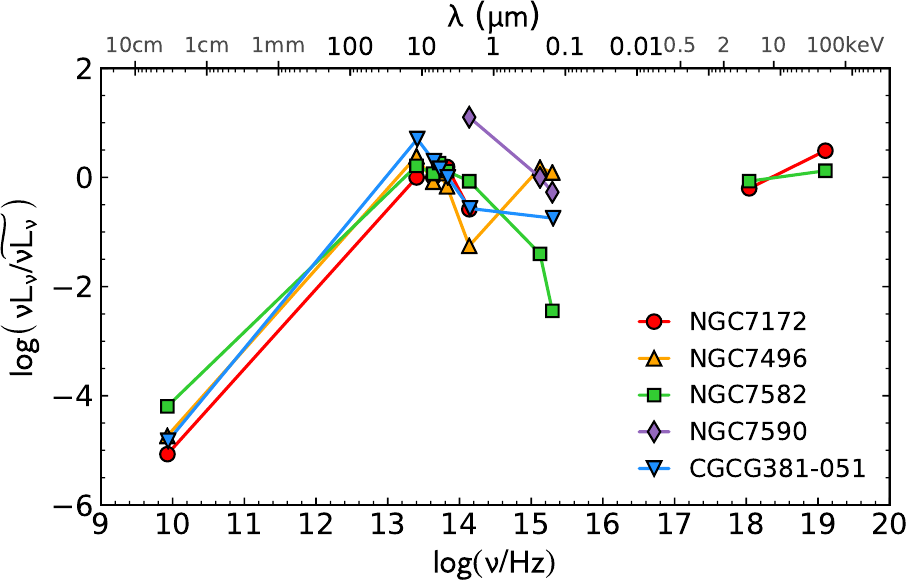}

\caption{Normalized rest-frame SEDs of Seyfert type 2. 
\label{fig:plot_sy2}}
\end{figure*}

\begin{figure*}[ht!]
\includegraphics[width = 0.5\textwidth]{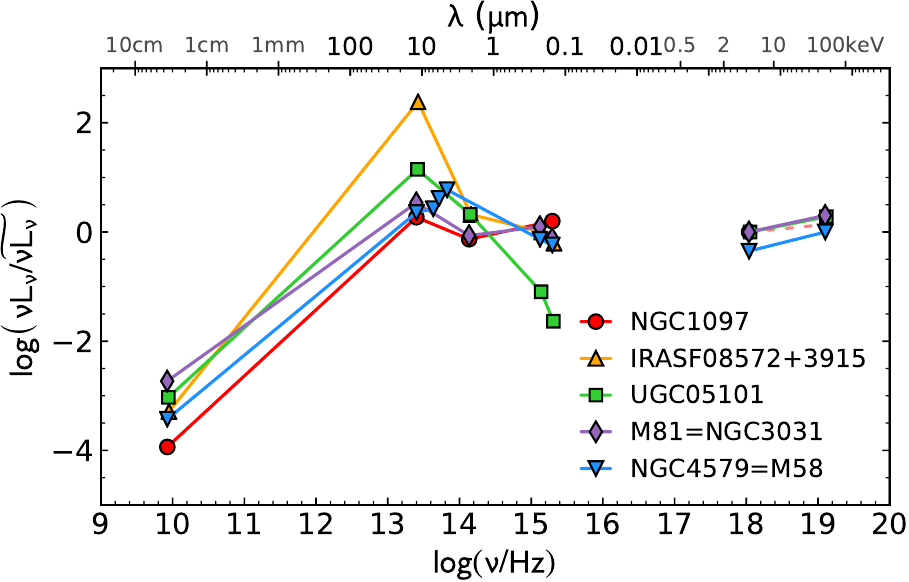}~
\includegraphics[width = 0.5\textwidth]{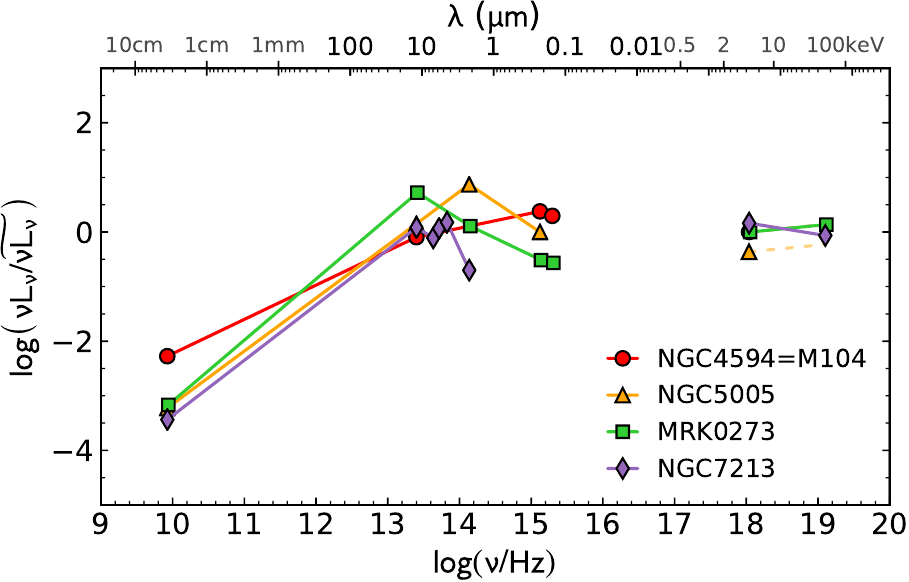}\\
\begin{center}
\includegraphics[width = 0.5\textwidth]{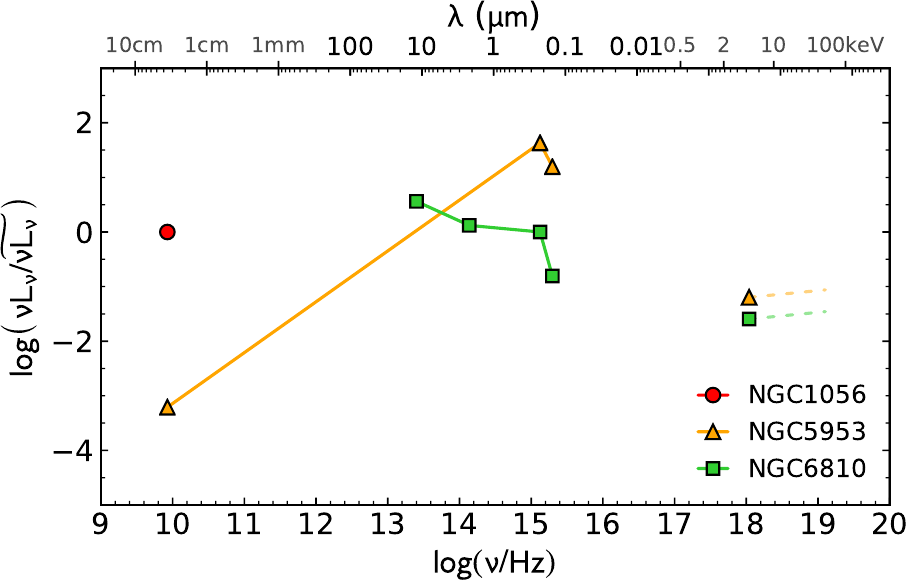}
\end{center}
\caption{Normalized rest-frame SEDs of LINERs, in the above two panels and {\bf two non-Seyfert galaxies (NGC\,1056 and NGC\,5953)}, one Starburst (NGC\,6810) galaxies in the lower panel. 
\label{fig:plot_nsy}}
\end{figure*}

 We present in Fig. \ref{fig:plot_sy1_1} and Fig. \ref{fig:plot_sy1_2} the individual SEDs of all Seyfert type 1 in our sample.
  We present in Fig. \ref{fig:plot_hbl} the individual SEDs of all HBL galaxies.
  We present in Fig. \ref{fig:plot_sy2} the individual SEDs of all Seyfert type 2.
   We present in Fig. \ref{fig:plot_nsy} the individual SEDs of the LINER in our sample (two panels above) and the SEDs of those classified an non-Seyfert (NGC\,1056) and Starburst (NGC\,6810) (lower panel).
 
\clearpage
\newpage

\section{Correlations of luminosities normalized to the Eddington luminosity}\label{corr_edd}

\begin{figure*}[ht!]
\includegraphics[width = 0.5\textwidth]{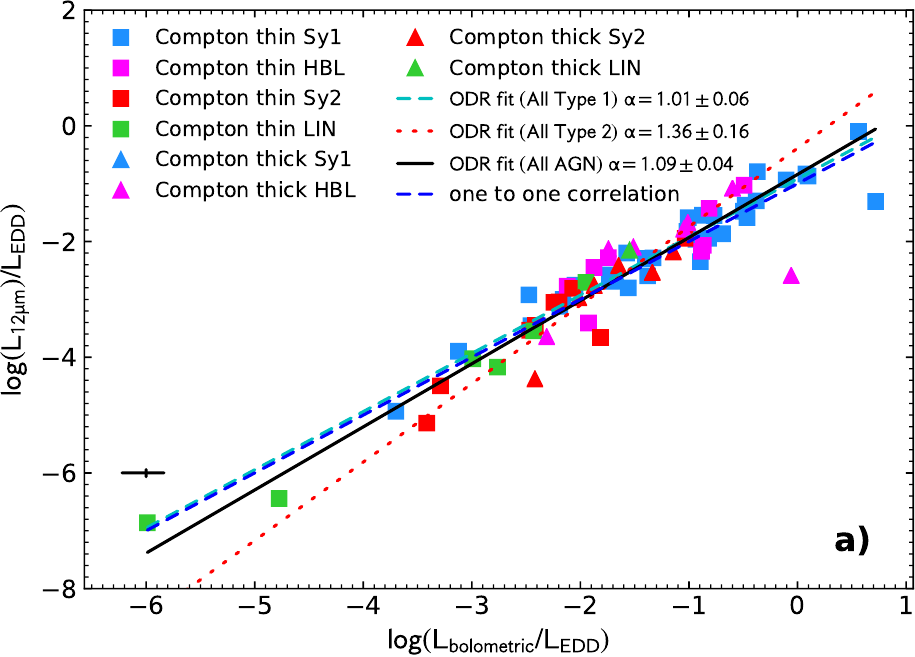}~
\includegraphics[width = 0.5\textwidth]{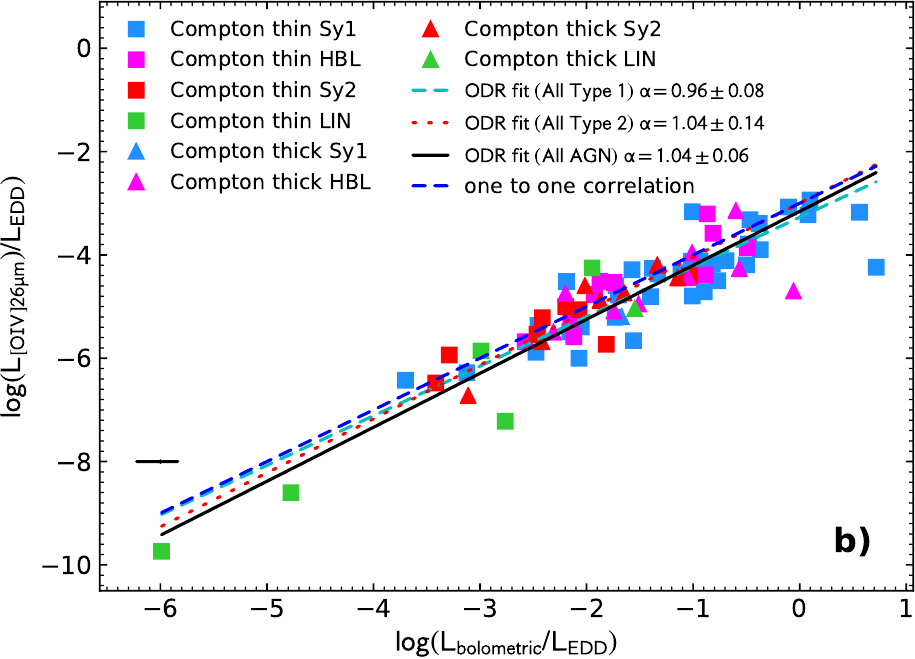}
\caption{In each figure representing the correlations between physical quantities, the error bar at the left corner has been computed using the median value of the relative errors of the plotted data (from Fig. \ref{fig:L12_Loiv_lboledd} to Fig. \ref{fig:F2-10_FK_total}).
{\bf (a:)} Ratio of the 12$\mu$m luminosity to the Eddington luminosity as a function of the ratio of the bolometric luminosity to the Eddington luminosity. {\bf (b:)} Ratio of the [OIV]26$\mu$m luminosity to the Eddington luminosity as a function of the ratio of the bolometric luminosity to the Eddington luminosity.
\label{fig:L12_Loiv_lboledd}}
\end{figure*}

\begin{figure*}[ht!]
\includegraphics[width = 0.5\textwidth]{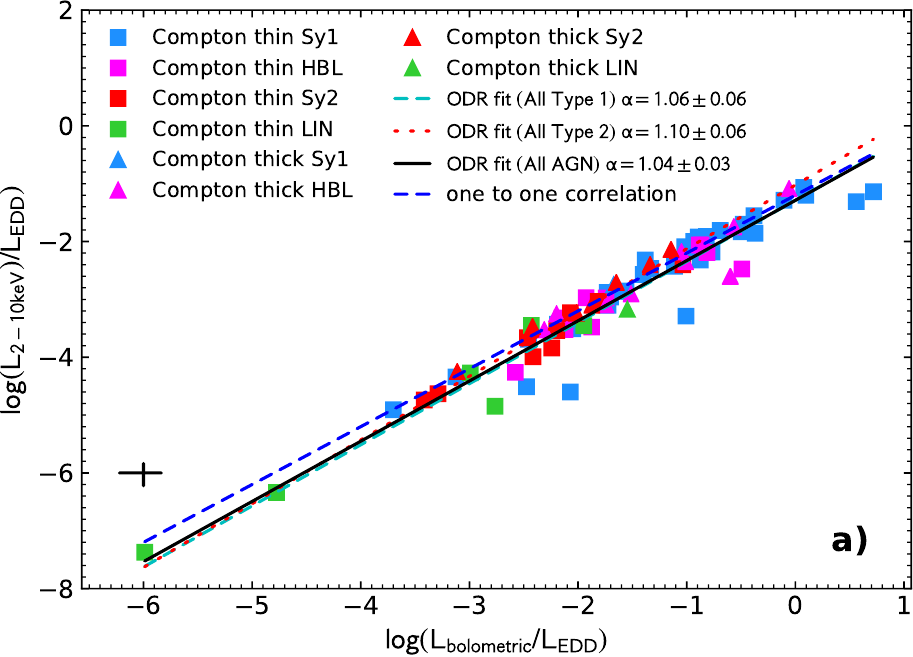}~
\includegraphics[width = 0.5\textwidth]{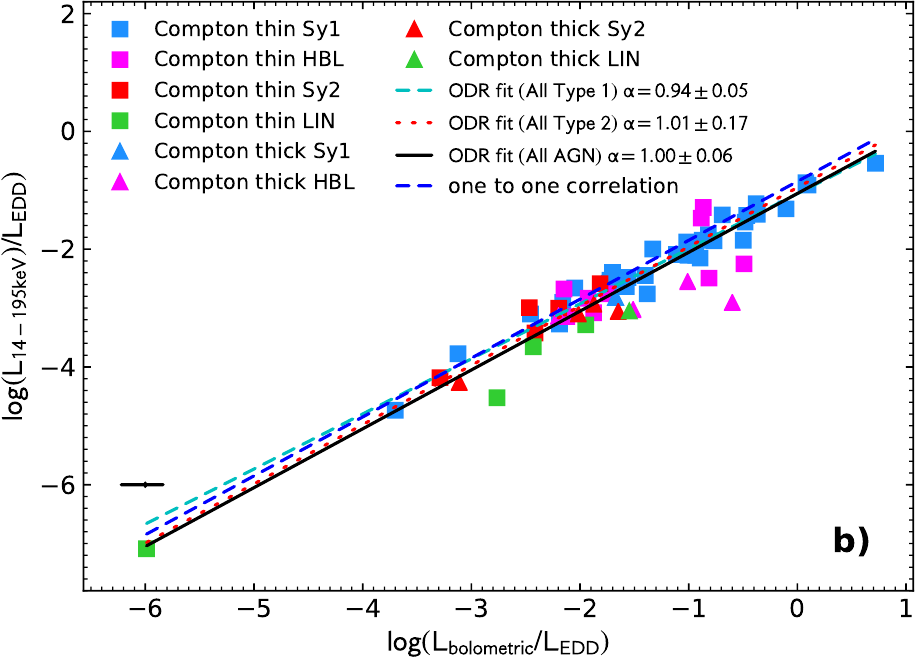}
\caption{{\bf (a:)} Ratio of the 2-10\,keV corrected luminosity to the Eddington luminosity as a function of the ratio of the bolometric luminosity to the Eddignton luminosity. {\bf (b:)} Ratio of the (10-195)\,keV observed luminosity to the Eddington luminosity as a function of the ratio of the bolometric luminosity to the Eddington luminosity.
\label{fig:x2-10_x14-195_lboledd}}
\end{figure*}

We show in Fig.\ref{fig:L12_Loiv_lboledd}a and b the correlations of the 12$\mu$m and [OIV]\,26$\mu$m line luminosities normalized to the Eddington luminosity (L$_{\rm EDD}$) as a function of the ratio of L$_{BOL}/L_{EDD}$.
We show in Fig.\ref{fig:x2-10_x14-195_lboledd}a and b the correlations of the 2-10\,keV and 14-195\,keV  luminosities normalized to the Eddington luminosity (L$_{EDD}$) as a function of the ratio of L$_{BOL}/L_{EDD}$.
We show in Fig.\ref{fig:L12_Loiv_lboledd}a and b the correlations of the composite (12$\mu$m and [OIV]26$\mu$m) and (12$\mu$m and [NeV]14.3$\mu$m) luminosities to the Eddington luminosity (L$_{EDD}$) as a function of the ratio of L$_{BOL}/L_{EDD}$.
We present in Table \ref{tab:cortot2} the results of the correlations of the 12$\mu$m, [OIV]\,26$\mu$m line, 2-10\,keV and 14-195\,keV luminosities normalized to the Eddington luminosity, as well as the combined continuum and line luminosities. 

\begin{figure*}[ht!]
\includegraphics[width = 0.5\textwidth]{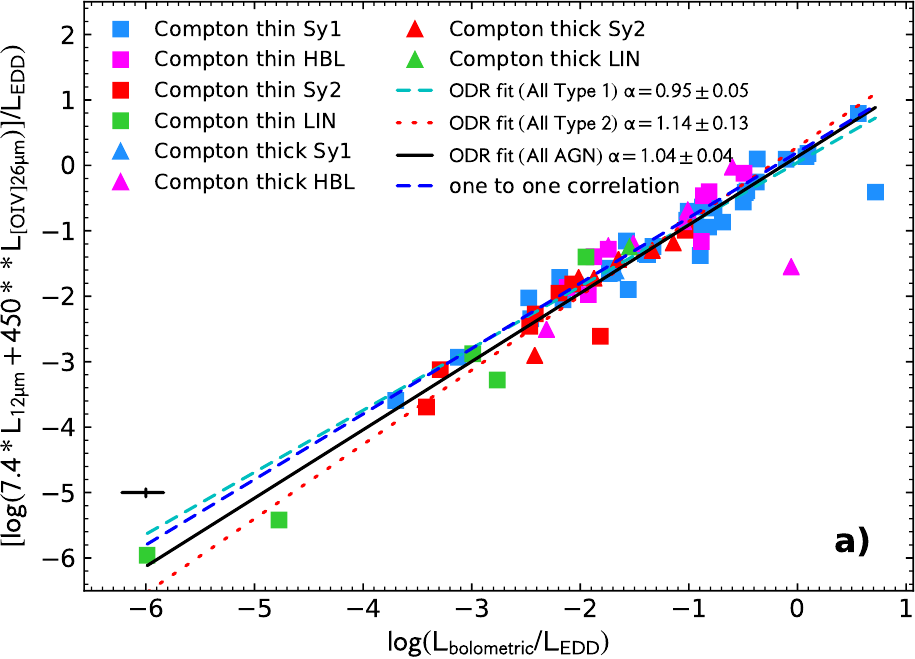}~
\includegraphics[width = 0.5\textwidth]{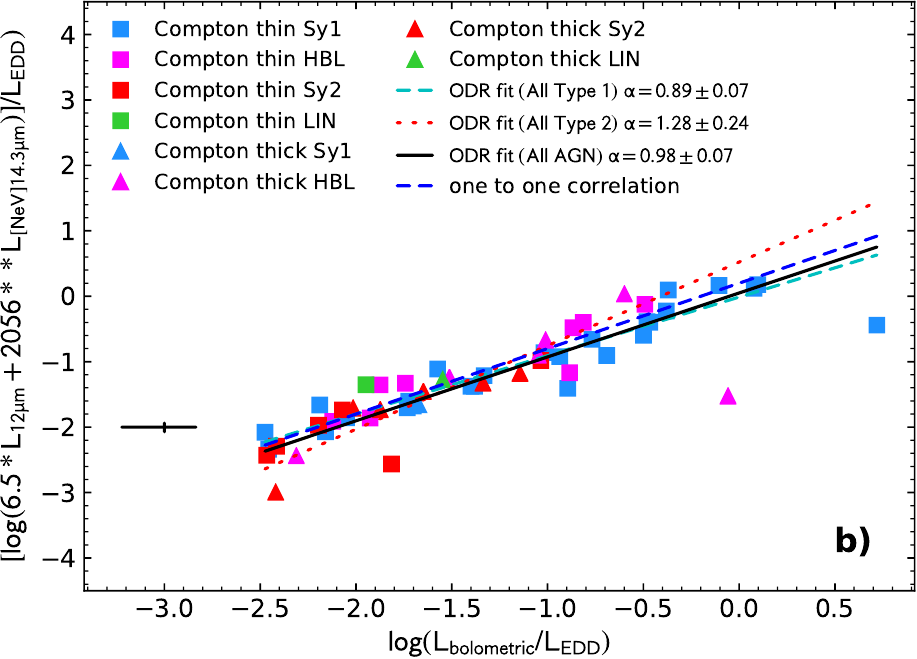}
\caption{{\bf (a:)} { Ratio of the composite 12$\mu$m  and [OIV]26$\mu$m luminosity to the Eddington luminosity as a function of the ratio of the bolometric luminosity to the Eddington luminosity. {\bf (b:)} Ratio of the composite 12$\mu$m and [NeV]14.3$\mu$m luminosity to the Eddington luminosity as a function of the ratio of the bolometric luminosity to the Eddington luminosity.}
\label{fig:L12+oiv_L12+nev1_lboledd}}
\end{figure*}

\begin{table*}[ht!!!]
\centering
\setlength{\tabcolsep}{3.pt}
\caption{Correlation of various bands luminosities normalized to the Eddington luminosity with the bolometric luminosities, normalized to the Eddington luminosity. 
}\label{tab:cortot2}
\footnotesize
\begin{tabular}{llclccc}
    \hline\\[-0.3cm]
\bf Considered variables &\bf Subset  &  \bf N &  \bf $\rho$~~~~~ (P$_{\rm null}$)  & \bf $\beta$ $\pm\ \Delta \beta$ & \bf $\alpha$ $\pm\ \Delta \alpha$  & $\sigma$ \\ [0.02cm]
 (1)            & (2)       & (3)           & (4)     & (5)      &  (6)   & (7)  \\[0.05cm]
\hline\\[-0.25cm]
$L^{\rm nuc}_{\rm 12 \mu m}/\rm{L}_{\rm EDD}$\ vs. $L_{\rm bol}/\rm{L}_{\rm EDD}$\  & All & 71  &  0.95 (9.6 $\times 10^{-36}$) & -0.84 $\pm$ 0.08 & {\bf 1.09 $\pm$ 0.04 } & 0.075 \\
 "~~~~~~~~~~~~~~~~~~~~"~~~~~~~~~~~~" & Type~1 & 41 &  0.94 (2.9 $\times 10^{-20}$) &  -0.92 $\pm$ 0.08 & {\bf 1.01 $\pm$ 0.05}  & 0.051 \\
 "~~~~~~~~~~~~~~~~~~~~"~~~~~~~~~~~~" & Type~2 & 15 &  0.91  (2.3  $\times 10^{-6}$) & -0.39 $\pm$ 0.36 & {\bf 1.36 $\pm$ 0.16}  & 0.056 \\
$L_{\rm [OIV]26 \mu m}/\rm{L}_{\rm EDD}$\ vs. $L_{\rm bol}/\rm{L}_{\rm EDD}$\  & all & 77 &  0.90  (8.5 $\times 10^{-29}$) &  -3.16  $\pm$ 0.11 & {\bf 1.04 $\pm$ 0.06}  & 0.131 \\
 "~~~~~~~~~~~~~~~~~~~~"~~~~~~~~~~~~" & Type~1 & 46 &  0.85 (6.4 $\times 10^{-14}$) &  -3.27 $\pm$ 0.13 & {\bf 0.96 $\pm$ 0.08}  & 0.125 \\
 "~~~~~~~~~~~~~~~~~~~~"~~~~~~~~~~~~" & Type~2 & 15 &  0.88 (1.2  $\times 10^{-5}$) & -3.00 $\pm$ 0.32 & {\bf 1.04 $\pm$ 0.14}  & 0.069 \\
$L_{\rm 2-10 keV}/\rm{L}_{\rm EDD}$\ vs. $L_{\rm bol}/\rm{L}_{\rm EDD}$\ & all & 79 &  0.96 (3.0 $\times 10^{-46}$) & -1.27 $\pm$ 0.06 & {\bf 1.04 $\pm$ 0.03}  & 0.046 \\
 "~~~~~~~~~~~~~~~~~~~~"~~~~~~~~~~~~" & Type~1 & 46 &  0.94 (4.0 $\times 10^{-22}$) & -1.29 $\pm$ 0.09 & {\bf 1.06 $\pm$ 0.06}  & 0.055 \\
 "~~~~~~~~~~~~~~~~~~~~"~~~~~~~~~~~~" & Type~2 & 16 &  0.98 (1.0  $\times 10^{-10}$) & -1.02 $\pm$ 0.15 & {\bf 1.10 $\pm$ 0.06}  & 0.014 \\
$L_{\rm 14-195 keV}/\rm{L}_{\rm EDD}$\ vs. $L_{\rm bol}/\rm{L}_{\rm EDD}$\  & all & 60 &  0.95 (1.0 $\times 10^{-30}$) & -1.06 $\pm$ 0.08 & {\bf1.00 $\pm$ 0.04}  & 0.059 \\
 "~~~~~~~~~~~~~~~~~~~~"~~~~~~~~~~~~" & Type~1 & 41 &  0.95 (5.3 $\times 10^{-21}$) & -1.06 $\pm$ 0.07 & {\bf 0.94 $\pm$ 0.05}  & 0.039 \\
 "~~~~~~~~~~~~~~~~~~~~"~~~~~~~~~~~~" & Type~2 &  9 &  0.90 (9.1  $\times 10^{-4}$) & -0.95 $\pm$ 0.41 & {\bf 1.01 $\pm$ 0.17}  & 0.037 \\
\hline
$(7.4 \times L^{\rm nuc}_{\rm 12 \mu m} + 450 \times L_{\rm [OIV]26 \mu m})/\rm{L}_{\rm EDD}$\ vs. $L_{\rm bol}/\rm{L}_{\rm EDD}$\ & all & 69 &  0.95 (7.2 $\times 10^{-36}$) & 0.13 $\pm$ 0.08 & {\bf 1.04 $\pm$ 0.04}  & 0.063 \\
 "~~~~~~~~~~~~~~~~~~~~"~~~~~~~~~~~~" & Type~1 & 41 &  0.95 (5.7 $\times 10^{-21}$) & 0.04 $\pm$ 0.07 & {\bf 0.95 $\pm$ 0.05}  & 0.044 \\
 "~~~~~~~~~~~~~~~~~~~~"~~~~~~~~~~~~" & Type~2 & 14 &  0.92 (3.5  $\times 10^{-6}$) & 0.28 $\pm$ 0.29 & {\bf 1.14 $\pm$ 0.13}  & 0.049 \\
$(6.5\times L^{\rm nuc}_{\rm 12 \mu m} + 2056 \times L_{\rm [NeV]14.3 \mu m})/\rm{L}_{\rm EDD}$\ vs. $L_{\rm bol}/\rm{L}_{\rm EDD}$\  & all & 55 &  0.88 (1.5 $\times 10^{-18}$) & 0.05 $\pm$ 0.10 & {\bf 0.98 $\pm$ 0.07}  & 0.074 \\
 "~~~~~~~~~~~~~~~~~~~~"~~~~~~~~~~~~" & Type~1 & 34 &  0.92 (2.4 $\times 10^{-14}$) & -0.016 $\pm$ 0.09 & {\bf 0.89 $\pm$ 0.07}  & 0.050 \\
 "~~~~~~~~~~~~~~~~~~~~"~~~~~~~~~~~~" & Type~2 & 12 &  0.83 (7.2  $\times 10^{-4}$) & 0.52 $\pm$ 0.47 & {\bf 1.28 $\pm$ 0.24}  & 0.053 \\
\hline
\end{tabular}\\[0.2cm] 
\begin{tablenotes}
\footnotesize
\item \textbf{Notes.} Fit results. The columns give for each correlation: (1) variables; (2) Subset of the sample on which the fit was computed: ``all'' indicates the entire sample and Type~1 and Type~2 the Seyfert type subsets; (3) Number of sources; (4) Pearson  correlation coefficient $\rho$ (1: completely correlated variables, 0: uncorrelated variables) with the relative null hypothesis (zero correlation) probability; (5) and (6): Parameters of the linear regression fit using the equation: 
${\rm log(L_y) = \beta + \alpha \times log(L_x)}$; (7): residual variance of the fit $\sigma$. 
\end{tablenotes}
\end{table*}

\clearpage
\newpage

\section{Flux-flux correlations}\label{corflux}

\begin{figure*}[ht!]
\includegraphics[width = 0.5\textwidth]{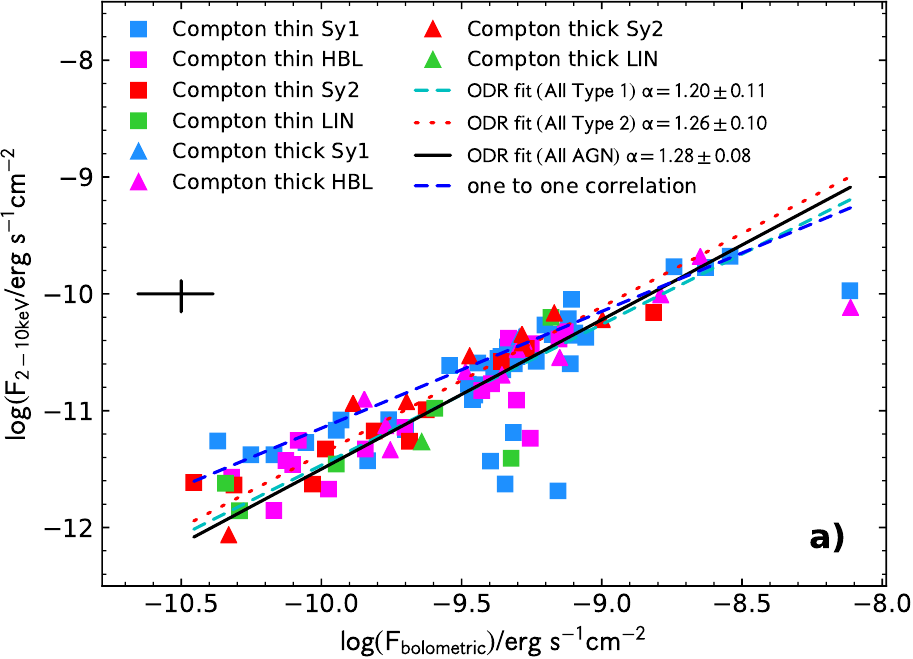}~
\includegraphics[width = 0.5\textwidth]{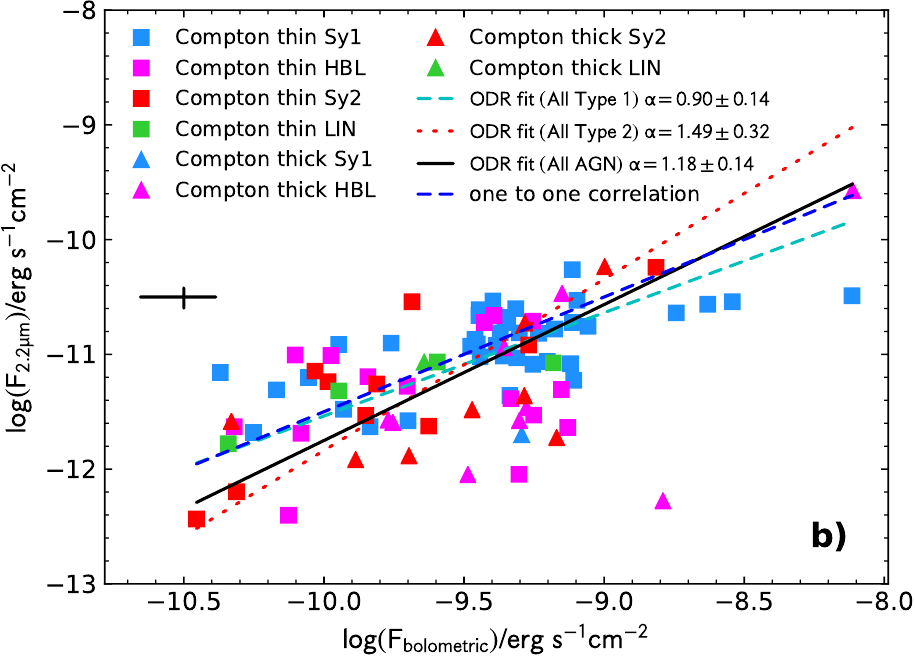}
\caption{{\bf (a:)} 2-10\,keV flux as a function of the  estimated bolometric flux. {\bf (b:)} Nuclear \textit{K} band flux as a function of the estimated bolometric flux.
\label{fig:F2-10_FK_total}}
\end{figure*}

We show in Fig.\ref{fig:F2-10_FK_total}a and b the correlations of the 2-10\,keV flux and the K-band 2.2$\mu$m flux density with the bolometric flux.

We present in Table \ref{tab:cortot} the results of the correlations of the 12$\mu$m, [OIV]\,26$\mu$m line, 2-10\,keV and K-band fluxes with the computed bolometric flux. 

\begin{table*}[ht!!!]
\centering
\setlength{\tabcolsep}{3.pt}
\caption{Correlation of various bands fluxes with the bolometric fluxes.
}\label{tab:cortot}
\footnotesize
\begin{tabular}{llclccc}
    \hline\\[-0.3cm]
\bf Considered variables &\bf Subset  &  \bf N &  \bf $\rho$~~~~~ (P$_{\rm null}$)  & \bf $\beta$ $\pm\ \Delta \beta$ & \bf $\alpha$ $\pm\ \Delta \alpha$  & $\sigma$ \\ [0.02cm]
 (1)            & (2)       & (3)           & (4)     & (5)      &  (6)   & (7)  \\[0.05cm]
\hline\\[-0.25cm]
F(12$\mu$m)$_{\rm nuc}$\ vs. F(bolometric)\ & All   & 83  &  0.69 (3.5 $\times 10^{-13}$) & 1.27 $\pm$ 1.11 & {\bf 1.23 $\pm$ 0.12 }  & 0.077\\
 "~~~~~~~~~~~~~~~~~~~~"~~~~~~~~~~~~" & Type~1 & 48 &  0.60 (7.4 $\times 10^{-6}$) &  -2.11 $\pm$ 1.21 & {\bf 0.87 $\pm$ 0.13}  &  0.056 \\
 "~~~~~~~~~~~~~~~~~~~~"~~~~~~~~~~~~" & Type~2 & 17 &  0.73 (8.9  $\times 10^{-4}$) &  3.25 $\pm$ 2.90 & {\bf 1.45 $\pm$ 0.30}  &  0.082 \\
F([OIV]26$\mu$m)\ vs. F(bolometric)\  & all & 94 &  0.51  (1.9 $\times 10^{-7}$) &  -0.36  $\pm$ 1.60 & {\bf 1.37 $\pm$ 0.17}  & 0.133 \\
 "~~~~~~~~~~~~~~~~~~~~"~~~~~~~~~~~~" & Type~1 & 57 &  0.44 (7.2 $\times 10^{-4}$) &  -0.97 $\pm$ 2.00 & {\bf 1.23 $\pm$ 0.21}  & 0.128 \\
 "~~~~~~~~~~~~~~~~~~~~"~~~~~~~~~~~~" & Type~2 & 18 &  0.62 (5.7  $\times 10^{-3}$) & -0.66 $\pm$ 2.91 & {\bf 1.24 $\pm$ 0.30}  & 0.097 \\
F(2-10keV)\ vs. F(bolometric)\       & all & 95   &  0.84 (2.3 $\times 10^{-26}$) &  1.29 $\pm$ 0.75 & {\bf 1.28 $\pm$ 0.08}  & 0.045 \\
 "~~~~~~~~~~~~~~~~~~~~"~~~~~~~~~~~~" & Type~1 & 57 &  0.81 (4.4 $\times 10^{-14}$) & 0.58 $\pm$ 1.01 & {\bf 1.20 $\pm$ 0.11}  & 0.047 \\
 "~~~~~~~~~~~~~~~~~~~~"~~~~~~~~~~~~" & Type~2 & 18 &  0.95 (1.7  $\times 10^{-9}$) & 1.21 $\pm$ 0.98 & {\bf 1.26 $\pm$ 0.10}  & 0.014 \\
 F(2.2$\mu$m)$_{\rm nuc}$\ vs. F(bolometric)\  & all & 88 &  0.56 (1.1 $\times 10^{-8}$) & 0.10 $\pm$ 1.29 & {\bf 1.18 $\pm$ 0.14} & 0.110 \\
 "~~~~~~~~~~~~~~~~~~~~"~~~~~~~~~~~~" & Type~1 & 54 &  0.54 (2.1 $\times 10^{-5}$) & -2.53 $\pm$ 1.29 & {\bf 0.90 $\pm$ 0.14}  & 0.089 \\
 "~~~~~~~~~~~~~~~~~~~~"~~~~~~~~~~~~" & Type~2 & 18 &  0.70 (1.2  $\times 10^{-3}$) & 3.09 $\pm$ 3.11 & {\bf 1.49 $\pm$ 0.32}  & 0.088 \\
\hline
\end{tabular}\\[0.2cm] 
\begin{tablenotes}
\footnotesize
\item \textbf{Notes.} Fit results. The columns give for each correlation: (1) variables; (2) Subset of the sample on which the fit was computed: ``all'' indicates the entire sample and Type~1 and Type~2 the Seyfert type subsets; (3) Number of sources; (4) Pearson  correlation coefficient $\rho$ (1: completely correlated variables, 0: uncorrelated variables) with the relative null hypothesis (zero correlation) probability; (5) and (6): Parameters of the linear regression fit using the equation: ${\rm log(F_y) = \beta + \alpha \times log(F_x)}$; (7): residual variance of the fit $\sigma$. 
\label{tab:fit}
\end{tablenotes}
\end{table*}


\end{document}